\documentclass[journal]{IEEEtran}
\usepackage{latexsym}
\usepackage{cite}
\usepackage{amssymb}
\usepackage{bbm}
\usepackage{textpos}
\usepackage{enumerate}
\usepackage{bm}
\usepackage{mathtools}
\usepackage[usenames,dvipsnames]{xcolor}
\usepackage{stmaryrd}
\usepackage{amsmath, amssymb,accents}
\usepackage{dsfont}
\usepackage{algorithmicx}
\usepackage{algorithm} 
\usepackage[noend]{algpseudocode} %
\usepackage{comment}
\usepackage{accents}
\usepackage[dvips]{graphics}
\DeclareMathOperator{\supp}{supp}
\usepackage{graphicx}
\usepackage{lipsum}
\usepackage{psfrag}
\usepackage{units}
\usepackage{subfig}
\usepackage{setspace}
\usepackage{theorem}
\usepackage{float}
\usepackage{mathtools}
\allowdisplaybreaks

\newtheorem{definition}{Definition}

\newtheorem{theorem}{Theorem}
\newtheorem{lemma}{Lemma}
\newtheorem{remark}{Remark}

\newtheorem{corollary}{Corollary}
\newtheorem{proposition}{Proposition}

\newcommand\numberthis{\addtocounter{equation}{1}\tag{\theequation}}
{\renewcommand{\arraystretch}{1.2}} 
\begin{document}

\makeatletter
\newcommand{\vasti}{\bBigg@{3.5 }}
\newcommand{\vast}{\bBigg@{4}}
\newcommand{\Vast}{\bBigg@{5}}
\newcommand{\Vastt}{\bBigg@{7}}
\makeatother
\newcommand{\be}{\begin{equation}}
\newcommand{\ee}{\end{equation}}
\newcommand{\ba}{\begin{align}}
\newcommand{\ea}{\end{align}}
\newcommand{\baa}{\begin{align*}}
\newcommand{\eaa}{\end{align*}}
\newcommand{\bea}{\begin{eqnarray}}
\newcommand{\eea}{\end{eqnarray}}
\newcommand{\beaa}{\begin{eqnarray*}}
\newcommand{\eeaa}{\end{eqnarray*}}
\newcommand{\p}[1]{\left(#1\right)}
\newcommand{\pp}[1]{\left[#1\right]}
\newcommand{\ppp}[1]{\left\{#1\right\}}
\newcommand{\ber}{$\ \mbox{Ber}$}
\newcommand{\mkv}{-\!\!\!\!\!\minuso\!\!\!\!\!-}
\newcommand{\fillater}[1] {{\Large \color{red} #1}}
\newcommand{\argmin}{\mathop{\mathrm{argmin}}}
\newcommand{\argmax}{\mathop{\mathrm{argmax}}}
\newcommand{\ubar}[1]{\underaccent{\bar}{#1}}

\makeatletter
\def\Ddots{\mathinner{\mkern1mu\raise\p@
\vbox{\kern7\p@\hbox{.}}\mkern2mu
\raise4\p@\hbox{.}\mkern2mu\raise7\p@\hbox{.}\mkern1mu}}
\makeatother

\title{Information Storage in the Stochastic Ising Model}

\author{Ziv Goldfeld, Guy Bresler, and Yury Polyanskiy

		\thanks{This work of Z. Goldfeld and Y. Polyanskiy was supported in part by the National Science Foundation CAREER award under grant agreement CCF-12-53205, by the Center for Science of Information (CSoI), an NSF Science and Technology Center under grant agreement CCF-09-39370, and a grant from Skoltech--MIT Joint Next Generation Program (NGP). The work of Z. Goldfeld was also supported by the National Science Foundation Grant CCF-1947801, the 2020 IBM Academic Award, and the Rothschild postdoc fellowship. The work of G. Bresler was supported by
		the ONR N00014-17-1-2147, DARPA W911NF-16-1-0551, and NSF CCF-1565516. 
		\newline This paper was presented in part at the 2018 IEEE International Symposium on Information Theory (ISIT-2018), Vail, Colorado, US, in part at the 2019 IEEE International Symposium on Information Theory (ISIT-2019), Online, and in part at the 2020 Beyond IID Conference (BIID-2020), Online. 
		\newline Z. Goldfeld is with the School of Electrical and Computer Engineering, Cornell University, Ithaca, NY 14850, USA (email: goldfeld@cornell.edu). G. Bresler and Y. Polyanskiy are with the Department of Electrical Engineering and Computer Science, Massachusetts Institute of Technology, Cambridge, MA 02139 USA (e-mails: guy@mit.edu, yp@mit.edu).}}

\maketitle



\begin{abstract}

Most information storage devices write data by modifying the local state of matter, in the hope that sub-atomic local interactions stabilize the state for sufficiently long time, thereby allowing later recovery. Motivated to explore how temporal evolution of physical states in magnetic storage media affects their capacity, this work initiates the study of information retention in locally-interacting particle systems. The system dynamics follow the stochastic Ising model (SIM) over a 2-dimensional $\sqrt{n}\times\sqrt{n}$ grid. The initial spin configuration $X_0$ serves as the user-controlled input. The output configuration $X_t$ is produced by running $t$ steps of Glauber dynamics. Our main goal is to evaluate the information capacity $I_n(t):=\max_{p_{X_0}}I(X_0;X_t)$ when time $t$ scales with the system's size $n$. While the positive (but low) temperature regime is our main interest, we start by exploring the simpler zero-temperature dynamics. 

We first show that at zero temperature, order of $\sqrt{n}$ bits can be stored in the system indefinitely by coding over stable, striped configurations. While $\sqrt{n}$ is order optimal for infinite time, backing off to $t<\infty$, higher orders of $I_n(t)$ are achievable. First, linear coding arguments imply that $I_n(t) = \Theta(n)$ for $t=O(n)$. To go beyond the linear scale, we develop a droplet-based achievability scheme that reliably stores $\Omega\left(n/\log n\right)$ for $t=O(n\log n)$ time ($\log n$ can be replaced with any $o(n)$ function). Moving to the positive but low temperature regime, two main results are provided. First, we show that an initial configuration drawn from the Gibbs measure cannot retain more than a single bit for $t\geq \exp(C\beta n^{1/4+\epsilon})$ time. On the other hand, when scaling time with the inverse temperature $\beta$, the stripe-based coding scheme (that stores for infinite time at zero temperature) is shown to retain its bits for $e^{c\beta}$.

\end{abstract}




\section{Introduction}\label{SEC:Intro}


\subsection{New Model for Storing Information Inside Matter}

As modern information processing systems are fueled by massive accumulation of data, the need for storage technologies with enhanced capacity is pressing. Key players in storage electronics, such as Western Digital and Seagate, are developing methods designed to meet these demands based on ideas such as shingled magnetic recording (SMR) \cite{wood2009feasibility}, heat assisted magnetic recording (HAMR) \cite{seigler2008integrated}, and bit-patterned media (BPM) \cite{terris2005nanofabricated}. The latter two, for instance, drastically reduce the area magnetic mediums allocate for storing each bit: from roughly 20-30 magnetic grains per bit in today's off-the-shelf hard-drives to a \emph{single} magnetic grain per bit via BPM or HAMR. However, while shrinkage of magnetization domains increases storage capacity, it also entails new challenges in stabilizing the written data long enough to allow later recovery. Specifically, in the time between writing and reading, the stored data (a configuration of states at which the particles comprising the medium are initialized) dissipates due to interparticle interactions driven by quantum/thermal fluctuations. Despite the significance of this physical phenomenon for future magnetic hard-drive designs, it is largely overlooked in typical models for studying reliable data storage.

		
\begin{figure}[t!]
	\begin{center}
		\begin{psfrags}
			\psfragscanon
			\psfrag{A}[][][1]{\ $m$}
			\psfrag{B}[][][1]{Enc}
			\psfrag{C}[][][1]{\ $X_0$}
			\psfrag{D}[][][0.8]{\ \ \ \ \ \ \ \ \ $\begin{array}{cc}
			     &  t\mbox{ steps of}\\
			     & \mbox{Glauber}\\
			     & \mbox{dynamics}
			\end{array}$}
			\psfrag{E}[][][1]{\ $X_t$}
			\psfrag{F}[][][1]{Dec}
			\psfrag{G}[][][1]{\ \ $\hat{m}$}
			\includegraphics[scale = .475]{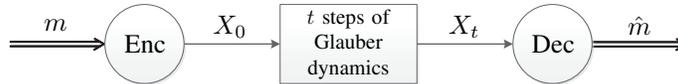}
			\caption{A new model for storage: The encoder maps the data $m$ into the initial Ising configuration $X_0$. The channel applies $t$ steps of Glauber dynamics. The decoder tries to recover the data from the resulting configuration~$X_t$.} \label{FIG:Ising_intro}
			\psfragscanoff
		\end{psfrags}
	\end{center}
\end{figure}
		




A common approach towards magnetic storage uses run-length limited (RLL) codes \cite{Franaszek_RLL1970,Immink_runlength1990} to mitigate unwanted patterns. On top of that, a certain (stationary, with respect to (w.r.t.) the sequence of input bits) error model is adopted to describe the relation between written and read data. Error-correcting codes are used to mitigate these errors. Maximum distance separable (MDS) codes, such as Reed-Solomon codes \cite{RS_Codes1960}, are preferable from several aspects, e.g., their resilience to erasures. This approach reduces the storage problem to that of coding over a memoryless noisy channel. While the conversion to the well-understood channel coding problem is helpful, it does not account for physical phenomena concerning the system's evolution in time.  

Motivated to explore how temporal evolution of physical states in magnetic storage media affects their capacity, this work initiates the study of information retention in locally-interacting particle systems. The Ising model is adopted to describe the magnetic substance \cite{Ising_Model1925}, and the system is evolved in time via Glauber dynamics \cite{Glauber_Dynamics1963}. This framework, known as the \emph{stochastic Ising model} (SIM), is an idealized description of nonequilibrium ferromagnetic statistical mechanics~\cite{Friedli_Statistical_Mechanics2017}. The proposed storage setup is illustrated in Fig. \ref{FIG:Ising_intro}. 



A graph is chosen to model the topology of the storage medium, i.e., the neighboring relations between the particles. A spin system at inverse temperature $\beta$ and nearest-neighbor interactions over that graph comprises the proposed model for the magnetic storage device. Each particle can take one of two states, spin up or down, labeled by $\pm 1$, respectively. The encoder controls the initial spin configuration $X_0$. Running $t$ steps of Glauber dynamics produces the output configuration $X_t$, from which the decoder tries to recover the written information. The fundamental quantity we consider is the information capacity
\begin{equation}
    I_n^{(\beta)}(t):=\max_{p_{X_0}}I(X_0;X_t).\label{EQ:Information_Capacity_Intro}
\end{equation}
This quantity has the desired operational meaning, as it approximately characterizes the number of bits that can be stored in the system for $t$ time. Our goal is to explore how \eqref{EQ:Information_Capacity_Intro} scales when $t$ and $n$ grow together at various rates.




\subsection{The Stochastic Ising Model}

The SIM at inverse temperature $\beta$ is a reversible Markov chain (MC) whose stationary distribution is the Gibbs measure \cite[Chapter 15]{Perez_Mixing_Times2017}. At each step, the discrete-time chain picks a site uniformly at random and refresh its spin according to the conditional Gibbs distributions given the rest of the system. Consequently, spins have a tendency to align, favoring similar values at adjacent sites. The lower the temperature is, the stronger the interactions between neighboring spins becomes. The SIM was extensively studied since its introduction by Glauber in 1963 \cite{Glauber_Dynamics1963}. Works on the topic are diverse, varying between the study of mixing times \cite{Lubetzky_Mixing2010,Diaconis_Mixing2011} (see also \cite[Chapter 15]{Perez_Mixing_Times2017}), phase transition \cite{Ising_Model1925,Onsager_Ising1944,Brush_Ising_History1967}, metastability \cite{Capocaccia_Metastability1974,Davies_Metastability1982,Neves_Metastability1991,Cirillo_Metastability1998}, and many more.

Of particular relevance to this work is the zero-temperature SIM on the two-dimensional (2D) square lattice. Taking the limit of $\beta\to\infty$, the transition rule amounts to a majority update: the updated site adopts the spin of the majority of its neighbors, if a majority exists, and flips a fair coin otherwise. This process has been studied in the physics literature as a model of `domain coarsening' (see, e.g., \cite{Spirin_Kinetic_Ising2001}): clusters of constant spin shrink, grow, split or coalesce, as their boundaries evolve. The particular question of the disappearance time of an all-plus droplet in a sea of minuses is tightly related to the storage problem considered herein.  In \cite{Lacoin_Anisotropic2014}, it was shown that any such convex droplet disappears within time proportional to its area. This result, known as the Lifshitz law \cite{lifshitz1962kinetics} will be used to analyze one of the storage schemes we later propose. 

\subsection{The Storage Problem and Contributions}


Although not practically meaningful as a model for particle interactions in physical matter, a natural first case-study is the complete graph $K_n$ on $n$ vertices, also known as the Curie-Weiss model. In the high-temperature regime, when $\beta<\frac{1}{n}$, the Glauber chain exhibits fast mixing of time $O(n\log n)$ \cite{Aizenman_FastMix_Complete1987}. Hence, not even a single bit can be stored in this system for a long (e.g., exponential) time. In the low temperature regime ($\beta>\frac{1}{n}$), at first glance, the situation seems better: an exponential mixing time was established in \cite{Griffiths_exponential_mixing1966}. The equilibrium distribution here concentrates around two values, corresponding to positive and negative magnetization\footnote{The magnetization is the normalized sum of all the spins.}. The exponential mixing time is a consequence of the bottleneck between these two phases. In \cite{Levin_Complete_Graph2010}, it was shown that if the chain is restricted to states of, e.g., non-negative magnetization, mixing happens in $O(n\log n)$ time. From the storage perspective, this means that while a single bit can be encoded in the magnetization for exponential time, any data beyond that dissolves after order of $n\log n$ updates. Once again, a pessimistic conclusion.


\subsubsection{\underline{Positive Temperature}}

Planar topologies capture the structure of real-life storage devices. We focus here on lattices, and in particular on the 2D square grid of $n$ vertices. At infinite temperature ($\beta=0$), interactions are eliminated and, upon selection, particles flip with probability $\frac{1}{2}$, independent of their surrounding. Taking $t=cn$, the grid essentially becomes an $n$-fold binary-symmetric channel (BSC) with flip probability $\frac{1}{2}\left(1-e^{-c/4}\right)$. As this flip probability is  arbitrarily close to $\frac{1}{2}$ when $c$ is large, the per-site capacity becomes negligible. One of our main interest is to understand whether interactions (i.e., $\beta>0$) enhance the system's storage capacity. Classical results on the 2D Ising model phase transition and mixing times~\cite{martinelli1999lectures} imply the following: for $\beta < \beta_c$, where $\beta_c=\frac{1}{2}\log\big(1+\sqrt{2}\big)$, we have $I_n^{(\beta)}\big(\mathrm{poly}(n)\big) = 0$. On the other hand, $I_n^{(\beta)}\big(\exp(\sqrt{n})\big) \geq 1$ when  $\beta>\beta_c$.\footnote{The 2D SIM on the grid mixes within $O(n\log n)$ time when $\beta<\beta_c$, and exhibits exponential mixing time of $e^{\Omega(\sqrt{n})}$, when $\beta>\beta_c$ \cite{martinelli1999lectures}.}

To understand whether anything beyond a single bit can be stored in the SIM on the 2D grid at $\beta>\beta_c$ consider two (stochastic) trajectories, one starting from an all-plus state and another one from some fixed $\sigma$. Evolving them jointly using the synchronous coupling, we let $p_t(\sigma)$ be the probability that the trajectories have coupled by time $t$. \cite{Martinelli_PhaseCoex1994} shows that $p_t(\sigma)$ averaged over all $\sigma$ sampled from the Gibbs distribution conditioned on positive magnetization converges to one for $t\geq \exp(C\beta n^{1/4+\epsilon})$. This time corresponds to the mixing time of a SIM on a 2D grid with a plus boundary condition, recently improved by \cite{Lubetzky_Mixing2010} to $n^{O(\log n)}$, and conjectured to be order $n$ (up to logarithmic terms). We point out that this does not resolve the question we posed, but suggests that if a state that does not couple with the all-plus trajectory exists, it should not be a typical one w.r.t. the Gibbs distribution. To quantify this statement, we show that for sufficiently large $\beta>0$ and $X_0$ drawn from the Gibbs distribution, $I(X_0;X_t)\leq \log2+o(1)$ for $t\geq \exp(C\beta n^{1/4+\epsilon})$. The positive temperature regime also enables scaling time with $\beta$ (instead of $n$). We show that $\sqrt{n}$ bits stored into monochromatic horizontal or vertical stripes, are decodable via majority after $t\sim \exp(c\beta)$. Key to this observation is a new result on the survival time of a single plus-labeled stripe in a sea of minuses. Even with these understandings, however, an order optimal characterization of the information capacity for general $\beta>0$ seems challenging. 

\begin{table*}[!t]
\begin{center}
\renewcommand\arraystretch{1.2}
\begin{tabular}{|c|c|c|}
\hline
Time & Information Capacity & Comments\\ \hline\hline
$t=0$ & $I_n(t)=n$ & Upper bound for all $t$
\\ \hline
$t=O(n)$ &  $I_n(t)=\Theta(n)$ & Linear discrete-time $=$ Constant `physical' time\\\hline
\begin{tabular}{c}
$t=a(n)\cdot n$\\where $a(n)$ is $o(n)$
\end{tabular}
& $I_n(t)=\Omega\left(\frac{n}{a(n)}\right)$ & \begin{tabular}{c}
$t=n\log n\ \implies\ I_n(t)=\Omega\left(\frac{n}{\log n}\right)$  \\$t=n^{1+\alpha},\ \alpha\in\left[0,\frac{1}{2}\right]\ \implies I_n(t)=\Omega(n^{1-\alpha})$
\end{tabular}
\\\hline
\begin{tabular}{c}
$t\to\infty$\\independent of $n$
\end{tabular}
& $\lim_{t\to\infty}I_n(t)=\Theta(\sqrt{n})$ & Lower bound for all $t$ \\\hline
\end{tabular}
\caption{Summary of main results for data storage in the zero-temperature SIM on the 2D grid}\label{TABLE:result_summary}
\end{center}
\end{table*}


\subsubsection{\underline{Zero Temperature}} The extreme case of the zero-temperature dynamics turns out to be more amenable for analysis. 
While the SIM is a reversible MC for any $\beta<\infty$, at zero temperature the chain becomes absorbing. This happens since the update rule simplifies to a majority vote, thereby giving rise to configurations that, once entered, cannot be left (e.g., the two ground states). This enables storing information for infinite time, and brings us to the information capacity $I_n(t):=\lim_{\beta\to \infty}I_n^{(\beta)}(t)$ at the limit of $t\to\infty$. This quantity not only answers the question of `how much information can be stored in the system indefinitely?', but it also lower bounds $I_n(t)$, for all $t$, due to the data processing inequality (DPI).

We establish an order-optimal characterization of $\lim_{t\to\infty}I_n(t)=\Theta(\sqrt{n})$. This is done by identifying the set of absorbing configurations as those with a (horizontal or vertical) striped pattern, with stripes of width at least two. Since the number of stripes grows as $\sqrt{n}$, the result follows by coding over striped configurations (achievability) and the absorbing nature of the chain (converse). Consequently, for any $t$, the information capacity is at least $\Omega(\sqrt{n})$ and (trivially) at most $n$. We then show that the $n$ upper bound is achievable up to a constant factor via linear codes so long that $t$ scales linearly with $n$. The argument is a direct consequence of the Gilbert-Varshamov bound. Consequently, $I_n(t)=\Theta(n)$ in this linear time scale, and the main challenge becomes quantifying $I_n(t)$ in the intermediate regime, i.e., when $t$ is superlinear but finite.\footnote{The importance of storage for superlinear time is evident when considering the `physical' time domain. 
`Physical' time corresponds to a continuous-time SIM where the spin at each site is refreshed according to a Poisson clock of rate 1, independent of the system's size $n$. Since, on average, the discrete-time dynamics updates each site once in every $n$ steps, discrete time is a stretched version of physical time by a factor of $n$. Therefore, linear time in the discrete-time SIM corresponds to \emph{constant} physical time. Storage for superlinear time, on the other hand, translates into a system that stores for longer physical times as its size increases. A storage medium whose stability benefits from increased size is highly desirable from a practical perspective.}

To that end, we develop a droplet-based achievability scheme that stores $\Omega\left(\frac{n}{a(n)}\right)$ for $t=O\big(a(n)n\big)$ time, where $a(n)$ is any $o(n)$ function. Taking $a(n)=n^{\alpha}$, for $\alpha\in[0,0.5]$, we see that this scheme interpolates between the order of $n$ bits achievable for linear time and the $\sqrt{n}$ achievability via the stripe-based scheme. The analysis decomposes the grid into $\Omega\left(\frac{n}{a(n)}\right)$ independent Z-channels and proves each has positive capacity by exploiting the recently established Lifshitz law of phase boundary movement \cite{Lacoin_Anisotropic2014}. We provide a new and simple proof for the Lifshitz law based on stochastic domination and Hopf's Umlaufsatz. Our main results for the zero-temperature grid dynamics are summarized in Table \ref{TABLE:result_summary}.

Finally, we highlight two modifications to the zero-temperature dynamics for which storage performance significantly improves. Introducing an external magnetic field to the grid dynamics results in a tie-braking rule when the neighborhood is balanced. This increases the size of the stable set from $\Theta(\sqrt{n})$ to $\Theta(n)$, and implies that $I_n(t)=\Theta(n)$, uniformly in $t$. The same holds, without an external field, when the grid is replaced with the honeycomb lattice. This is since sites in its interior all have an odd degree (namely, 3), which prohibits ties gives rise to a stable set of size $\Theta(n)$ once more. We conclude that for zero-temperature storage, the honeycomb lattice is favorable over the grid when no external field is applied. If, on the other hand, the grid dynamics are endowed with an (however small) external field, $\lim_{t\to\infty}I_n(t)$ abruptly grows from $\Theta(\sqrt{n})$ to $\Theta(n)$. We stress that this instability of the storage capacity w.r.t. the magnetic field or the lattice strongly relies on the temperature being zero. At any positive temperature ($\beta<\infty$), we expect that $I_n^{(\beta)}(t)$ is of the same order is all three cases.

\subsection{Organization}

The remainder of this paper is organized as follows. Section \ref{SEC:Model} provides notations and defines the SIM. Section \ref{SEC:Operation_Definiton} sets up the operational storage problem and connects the maximal size of reliable codes to the information capacity. For ease of presentation, we start from the zero-temperature dynamics and move to positive temperature towards the end. Thus, Section \ref{SEC:Infinite_Time} studies the asymptotics of $I^{(\infty)}_n(t)$ when $t\to\infty$ independently of $n$. In Section \ref{SEC:Grid_droplet} we focus on reliable storage for superlinear times, which includes the construction and analysis of the droplet-based achievability scheme. This section also states the droplet disappearance time result, for which a simple and new proof is given in Appendix \ref{APPEN:Droplet_Erosion_Time_proof}. Section \ref{SEC:tight_results} considers the grid dynamics with an external field and the dynamics over the honeycomb lattice; in both cases, an order optimal characterization of $I_n(t)=\Theta(n)$, for all $t$, is given. Sections \ref{SEC:positive_temp1} and \ref{SEC:positive_temp2} give preliminary results on information capacity at positive but low temperatures. Finally, Section \ref{SEC:summary} discusses the main insights of the paper and appealing future directions.

\section{The Model}\label{SEC:Model}


\subsection{Notation}\label{SUBSEC:Definitions}

Given integers $k\leq\ell$, set $[k\mspace{-3mu}:\mspace{-3mu}\ell]:=\big\{i\in\mathbb{Z}:\, k\leq i \leq \ell\big\}$; if $k=1$, the shorthand $[\ell]$ is used. Let $G_n=(\mathcal{V}_n,\mathcal{E}_n)$ be a graph with a vertex set $\mathcal{V}_n$, of size $n$, and an edge set $\mathcal{E}_n$. We write $v\sim w$ if $\{v,w\}\in\mathcal{E}_n$, and denote the neighborhood of $v$ by $\mathcal{N}_v:=\{w\in\mathcal{V}_n:\, w\sim v\}$. The graph distance in $G_n$ is denoted by $d:\mathcal{V}_n\times\mathcal{V}_n\to\mathbb{N}_0$, where $\mathbb{N}_0:=\mathbb{N}\cup\{0\}$. For two sets $\mathcal{U},\mathcal{W}\subseteq\mathcal{V}_n$, we define $d(\mathcal{U},\mathcal{W}):=\min_{\substack{u\in\mathcal{U},\\w\in\mathcal{W}}}d(u,w)$.



\subsection{The Stochastic Ising Model}\label{SUBSEC:General_Def_Glauber}

Fix $n\in\mathbb{N}$ and let $\Omega_n:=\{-1,+1\}^{\mathcal{V}_n}$. For every $\sigma\in\Omega_n$ and $v\in\mathcal{V}_n$, $\sigma(v)$ is the value of $\sigma$ at $v$. The Hamiltonian associated with the Ising model on $G_n$ is
\begin{equation}
    \mathcal{H}(\sigma):=-\sum_{\{u,v\}\in\mathcal{E}_n}\sigma(u)\sigma(v),\quad\sigma\in\Omega_n.\label{EQ:Hamiltonian}
\end{equation}
The Gibbs measure at inverse temperature $\beta>0$ and free boundary conditions is
\begin{equation}
\pi_\beta(\sigma)=\frac{1}{Z(\beta)}e^{-\beta \mathcal{H}(\sigma)},\label{EQ:Gibbs_measure}  \end{equation}
where $Z(\beta)$ is the partition function. 

The Glauber dynamics for the Ising model is a discrete-time MC\footnote{The chain can also be set up in continuous-time, which we do in Section \ref{SUBSEC:Droplet_Erosion_Time} to simplify some derivations.} on the state space $\Omega_n$, reversible w.r.t. $\pi_\beta$. At each time step, a vertex $v\in\mathcal{V}_n$ is chosen uniformly at random; the spin at $v$ is refreshed by sampling a new value from $\pi_\beta\big(s\big|\{\sigma(u)\}_{u\neq v}\big)$, where $s\in\{-1,+1\}$. Denote the corresponding transition kernel by $P$ and let $(X_t)_{t\in\mathbb{N}_0}$ be the Glauber chain. We use $\mathbb{P}$ and $\mathbb{E}$ for the corresponding probability measure and expectation, respectively, while $\mathbb{P}_\sigma$ and $\mathbb{E}_\sigma$ indicate a conditioning on $X_0=\sigma$. Given an initial distribution $X_0\sim p_{X_0}$ and any $t\in\mathbb{N}_0$, the pair $(X_0,X_t)$ is distributed according to 
\begin{equation}
    p_{X_0,X_t}(\sigma,\eta):=\mathbb{P}(X_0=\sigma,X_t=\eta)=p_{X_0}(\sigma)P^t(\sigma,\eta),
\end{equation}
where $P^t$ is the $t$-step transition kernel. The $X_t$-marginal of $p_{X_0,X_t}$ is denoted by $p_{X_t}$. To stress that a tuple $(X_r,X_s,X_t)$, with $0\leq r<s<t$, forms a MC we write $X_r\leftrightarrow X_s\leftrightarrow X_t$. The mutual information $I(X_0;X_t)$ is taken w.r.t. $p_{X_0,X_t}$. The entropy of $X_t$ is $H(X_t)$.

Define a \emph{path} w.r.t. $P$ as $\omega:=(\omega_1,\omega_2,\ldots,\omega_k)$, where $k\in\mathbb{N}$, $\omega_i\in\Omega_n$ for $i\in[k]$, such that $P(\omega_i,\omega_{i+1})>0$, for all $i\in[k-1]$. For $\sigma,\sigma'\in\Omega_n$, we write $\sigma \leadsto\sigma'$ if $\sigma'$ is reachable from $\sigma$, i.e., if there exists a path $\omega$ (say of length $k$) with $\omega_1=\sigma$ and $\omega_k=\sigma'$. To specify that a path $\omega$ goes from $\sigma$ to $\sigma'$ we write $\omega:\sigma\leadsto\sigma'$. For $\sigma\in\Omega_n$ and $\mathcal{A}\subseteq\Omega_n$, we write $\omega:\sigma\leadsto\mathcal{A}$ if $\omega:\sigma\leadsto\sigma'$ for some $\sigma'\in\mathcal{A}$.


As a preliminary result, the next proposition states that when $G_n=K_n$ is the complete graph, the SIM has poor storage capability regardless of whether temperature is high or low.


\begin{proposition}[Storage in the Complete Graph]\label{PROP:complete}
Let $G_n=K_n$ be the complete graph on $n$ vertices, denote by $(X_t)_{t\in\mathbb{N}_0}$ the Glauber chain for the Ising model over $K_n$ at inverse temperature $\beta=\alpha/n$, and let $I_n^{(\alpha)}(t):=I^{(\beta n)}_n(t)=\max_{p_{X_0}}I(X_0;X_t)$. We have:
\begin{enumerate}
    \item If $\alpha <1$, then there exists $c>0$ such that $I_n^{(\alpha)}(t)= o(1)$, for any $t\geq c\cdot n\log n$;
    \item If $\alpha >1$, then there exists $c'>0$ such that $I_n^{(\alpha)}(t)\leq \log(2)+o(1)$, for any $t\geq c'\cdot n(\log n)^2$.
\end{enumerate}
\end{proposition}

\begin{IEEEproof}
Proposition \ref{PROP:complete} is a consequence of mixing time results for Curie-Weiss model. The first claim trivially follows because $(X_t)_{t\in\mathbb{N}_0}$ mixes within $O(n\log n)$ time when temperature is high ($\alpha<1$)~\cite{Aizenman_FastMix_Complete1987}.

For the second claim, Theorem 3 of \cite{Levin_Complete_Graph2010} shows that even at low temperature ($\alpha>1$), the mixing time is $O(n\log n)$ if the chain is restricted to configurations of positive magnetization. The restricted dynamics evolves on $\Omega_n^+ :=\left\{\sigma\in\Omega_n:\, m(\sigma)\geq 0\right\}$, where $m(\sigma):=\frac{1}{n}\sum_{i=1}^n\sigma(i)$, as described next. At each step, generate a candidate move~$\sigma$ according to the usual Glauber dynamics. If $m(\sigma)\geq 0$, accept the move; otherwise, move to $-\sigma$. To relate the restricted dynamics to $(X_t)_{t\in\mathbb{N}_0}$, let $\tilde{X}_t:=\{X_t,\bar{X}_t\}$, where $\bar{X}_t$ is obtained from $X_t$ by a global spin flip. In words, $\tilde{X}_t$ identifies $X_t$ and $\bar{X}_t$ with one another. One may verify that $(\tilde{X}_t)_{t\in\mathbb{N}_0}$ is also a MC with the same law as the restricted dynamics.\footnote{This relies on the $G_n$ being the complete graph, which implies the invariance of the update probabilities to permutations.} Since $\tilde{X}_t$ is deterministically specified by $X_t$, for any $X_0\sim p_{X_0}$, we have
\begin{equation}
    I(X_0;X_t)=I(X_0;X_t,\tilde{X}_t)\leq \log(2) + I(\tilde{X}_0;\tilde{X}_t),\label{EQ:tilde_MI_UB1}
\end{equation}
where the inequality is because given $\tilde{X}_t$, $X_t$ takes only two possible values, and since $X_0\leftrightarrow\tilde{X}_0\leftrightarrow\tilde{X}_t$ forms a~MC. 

To control $I(\tilde{X}_0;\tilde{X}_t)$ we use Proposition 12 from \cite{polyanskiy2015dissipation}, which states that 
\begin{align*}
    I(A;B)\leq \big(\log|\mathcal{A}|-1\big)&\big\|P_{A,B}-P_A\otimes P_B\big\|_{\mathsf{TV}}\\&+h\Big(\big\|P_{A,B}-P_A\otimes P_B\big\|_{\mathsf{TV}}\Big),
\end{align*}
where $A$ is supported on the finite set $\mathcal{A}$, $\|P-Q\|_{\mathsf{TV}}$ is the total variation distance between the distributions $P$ and $Q$, and $h:[0,1]\to[0,1]$ is the binary entropy function. Note that $\|P_{\tilde{X}_0,\tilde{X}_t}-P_{\tilde{X}_0}\otimes P_{\tilde{X_t}}\big\|_{\mathsf{TV}}\leq 2d_n(t)$, where $d_n(t):=\max_{\sigma_0\in\Omega_n^{+}}\big\|\tilde{P}^t(\sigma_0,\cdot)-\tilde{\pi}\big\|_{\mathsf{TV}}$, with $\tilde{P}$ and $\tilde{\pi}$ being the transition kernel and the equilibrium distribution of the restricted dynamics. We obtain
\begin{equation}
    I(\tilde{X}_0;\tilde{X}_t)\leq 2nd_n(t)\log(2)+h\big(\min\{2d_n(t),1\}\big)\label{EQ:tilde_MI_UB2}
\end{equation}
By \cite[Theorem 3]{Levin_Complete_Graph2010}, there exists $c'>0$, such that $d_n(t)\leq 1/4$, for all $t>c'n\log n$. This further implies that $d_n(bt)\leq(1/4)^b$, for any $b\in\mathbb{N}$. Thus, if $t>c'n(\log n)^2$, then $d_n(t)\leq n^{-2}$, which together with \eqref{EQ:tilde_MI_UB1} and \eqref{EQ:tilde_MI_UB2} completes the proof.
\end{IEEEproof}

\begin{remark}[Refinement of Low Temperature Result] 
The proof of \cite[Theorem 3]{Levin_Complete_Graph2010} does not seem to establish the convergence rate of $d_n(t)$ (just that $d_n(t)\leq 1/4$, for $t\gtrsim n\log n$). This is the reason we used the $d_n(bt)\leq(1/4)^b$ trick for $b=\log n$, producing the $\log(2)$ upper bound for $t>c'n(\log n)^2$. Provided a refined mixing time result for the restricted dynamics that shows any superlinear rate of decay for $d_n(t)$, Claim (2) from Proposition \ref{PROP:complete} is immediately strengthened from $t>c'\cdot n(log n)^2$ to $t>c'\cdot n\log n$.
\end{remark}

\subsection{Zero-Temperature Dynamics on the Grid}\label{SUBSEC:Glauber_Def}

From now on, unless stated otherwise, let $G_n=(\mathcal{V}_n,\mathcal{E}_n)$ be a square grid of side length $\sqrt{n}\in\mathbb{N}$, where $\mathcal{V}_n=\big\{(i,j)\big\}_{i,j\in[\sqrt{n}]}$. \footnote{For simplicity, we assume $\sqrt{n}\in\mathbb{N}$; if $\sqrt{n}\notin\mathbb{N}$, simple modification of some of the subsequent statements using ceiling and/or floor operations are needed. Regardless, our focus is on the asymptotic regime as $n\to\infty$, where the assumption that $\sqrt{n} \in \mathbb{N}$ has no effect.} We interchangeably use $v\in\mathcal{V}_n$ or $(i,j)\in\mathcal{V}_n$ for $v\in\mathcal{V}_n$. The zero-temperature SIM is obtained by taking $\beta\to\infty$ in the corresponding Glauber dynamics. This results in a majority update of the uniformly chosen site. To give a formal description we need some more definitions. 

For $\sigma\in\Omega_n$ and $v\in\mathcal{V}_n$, let $\sigma^v$ denote the configuration that agrees with $\sigma$ everywhere except $v$: 
\begin{equation*}
    \sigma^v(u):=\begin{cases}\sigma(u),\quad\ \ u\neq v \\ -\sigma(u),\quad u=v\end{cases}.
\end{equation*}
The all-plus and the all-minus configurations are denoted by $\boxplus$ and $\boxminus$, respectively. Let $m_v(\sigma):=\big|\big\{w\in\mathcal{N}_v:\,\sigma(w)=\sigma(v)\big\}\big|$ be the number of $v$'s neighbors whose spin agrees with $\sigma(v)$, and set $\ell_v(\sigma):= |\mathcal{N}_v|-m_v(\sigma)$. 

When at state $\sigma\in\Omega_n$, the zero-temperature SIM on $G_n$ with free boundary conditions transitions as follows:
\begin{enumerate}
    \item Pick a vertex $v\in\mathcal{V}$ uniformly at random.
    \item Modify $\sigma$ at $v$ as follows:
    \begin{itemize}
        \item If $m_v(\sigma)>\ell_v(\sigma)$, keep the value of $\sigma(v)$;
        \item If $m_v(\sigma)<\ell_v(\sigma)$, flip the value of $\sigma(v)$;
        \item Otherwise, draw $\sigma(v)$ uniformly from $\{-1,+1\}$ independently of everything else. 
    \end{itemize}
\end{enumerate}
Sections \ref{SEC:Operation_Definiton} to \ref{SEC:tight_results} study the zero-temperature dynamics, for which, with some abuse of notation, we recast $P$, $(X_t)_{t\in\mathbb{N}_0}$, $\mathbb{P}$, $\mathbb{E}$, $\mathbb{P}_\sigma$, $\mathbb{E}_\sigma$ as their transition kernel, Glauber chain, probability measure, expectation, and conditional versions of the latter two. When we move to discuss the positive temperature regime in Sections \ref{SEC:positive_temp1} and \ref{SEC:positive_temp2}, notation will be set up anew.

\section{Operational versus Information Capacity}\label{SEC:Operation_Definiton}

We mainly focus is on the asymptotic dependence on $n$ and $t$ of the information capacity 
\begin{equation}
    I_n(t)= \max_{p_{X_0}}I(X_0;X_t).\label{EQ:Information_Capacity}
\end{equation}
As the problem is motivated by coding for storage, this section describes the operational setup and establishes $I_n(t)$ as a fundamental quantity for its study. The rest of the paper deals with $I_n(t)$ without referring back to the operational setup.



\begin{figure}[t!]
	\begin{center}
		\begin{psfrags}
			\psfragscanon
			\psfrag{A}[][][1]{\ $m$}
			\psfrag{B}[][][1]{$\mspace{3mu}{}f^{(t)}_n$}
			\psfrag{C}[][][1]{\ $X_0$}
			\psfrag{D}[][][1]{\ \ \ \ \ \ \ \ \ \ \ \ $\mathsf
			{SIC}_n(t)$ }
			\psfrag{E}[][][1]{\ $X_t$}
			\psfrag{F}[][][1]{$\mspace{3mu}\phi_n^{(t)}$}
			\psfrag{G}[][][1]{\ \ $\hat{m}$}
			\includegraphics[scale = .475]{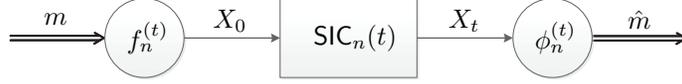}
			\caption{The stochastic Ising channel  $\mathsf{SIC}_n(t)$: The channel transition probability is $P^t$, where $P$ is the one step transition kernel of the SIM on the $\sqrt{n}\times\sqrt{n}$ two-dimensional grid at zero-temperature. The functions $f_n^{(t)}$ and $\phi_n^{(t)}$ are, respectively, the encoder and the decoder.} \label{FIG:Stochastic_Ising_channel}
			\psfragscanoff
		\end{psfrags}
	\end{center}
\end{figure}

For fixed $\sqrt{n}\in\mathbb{N}$ and $t\in\mathbb{N}_0$, $P^t$ constitutes a channel from $X_0$ to $X_t$. We call this channel the Stochastic Ising Channel of time $t$, abbreviated as $\mathsf{SIC}_n(t)$. As shown in Fig. \ref{FIG:Stochastic_Ising_channel}, the input $X_0$ is controlled by the encoder, while $X_t$ is the output observed by the decoder. This models a storage device of size $n$ (i.e., with $n$ cells) and $t$ time between writing and reading. Storing a message $m\in[M]$ in this setup is equivalent to reliably communicating it over the $\mathsf{SIC}_n(t)$. Thus, the fundamental limit of reliable storage is defined as the largest possible alphabet size $M$.





\begin{definition}[Code]\label{DEF:SIC_code}
An $(M,n,t,\epsilon)$-code $c_n^{(t)}$ for
the $\mathsf{SIC}_n(t)$ is a pair of maps, the encoder $f_n^{(t)}:[M]\to\Omega_n$ and the decoder $\phi_n^{(t)}:\Omega_n\to[M]$, satisfying
\begin{equation*}
    \frac{1}{M}\sum_{m\in[M]}\mathbb{P}_{f_n^{(t)}(m)}\left(\phi_n^{(t)}(X_t)\neq m\right)\leq \epsilon.
\end{equation*}
\end{definition}

\begin{definition}[Maximal Code Size]
The maximal code size for the $\mathsf{SIC}_n(t)$ attaining error probability at most $\epsilon$ is
\begin{equation*}
    M^\star(n,t,\epsilon)\mspace{-2mu}=\mspace{-2mu}\max\mspace{-2mu}\Big\{M\mspace{-2mu}: \exists \mbox{ an }(M,n,t,\epsilon)\mbox{-code for }\mathsf{SIC}_n(t) \Big\}\mspace{-2mu}.
\end{equation*}
\end{definition}

The next proposition relates $M^\star(n,t,\epsilon)$ and $I_n(t)$; see Appendix \ref{APPEN:operational_informational_proof} for the proof.

\begin{proposition}[Operational vs. Information Capacity]\label{PROP:operational_informational}
The following bounds on $M^\star(n,t,\epsilon)$ hold:
\begin{enumerate}
    \item \underline{Upper Bound:} For any $n\in\mathbb{N}$, $t\geq 0$ and $\epsilon>0$, we have
\begin{equation*}
\log M^\star(n,t,\epsilon)\leq \frac{1}{1-\epsilon}\Big(I_n(t)+h(\epsilon)\Big),
\end{equation*}
where $h:[0,1]\to[0,1]$ is the binary entropy function.

\item \underline{Lower Bound:} Let $n_1=o(n)$. For any $t\geq 0$ and $\epsilon>0$, we have
\begin{equation*}
\frac{1}{n}\mspace{-2mu}\log\mspace{-3mu} M^\star\mspace{-2mu}\left(\mspace{-2mu}n\mspace{-2mu}+\mspace{-2mu}o\mspace{-2mu}\left(\frac{n}{\sqrt{n_1}}\right)\mspace{-2mu},t,\epsilon\right)\mspace{-2mu}\geq\mspace{-2mu} \frac{I_{n_1}(t)}{n_1}-\sqrt{\frac{n_1}{n(1-\epsilon)}}.
\end{equation*}
\end{enumerate}
\end{proposition}
To interpret item (2), let $\alpha\in(0,1)$ and $n_1=n^{1-\alpha}$. This gives
\begin{equation*}
\frac{1}{n}\mspace{-2mu}\log\mspace{-2mu} M^\star\mspace{-2mu}\left(\mspace{-2mu}n\mspace{-2mu}+\mspace{-2mu}o\left(n^{\frac{1+\alpha}{2}}\right)\mspace{-2mu},t,\epsilon\right)\mspace{-2mu}\geq\mspace{-2mu}\frac{I_{n^{1-\alpha}}(t)}{n^{1-\alpha}}-\sqrt{\frac{1}{n^\alpha(1-\epsilon)}}
\end{equation*}
and approximates the operational capacity $M^\star_n$ in terms of the information capacity of $G_{n^{1-\alpha}}$, for $\alpha$ however small. 

\begin{remark}[Channel Degradation with Time]
For any $s\leq t$, $\mathsf{SIC}_n(t)$ is a degraded version of $\mathsf{SIC}_n(s)$. Therefore, $I_n(t)\leq I_n(s)$, and any storage scheme designed to retain the date for $t$ time is also valid for times~$s\leq t$. 
\end{remark}


\section{Infinite-Time Capacity}\label{SEC:Infinite_Time}


We first consider the infinite-time capacity  $I_n(\infty):=\lim_{t\to\infty}I_n(t)$, for fixed $n\in\mathbb{N}$. The DPI implies that $I_n(t)\geq I_n(s)$, for any $0\leq t<s$. Thus, $I_n(\infty)$ lower bounds~$I_n(t)$ uniformly in $t$, and serves as a benchmark for finite-time storage schemes proposed later on.


To characterize $I_n(\infty)$, we identify $(X_t)_{t\in\mathbb{N}_0}$ as an absorbing MC. The absorbing/stable set is shown to comprise striped configurations, of which there are $2^{\Theta(\sqrt{n})}$. This implies that $I_n(\infty)=\Theta(\sqrt{n})$, where achievability follows by coding only over stripes, while the converse relies on the being absorbing.

\subsection{Characterization of Stable Configurations}\label{SUBSEC:Stable_Configuration}


\begin{definition}[Stable Configurations]
A configuration $\sigma\in\Omega_n$ is stable if $P(\sigma,\sigma)=1$. The set of all stable configurations is denoted by $\mathcal{S}_n$.
\end{definition}

Clearly, $\mathcal{S}_n\neq\emptyset$ since $\{\boxminus,\boxplus\}\subset \mathcal{S}_n$. However, $\mathcal{S}_n$ contains many more configurations than the ground states. We show that $\mathcal{S}_n$ equals the set of all (vertically or horizontally) striped configurations, i.e., configurations that partition the grid into monochromatic stripes of width at least 2. The formal definition is as follows.

\begin{definition}[Striped Configuration]\label{DEF:striped}
A configuration $\sigma\in\Omega_n$ is \emph{horizontally striped} if there exist $\ell\in\mathbb{N}$ and integers $1< j_1< j_2<\ldots<j_{\ell-1}< j_\ell= \sqrt{n}$, with $j_{k+1}\geq j_k+2$ for every $k\in[\ell-1]$, such that
\begin{equation*}
\sigma\big((i,j)\big)\mspace{-2mu}=\mspace{-2mu}\sigma\big((1,j_k)\big),\ \forall i\mspace{-2mu}\in\mspace{-2mu}\left[\mspace{-2mu}\sqrt{n}\right],\, j_{k-1}\mspace{-1mu}+\mspace{-1mu}1\mspace{-1mu}\leq\mspace{-1mu} j\mspace{-1mu}\leq\mspace{-1mu} j_k,\, k\mspace{-2mu}\in\mspace{-2mu}[\ell],
\end{equation*}
where $j_0=0$. 

Similarly, a configuration $\sigma\in\Omega_n$ is \emph{vertically striped} if there exist $\ell\in\mathbb{N}$ and integers $1< i_1< i_2<\ldots<i_{\ell-1}< i_\ell= \sqrt{n}$, with $i_{k+1}\geq i_k+2$ for every $k\in[\ell-1]$, such that
\begin{equation*}
\sigma\big((i,j)\big)\mspace{-2mu}=\mspace{-2mu}\sigma\big((i_k,1)\big),\,\forall i_{k-1}\mspace{-1mu}+1\leq\mspace{-1mu} i\leq \mspace{-1mu}i_k\mspace{-1mu},\, j\mspace{-2mu}\in\mspace{-2mu}\left[\sqrt{n}\right],\, k\mspace{-2mu}\in\mspace{-2mu}[\ell],
\end{equation*}
where $i_0=0$. 

Finally, a \emph{striped} configuration is either vertically or horizontally striped; $\mathcal{A}_n$ is the set of all striped configurations. 
\end{definition}

The next proposition (proven in Appendix \ref{APPEN:number_stable_proof}) counts~$|\mathcal{A}_n|$. 

\begin{proposition}[Number of Striped Configurations]\label{Prop:number_stable}
Fix $\sqrt{n}\in\mathbb{N}$. We have $|\mathcal{A}_n|=4f_{\sqrt{n}-1}$, where $\{f_k\}_{k\in\mathbb{N}_0}$ is the Fibonacci sequence on the indexing where $f_0=0$ and $f_1=1$. Namely, setting $\phi=-1/\psi=(1+\sqrt{5})/2$, we have
\begin{equation}
    |\mathcal{A}_n|=\frac{4}{\sqrt{5}}\left(\phi^{\sqrt{n}-1}-\psi^{\sqrt{n}-1}\right).\label{EQ:num_of_stripes}
\end{equation}
\end{proposition}

The main result of this subsection, proven in Appendix \ref{APPEN:stable_iff_striped_proof}, characterizes the stable configurations as striped:

\begin{theorem}[Stable $\iff$ Striped]\label{TM:stable_iff_striped}
A configuration $\sigma\in\Omega_n$ is stable if and only if it is striped, i.e., $\mathcal{S}_n=\mathcal{A}_n$.
\end{theorem}


\subsection{Characterization of Infinite-Time Capacity}\label{SUBSEC:inifite_time}

Having Theorem \ref{TM:stable_iff_striped} we now state the main result for $I_n(\infty)$. 

\begin{theorem}[Infinite Time]\label{TM:Grid_capacity}
For the $\sqrt{n}\times\sqrt{n}$ zero-temperature SIM, we have $I_n(\infty)=\log|\mathcal{A}_n|$, where $|\mathcal{A}_n|$ is given in \eqref{EQ:num_of_stripes}. In particular, $I_n(\infty)=\Theta(\sqrt{n})$.
\end{theorem}

The proof of Theorem \ref{TM:Grid_capacity} relies on $(X_t)_{t\in\mathbb{N}_0}$ being an absorbing MC. Namely, if $\mathcal{T}_n:=\Omega_n\setminus\mathcal{S}_n$, we say that $(X_t)_{t\in\mathbb{N}_0}$ is an absorbing MC if for any $\rho\in\mathcal{T}_n$ there exist $t(\rho)\in\mathbb{N}$ such that $P^{t(\rho)}(\rho,\mathcal{S}_n)=\sum_{\sigma\in\mathcal{S}}P^{t(\rho)}(\rho,\sigma)>0$. 

\begin{lemma}[Absorbing Markov Chain]\label{LEMMA:absorbing_MC}
$(X_t)_{t\in\mathbb{N}_0}$ is an absorbing MC, and consequently,
\begin{equation}
    \lim_{t\to\infty}\max_{\sigma\in\Omega_n}\ \mathbb{P}_\sigma\big(X_t\notin\mathcal{S}_n\big)=0.\label{EQ:absorbed_wp1}
\end{equation}
\end{lemma}
We prove that $(X_t)_{t\in\mathbb{N}_0}$ is absorbing (Appendix \ref{APPEN:absorbing_MC_proof}) by constructing a path from any $\rho\in\mathcal{T}_n$ to some $\sigma\in\mathcal{S}_n$. Having this, \eqref{EQ:absorbed_wp1} is immediate (see, e.g., \cite[Chapter 11]{Into_Probability_Grinstead2012}). We are now ready to prove Theorem \ref{TM:Grid_capacity}.
\ \\

\begin{IEEEproof}[Proof of Theorem \ref{TM:Grid_capacity}]
For the lower bound, let $X_0\sim\mathsf{Unif}(\mathcal{S}_n)$. Since $P(\sigma,\sigma)=1$ for all $\sigma\in\mathcal{S}_n$, we have
\begin{equation}
    I_n(t)\geq H(X_0)=\log|\mathcal{S}_n|,\quad\forall t\in\mathbb{N}_0.\label{EQ:mutual_information_LB}
\end{equation}

To upper bound $I_n(t)$, for any $X_0\sim p_{X_0}$, let $E_t:=\mathds{1}_{\{X_t\in\mathcal{S}_n\}}$ and denote $p_t:=\max_\sigma\mathbb{P}(X_t\notin\mathcal{S}_n)$. Let $t\in\mathbb{N}_0$ be sufficiently large so that $p_t\in[0,1/2]$ and consider:
\begin{align*}
    I(X_0;X_t)&= I(X_0;E_t,X_t)\\
              &\leq H(E_t)+\mathbb{P}(X_t\in\mathcal{S}_n)I(X_0;X_t|X_t\in\mathcal{S}_n)\\
              &\qquad\qquad\qquad +\mathbb{P}(X_t\notin\mathcal{S}_n)I(X_0;X_t|X_t\notin\mathcal{S}_n)\\
              &\stackrel{(a)}\leq h(p_t)+\log|\mathcal{S}_n|+np_t\numberthis\label{EQ:mutual_information_UB}
    \end{align*}
where (a) uses $H(X_t|X_t\in\mathcal{S}_n)\leq\log|\mathcal{S}_n|$, $H(X_t|X_t\notin\mathcal{S}_n)\leq\log|\Omega_n|$ and the fact that the binary entropy $h$ is monotonically increasing inside $[0,1/2]$. Combining \eqref{EQ:mutual_information_LB}-\eqref{EQ:mutual_information_UB} and taking the limit as $t\to\infty$, while using \eqref{EQ:absorbed_wp1} and Proposition \ref{Prop:number_stable}, concludes the proof.\end{IEEEproof}



\section{Storage for Superlinear Time}\label{SEC:Grid_droplet}

The previous section showed that $I_n(t)=\Omega(\sqrt{n})$ for all $t\in\mathbb{N}_0$ by coding over striped configurations. We next address the question of whether one can achieve higher information capacity, and if so, for which time scales? We start with the following observation.

\begin{proposition}[Order Optimality for Linear Time]\label{PROP:linear_time}
Fix $\epsilon>0$. For any $t<(1/4-\epsilon)n$, we have $I_n(t)=\Theta(n)$.
\end{proposition}

\begin{IEEEproof}
The upper bound follows by $I_n(t)\mspace{-2mu}\leq\mspace{-2mu}I_n(0)\mspace{-2mu}=\mspace{-2mu}n$, for all $t\geq 0$. To show that $I_n(t)=\Omega(n)$, for any $t<(1/4-\epsilon)n$ and $\epsilon>0$, we use the Gilbert-Varshamov bound. It states that there exist error-correcting codes of rate $1-h(1/4-\epsilon)>0$ and minimum distance $d>(1/4+\epsilon)n$. This minimum distance is greater than the number of elapsed time-steps and hence also greater than the number of spin flips, making decoding possible. Thus, so long that $t<(1/4-\epsilon)n$, we have $I_n(t)>n\cdot \big[1-h(1/4-\epsilon)-o(1)\big]=\Omega(n)$. \end{IEEEproof}


To go beyond linear time, we propose a storage scheme that decomposes $G_n$ into independent sub-squares (droplets), each capable of reliably storing a bit for $t=\omega(n)$. Separating the droplets by all-minus stripes of width 2, we decorrelate the dynamics inside them, enabling a tensorization-based analysis. 

Consider a square droplet of positive spins with side-length $\ell\in\mathbb{N}$ surrounded by a sea of minuses (see Fig. \ref{FIG:droplet}). Evolving the SIM from this configuration, we show that at least one positive spin survives for $t\sim n\ell^2$. For instance, if $\ell= n^{\frac{\alpha}{2}}$, for $\alpha\in(0,1)$, the corresponding erosion time is $\Theta(n^{1+\alpha})$. Since the $\sqrt{n}\times\sqrt{n}$ grid can be padded with roughly $n^{1-\alpha}$ properly spaced droplets, we will deduce that $I_n(t)=\Omega(n^{1-\alpha})$, for $t=O(n^{1+\alpha})$. Other growth rates are of interest too.   


\subsection{Monochromatic Droplets}


		
\begin{figure}[t!]
	\begin{center}
		\begin{psfrags}
			\psfragscanon
			\psfrag{A}[][][1]{$\sqrt{n}$}
			\psfrag{B}[][][1]{$\ell$}
			\psfrag{C}[][][1]{\ }
			\psfrag{D}[][][1]{\ }
			\includegraphics[scale = .4]{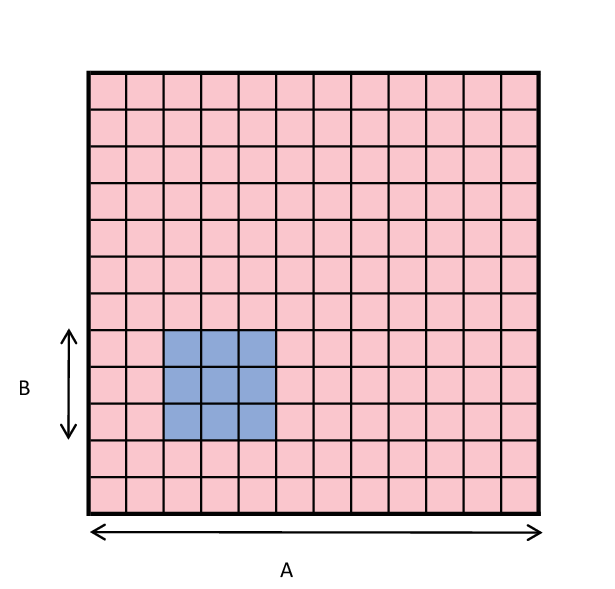}
			\caption{A square droplet of size $\ell\times\ell$. The spins inside the square are $+1$ (shown in blue) and those outside are $-1$ (red).} \label{FIG:droplet}
			\psfragscanoff
		\end{psfrags}
	\end{center}
\end{figure}
		

Fix $\ell\leq \sqrt{n}-2$, and let $\mathcal{R}_\ell\subset\Omega_n$ be the set of configurations with a single $\ell\times\ell$ square droplet (Fig.~\ref{FIG:droplet}). Namely, all spins are -1 except for those inside the square, which are +1. The graph distance between the square and the boundary of the grid is assumed to be at least one.


Consider a system initiated at $X_0=\rho\in\mathcal{R}_\ell$ and let $\tau:=\inf\{t\in\mathbb{N}_0:\,X_t=\boxminus\}$ be the hitting time of the all-minus configuration $\boxminus$. Observe that the plus-labeled vertices cannot exceed the original $\ell\times\ell$ box as the system evolves. Thus, the only stable configuration reachable from $\mathcal{R}_\ell$ is $\boxminus$. 
To approximate $\tau$ (with high probability), it is convenient to consider the continuous-time dynamics. As described next, continuous-time results translate back to discrete time via Poisson approximation. 



\vspace{2mm}
\subsubsection{\underline{Continuous-Time Zero-Temperature Dynamics}}\label{SUBSEC:Continuous_Time_Dynamics_Def}

We present the standard setup for this chain \cite[Chapter 20]{Perez_Mixing_Times2017}. Recall that $P$ is the transition kernel of the discrete-time dynamics $(X_t)_{t\in\mathbb{N}_0}$. The continuous-time process $\big(X_t^{(c)}\big)_{t\geq 0}$ is defined by setting $X_0^{(c)}=X_0$ and $X_t^{(c)}=X_{N_t}$, for $t>0$, where $(N_t)_{t\geq 0}$ is a rate 1 Poisson process independent of $(X_t)_{t\in\mathbb{N}_0}$. The transition probabilities of $\big(X_t^{(c)}\big)_{t\geq 0}$ are described by its heat-kernel:
\begin{equation*}
    H_t(\sigma,\eta):=\sum_{k=0}^\infty\frac{e^{-t}t^k}{k!}P^k(\sigma,\eta)=\mathbb{P}_\sigma\Big(X_t^{(c)}=\eta\Big),
\end{equation*}
where the 2nd step uses $\mathbb{P}_\sigma\left(X_t^{(c)}=\eta\middle|N_t=k\right)=P^k(\sigma,\eta)$.


For any $\sqrt{n}\in\mathbb{N}$ and $t\geq 0$, let $I_n^{(c)}(t):=\max_{p_{X^{(c)}_0}} I\left(X_0^{(c)};X_t^{(c)}\right)$. The following proposition states that $I_n^{(c)}(t)$ and $I_n(t)$ are of the same order, so long as $t$ scales super-logarithmically with $n$.

\begin{proposition}[Discrete- vs. Continuous-Time Dynamics]\label{PROP:discrete_continuous_relation} For any $\epsilon\in(0,1)$, we have
\begin{align*}
    &I_n^{(c)}\big((1+o(1))t\big)-n\cdot e^{-\Theta(t^{\epsilon})}\leq I_n(t) \\&\leq\left(1-e^{-\Theta(t^\epsilon)}\right)^{-1}\Big[I_n^{(c)}\big((1-o(1))t\big)+h\left(e^{-\Theta(t^{\epsilon})}\right)\Big],
\end{align*}
where $h:[0,1]\to[0,1]$ is the binary entropy function.
\end{proposition}
The proof is relegated to Appendix \ref{APPEN:discrete_continuous_relation_proof}. Since the droplet scheme is designed for $t=\omega(n)$, Proposition \ref{PROP:discrete_continuous_relation} allows analyzing $I_n^{(c)}(t)$ instead of $I_n(t)$.


\begin{remark}[Equivalent Representation]
The MC $\big(X_t^{(c)}\big)_{t\geq 0}$ can be equivalently defined through its generator $\mathfrak{L}$, which operates on functions $f:\Omega_n\to\mathbb{R}$ as
\begin{equation}
    \mathfrak{L}f(\sigma)=\sum_{v\in\mathcal{V}_n}c_{v,\sigma}\big[f(\sigma^v)-f(\sigma)\big],\label{EQ:Droplet_generator}
\end{equation}
where $c_{v,\sigma}:= P(\sigma,\sigma^v)$ (see Appendix \ref{APPEN:continuout_time_equivalence} for details). The corresponding interpretation is that, when in state $\sigma\in\Omega_n$, each site $v\in\mathcal{V}_n$ is assigned with an independent Poisson clock of rate $c_{v,\sigma}$. When the clock at site $v$ rings, the spin at this site flips. Under this description $I_n^{(c)}(t)$ is easier to analyze because the independent Poisson clocks decorrelate non-interacting pieces of the grid, which in turn, tensorizes mutual information. 
\end{remark}

\begin{remark}[Flip Rates Speedup or Slowdown]\label{REM:Speedup_Slowdown} Let $\big(\bar{X}_t^{(c)}\big)_{t\geq 0}$ be a continuous-time zero-temperature dynamics with flip rate $r_{v,\sigma}=n\cdot c_{v,\sigma}$, $v\in\mathcal{V}_n$ and $\sigma\in\Omega_n$. In words, $\big(\bar{X}_t^{(c)}\big)_{t\geq 0}$ is a speedup of $\big(X_t^{(c)}\big)_{t\geq 0}$ by a factor $n$. An immediate consequence of Poisson process properties is that for any $p_{X_0}$, where $X_0^{(c)},\bar{X}_0^{(c)}\sim p_{X_0}$, we have
\begin{equation}
    I\left(\bar{X}_0^{(c)};\bar{X}_t^{(c)}\right)=I\left(X_0^{(c)};X_{nt}^{(c)}\right).\label{EQ:_speedup_slowdown}
\end{equation}
Furthermore, the flip rates of $\big(\bar{X}_t^{(c)}\big)_{t\geq 0}$ are independent of $n$, given by $r_{v,\sigma}\in\left\{0,0.5,1\right\}$, for all $v\in\mathcal{V}_n$ and $\sigma\in\Omega_n$.
\end{remark}

\subsubsection{\underline{Erosion Time of a Square Droplet}}\label{SUBSEC:Droplet_Erosion_Time}

Consider $\big(\bar{X}^{(c)}_t\big)_{t\geq 0}$ defined in Remark \ref{REM:Speedup_Slowdown} initiated at  $\bar{X}_0^{(c)}=\rho\in\mathcal{R}_\ell$. Since the graph distance between the droplet and the borders of the grid is at least 2, the dynamics are agnostic to the system size. Thus, with some abuse of notation, we assume that $\big(\bar{X}^{(c)}_t\big)_{t\geq 0}$ has state space $\Omega_{\mathbb{Z}^2}:=\{-1,+1\}^{\mathbb{Z}^2}$. Define the hitting time to the all-minus configuration $\bar{\tau}:=\inf\big\{t\geq 0:\,\bar{X}^{(c)}_t=\boxminus\big\}$.

A recent result from the zero-temperature SIM literature (known as the Lifshitz law) states that with high probability, $\bar{\tau}$ is of order $\ell^2$ \cite[Theorem 2.2]{Lacoin_Anisotropic2014}. The result from \cite{Lacoin_Anisotropic2014} is more general, showing that the disappearance time of any convex droplet is proportional to its area. In Appendix \ref{APPEN:Droplet_Erosion_Time_proof} we give a simple new proof for the square droplet case (stated next) based on stochastic domination and Hopf's Umlaufsatz. 

\begin{theorem}[Erosion Time of a Square Droplet]\label{TM:Droplet_Erosion_Time} There exist positive constants $c,C,\gamma>0$, such that
\begin{equation}
     \mathbb{P}_\rho\Big(c\ell^2\leq \bar{\tau}\leq C\ell^2\Big)\geq 1 - e^{-\gamma\ell},\quad\ell\geq 1. \label{EQ:Droplet_erosion_time}
\end{equation}
\end{theorem}
\vspace{2mm}
Let $\tau^{(c)}:=\inf\big\{t\geq 0:\,X^{(c)}_t=\boxminus\big\}$ be the hitting time of the all-minus configuration in the original $\big(X_t^{(c)}\big)_{t\geq 0}$ dynamics. Combining Theorem \ref{TM:Droplet_Erosion_Time} and \eqref{EQ:_speedup_slowdown} gives
\begin{equation*}
     \mathbb{P}_\rho\left(cn\ell^2\leq \tau^{(c)}\leq Cn\ell^2\right)\geq 1 - e^{-\gamma\ell}.
\end{equation*}
This inequality key for the analysis of the droplet-based coding scheme presented next.

\begin{remark}[Droplets with Boundary Conditions]\label{REM:Droplet_cases}
Since the bounds from \eqref{EQ:Droplet_erosion_time} depend only on droplet's size, the result of Theorem \ref{TM:Droplet_Erosion_Time} applies also for droplets with boundary conditions. Consider an $\ell\times\ell$ all-plus droplet (that constitutes the entire system) with an all-minus boundary condition around it. Define $\mathcal{U}_\ell:=[0:\ell+1]\times[0:\ell+1]$,  $\mathcal{V}_\ell=[\ell]\times[\ell]$ and let $\mathcal{B}_\ell:=\mathcal{U}_\ell\setminus\Big\{\mathcal{V}_\ell\cup\big\{(0,0),(0,\ell+1),(\ell+1,0),(\ell+1,\ell+1)\big\}\Big\}$ be the boundary. Consider the continuous-time chain with state space $\Omega_\ell:=\{-1,+1\}^{\mathcal{V}_\ell}$, and boundary condition $\tau\in\Gamma_\ell:=\{-1,+1\}^{\mathcal{B}_\ell}$ with $\tau(v)=-1$, for all $v\in\mathcal{B}_\ell$. The flip rate at $v\in\mathcal{V}_\ell$ when in configuration $\sigma\in\Omega_\ell$ and boundary condition $\tau\in\Gamma_\ell$~is
    \begin{equation*}
        b^{(\tau)}_{v,\sigma}=\begin{cases}1, &m_v(\sigma,\tau)<\ell_v(\sigma,\tau)\\\frac{1}{2},& m_v(\sigma,\tau)=\ell_v(\sigma,\tau)\\0,& m_v(\sigma,\tau)>\ell_v(\sigma,\tau)\end{cases}
    \end{equation*}
    where $m_v(\sigma,\tau)$ and $\ell_v(\sigma,\tau)$ are defined analogously to $m_v(\sigma)$ and $\ell_v(\sigma)$, respectively, but with the neighborhood of $v$ being $\mathcal{N}_v:=\big\{w\in\mathcal{V}_\ell\cup\mathcal{B}_\ell:\,\|v-w\|_1=1\big\}$, where $\|\cdot\|_1$ is the $L^1$ norm in $\mathbb{R}^2$. The generator of these dynamics is the same as \eqref{EQ:Droplet_generator}, but with $b^{(\tau)}_{v,\sigma}$ instead of $c_{v,\sigma}$. 
\end{remark}


\subsection{Superlinear Time Storage Scheme}\label{SUBSEC:SupLin_Time}

The following is our main result for superlinear time storage in zero temperature.

\begin{theorem}[Storing for Superlinear Time]\label{TM:SupLin_Time}
Let $a(n)=o(n)$. Then there exists $c>0$, such that for all $t\leq c\cdot a(n)\cdot n$, we have $I_n(t)=\Omega\big(n/a(n)\big)$.
\end{theorem}

\begin{IEEEproof}
We analyze the continuous-time information capacity $I_n^{(c)}(t)$, which by Proposition \ref{PROP:discrete_continuous_relation} is of the same order as $I_n(t)$. The distribution of $X_0^{(c)}$ is constructed as follows. Tile the $\sqrt{n}\times\sqrt{n}$ grid with monochromatic sub-squares of side length $\sqrt{a(n)}$ (whose spins are to be specified later) separated by all-minus stripes of width 2. The tiling is illustrated in Fig. \ref{FIG:Droplet_Scheme} and contains $\Theta\big(n/a(n)\big)$ sub-squares.

\begin{figure}[t!]
	\begin{center}
		\begin{psfrags}
		\psfragscanon
		\psfrag{A}[][][1]{$\mspace{-40mu}\sqrt{a(n)}$}
		\psfrag{B}[][][1]{$\sqrt{n}$}
		\psfrag{C}[][][0.5]{$\ldots$}
		\psfrag{D}[][][0.5]{$\vdots$}
		\psfrag{E}[][][0.7]{$\Ddots$}
		\includegraphics[scale = .35]{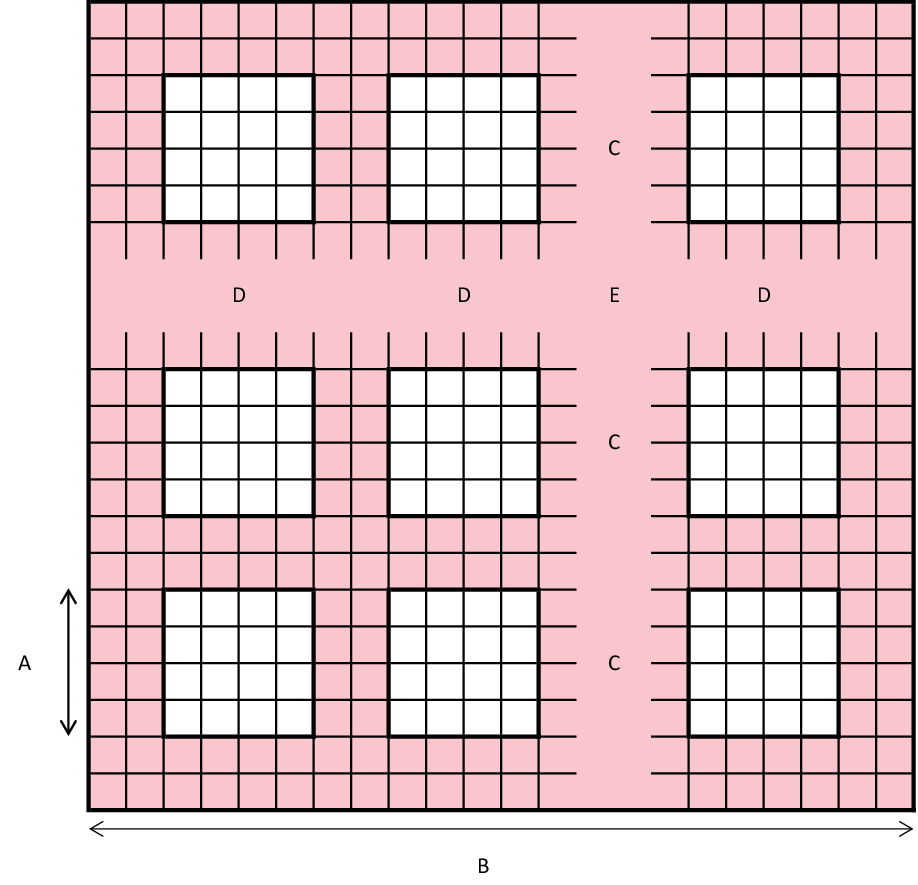}
		\caption{A tiling of the $\sqrt{n}\times\sqrt{n}$ grid by sub-squares of side $\sqrt{a(n)}$ separated by stripes of negative spins of width 2. Red squares represent negative spins, while white squares stand for unspecified spins. Each sub-square will encode a bit by setting all its spins to +1 or -1.} \label{FIG:Droplet_Scheme}
		\psfragscanoff
		\end{psfrags}
	\end{center}
	\vspace{-2mm}
\end{figure}
		
 
Let $\mathcal{C}_n$ be the collection of all configurations with the structure shown in Fig. \ref{FIG:Droplet_Scheme} and a monochromatic spin assignment to each of the sub-squares. The formal definition is as follows. Set $K:=\left\lfloor(\sqrt{n}-2)/(\sqrt{a(n)}+2)\right\rfloor$ and for $i,j\in[K]$, let $\rho_{i,j}\in\mathcal{R}_{\sqrt{a(n)}}$ be the configuration with positive spins inside the $\sqrt{a(n)}\times\sqrt{a(n)}$ sub-square whose top-right corner is $\Big(1+i\big(\sqrt{a(n)}+2\big),1+j\big(\sqrt{a(n)}+2\big)\Big)\in\mathcal{V}_n$; the rest of the spins are -1. Namely, $\rho_{i,j}\big((x,y)\big)=+1$ if and only if $(x,y)\in\mathcal{D}_{i,j}$, where
\begin{equation*}
    \mathcal{D}_{i,j}:=\left\{(x,y)\mspace{-2mu}\in\mspace{-2mu}\mathcal{V}_n:\mspace{-4mu}\begin{array}{c}
    1+i\big(\sqrt{a(n)}+2\big)-\sqrt{a(n)}\leq x\\
    \qquad\qquad\ \ \leq 1+i\big(\sqrt{a(n)}+2\big),\\ 
        1+j\big(\sqrt{a(n)}+2\big)-\sqrt{a(n)}\leq y\\
    \qquad\qquad\ \ \leq 1+j\big(\sqrt{a(n)}+2\big)\end{array}\mspace{-6mu}\right\}\mspace{-2mu}.
\end{equation*}
The external boundary of $\mathcal{D}_{i,j}$ is denoted by
\begin{equation*}
    \mathcal{B}_{i,j}:=\big\{v\in\mathcal{V}_n\setminus\mathcal{D}_{i,j}:\,d(v,\mathcal{D}_{i,j})=1\big\}.
\end{equation*}
Letting $\Xi:=\bigcup_{i,j\in[K]}\{\rho_{i,j}\}$, define
\begin{equation}
    \mathcal{C}_n:=\bigcup_{\mathcal{A}\in 2^{\Xi}}\bigvee_{\rho\in\mathcal{A}}\rho,\label{EQ:droplet_codebook}
\end{equation}
where $2^{\Xi}$ is the power set of $\Xi$, while $\bigvee_{\sigma\in\mathcal{A}}\sigma$ is the `or' operation of all the configurations in $\mathcal{A}$ (for $\sigma,\eta\in\Omega_n$, $(\sigma\vee\eta)(v)=-1$ if and only if $\sigma(v)=\eta(v)=-1$). The collection $\mathcal{C}_n$ serves as our codebook, with each sub-square storing a bit. Accordingly, one may encode $K^2=\Theta\big(n/a(n)\big)$ bits into the initial configuration. We next show that $I^{(c)}_n(t)=\Omega(K^2)$, for $t\leq C\cdot a(n)\cdot n$, with an absolute constant $C>0$.




Let $X_0^{(c)}\sim p_{X_0}$ with $\supp(p_{X_0})=\mathcal{C}_n$ such that $p_{X_0}$ is an identically and independently distributed (i.i.d.) Bernoulli measure in each sub-square (the parameter will be specified later). With this choice, $X_0^{(c)}$ and $X_t^{(c)}$ decompose into $K^2$ independent components, each corresponding to a different sub-square surrounded by all-minus boundary conditions. 
Letting $X_0^{(c)}(i,j)$ and $X_t^{(c)}(i,j)$ be the restriction of $X_0$ and $X_t$ to $\mathcal{D}_{i,j}\cup\mathcal{B}_{i,j}$, we have that $\big\{\big(X_0^{(c)}(i,j),X_t^{(c)}(i,j)\big)\big\}_{i,j\in[K]}$ are i.i.d. Consequently,
\begin{align*}
    I^{(c)}_n(t)&\geq \sum_{i,j\in[K]}\max_{p_{i,j}}I\left(X_0^{(c)}(i,j);X_t^{(c)}(i,j)\right)\\
    &=K^2\cdot\max_{p_{1,1}}I\left(X_0^{(c)}(1,1);X_t^{(c)}(1,1)\right).\numberthis\label{EQ:MI_tenzorization}
\end{align*}
Denote the restriction of $X_0^{(c)}(i,j)$ to $\mathcal{A}\subseteq\mathcal{V}_n$ by $X_0^{(c)}(i,j)\big|_\mathcal{A}$. Each mutual information term in the above sum corresponds to a continuous-time dynamics with state space $\{-1,+1\}^{\mathcal{D}_{i,j}}$, an initial configuration $X_0^{(c)}(i,j)\big|_{\mathcal{D}_{i,j}}\sim p_{i,j}$ with $\supp\left(p_{i,j}\right)=\{\boxminus,\boxplus\}\subset\{-1,+1\}^{\mathcal{D}_{i,j}}$, and boundary condition $X_0^{(c)}(i,j)\big|_{\mathcal{B}_{i,j}}=\boxminus\in\{-1,+1\}^{\mathcal{B}_{i,j}}$. 

Based on \eqref{EQ:MI_tenzorization}, to prove Theorem \ref{TM:SupLin_Time} it suffices to show that there exists $c>0$ such that  
\begin{equation*}
    \max_{p_{1,1}}I\left(X_0^{(c)}(1,1);X_t^{(c)}(1,1)\right)>0,\quad \forall t\leq c\cdot a(n)\cdot n.
\end{equation*}
This indeed holds since $X_0^{(c)}(1,1)$ and $X_t^{(c)}(1,1)$ are related through a Z-channel of positive capacity. To see this, define the map $\phi:\{-1,+1\}^{\mathcal{D}_{1,1}}\to\{-1,+1\}^{\mathcal{D}_{1,1}}$~as
\begin{equation*}
    \phi(\sigma)=\boxplus\cdot\mathds{1}_{\{\sigma\neq\boxminus\}}+\boxminus\cdot\mathds{1}_{\{\sigma=\boxminus\}}.
\end{equation*}
Namely, for a sub-square with at least one positive spin, $\phi$ sets all its spins to $+1$; otherwise, $\phi$ is the identity.

If $X_0^{(c)}(1,1)\big|_{\mathcal{D}_{i,j}}=\boxminus$, then $\phi\big(X_t^{(c)}(1,1)\big)=X_t^{(c)}(1,1)$ $=X_0^{(c)}(1,1)$. On the other hand, if $X_0^{(c)}(1,1)\big|_{\mathcal{D}_{i,j}}=\boxplus$, then $\phi\big(X_t^{(c)}(1,1)\big)=X_0^{(c)}(1,1)$ with probability $q_n:=\mathbb{P}\big(\tau^{(c)}_{1,1}\leq c\cdot a(n)\cdot n\big)$, and $\phi\big(X_t^{(c)}(1,1)\big)=\boxminus$ with the complement probability, where $\tau^{(c)}_{1,1}:=\inf\big\{t\geq 0:\,X^{(c)}_t(1,1)=\boxminus\big\}$. The correspondence to the Z-channel is shown in Fig.~\ref{FIG:Z-channel}.


\begin{figure}[t!]
	\begin{center}
		\begin{psfrags}
		\psfragscanon
		\psfrag{A}[][][1]{\ \ \ $1$}
		\psfrag{B}[][][1]{\ \ \ \ $1-q_n$}
		\psfrag{C}[][][1]{\ \ $q_n$}
		\psfrag{X}[][][1]{$\mspace{5mu}X_0^{(c)}(1,1)$}
		\psfrag{Y}[][][1]{\ \ \ \ \  \ \ \ \ \ \ $\phi\left(X_t^{(c)}(1,1)\right)$}
		\hspace{-8mm}\includegraphics[scale = .3]{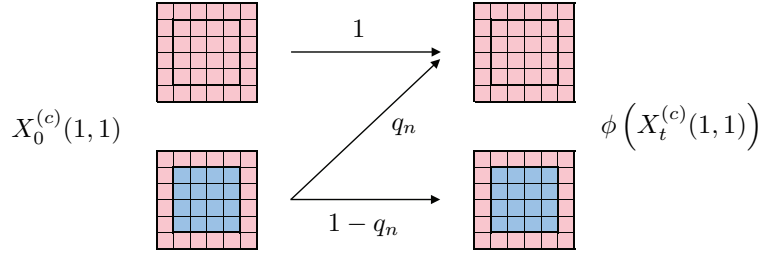}
		\caption{The Z-channel between $X_0^{(c)}(1,1)$ and $\phi\big(X_t^{(c)}(1,1)\big)$, whose crossover probability is $q_n:=\mathbb{P}\big(\tau^{(c)}_{1,1}\leq c\cdot a(n)\cdot n\big)$. Red and blue squares correspond to negative and positive spins, respectively.} \label{FIG:Z-channel}
		\psfragscanoff
		\end{psfrags}
	\end{center}
\end{figure}
		
 
By Theorem \ref{TM:Droplet_Erosion_Time}, there exist $c,\gamma>0$, such that $q_n\leq e^{-\gamma\cdot \sqrt{a(n)}}$. The capacity of the Z-channel with crossover probability $q_n$ is
\begin{equation*}
\max_{p_{1,1}}I\left(X_0^{(c)}(1,1);X_t^{(c)}(1,1)\right)=\log\Big(1+(1-q_n)q_n^{\frac{q_n}{1-q_n}}\Big),    
\end{equation*}
which is non-zero for all $t<c\cdot a(n)\cdot n$.\end{IEEEproof}


\section{Tight Information Capacity Results}\label{SEC:tight_results}


We consider two additional zero-temperature models: the 2D grid with an external field and the honeycomb lattice. In both cases we show that $I_n(t)=\Theta(n)$, for all $t$, which is orders of magnitude higher than the $\Theta(\sqrt{n})$ asymptotic storage capability of the gird without of external field (Theorem \ref{TM:Grid_capacity}). This abrupt growth of $I_n(t)$ strongly relies on the zero-temperature dynamics majority update rule. No such phenomenon is expected to occur for any positive temperature. 



\subsection{2D Grid with External Field}

In the presence of an external field $h>0$, the Hamiltonian from \eqref{EQ:Hamiltonian} is replaced with
\begin{equation*}
    \mathcal{H}(\sigma):=-\sum_{\{u,v\}\in\mathcal{E}_n}\sigma(u)\sigma(v)-h\sum_{v\in\mathcal{V}_n}\sigma(v).
\end{equation*}
The magnetic field serves as a tie-breaker in the zero-temperature grid dynamics. Namely, for $h>0$, updated sites with a balanced neighborhood ($m_v(\sigma) = \ell_v(\sigma)$) are assigned with a positive spin, with probability 1. We have the following order optimal characterization of the information capacity.




\begin{theorem}[Information Capacity with External Field]\label{TM:Info_Cap_external}
For the zero-temperature SIM on $G_n$ with an external field $h>0$, we have $I_n(t)=\Theta(n)$, for all $t\geq 0$.
\end{theorem}

\begin{IEEEproof}
The upper bound follows from $I_n(t)\leq I_n(0)=n$, for all $t\geq 0$. For the lower bound, note that in the presence of a positive external field, the droplet configurations comprising $\mathcal{C}_n$ from \eqref{EQ:droplet_codebook}, with size $a(n)=4$, are all stable. Taking $p_{X_0}$ as the uniform distribution over $\mathcal{C}_n$, we get $I_n(t)=\Omega(n)$.\end{IEEEproof}


\subsection{The honeycomb Lattice}

\begin{figure}[t!]
	\begin{center}	\subfloat[]{\includegraphics[scale = 0.55]{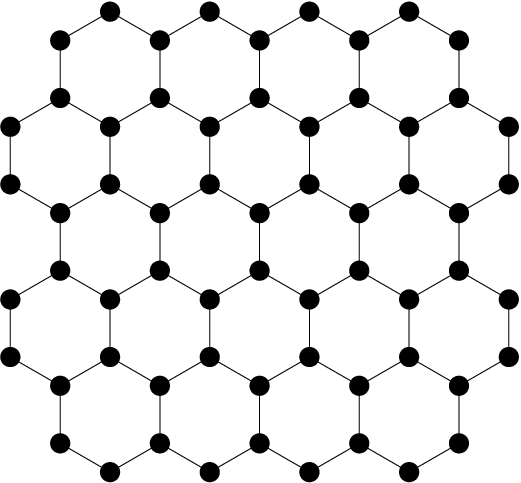}}\quad\quad\quad\quad\quad\quad
    \subfloat[]{\includegraphics[scale = 0.5]{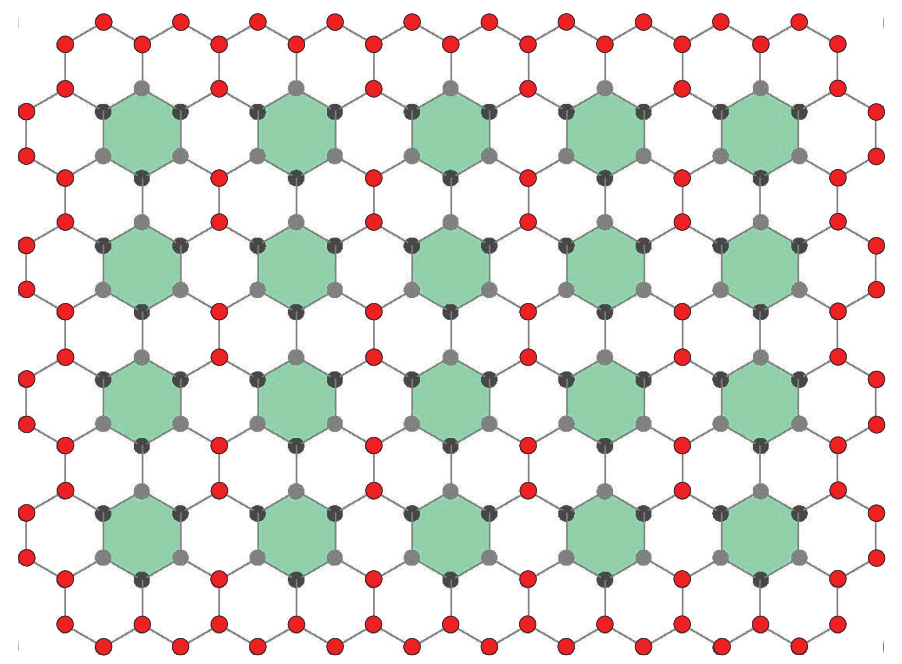}}\quad\quad
    \caption{(a) The 2D honeycomb lattice; (b) Storing $\Omega(n)$ stable bits in the honeycomb lattice: Each green region stores a bit via a monochromatic spin assignment to all the vertices on the region's border. The rest of the spins are set to $-1$ (shown in red). Any such configuration is stable.} \label{FIG:honeycomb}
	\end{center}
\end{figure}

Another instance of a zero-temperature SIM for which $I_n(t)=\Theta(n)$ is when the graph is the honeycomb Lattice on $n$ vertices (Fig. \ref{FIG:honeycomb}(a)), denoted by $H_n$.


\begin{theorem}[Information Capacity of Honeycomb Lattice]\label{TM:Info_Cap_honeycomb}
For the zero-temperature SIM on the honeycomb lattice $H_n$, we have $I_n(t)=\Theta(n)$, for all $t\geq 0$.
\end{theorem}

\begin{IEEEproof}
Key to this result is that all the vertices in the interior of $H_n$ have odd degrees, which makes ties impossible. 
The upper bound is immediate. For the lower bound, consider the tiling of $H_n$ shown in \ref{FIG:honeycomb}(b). 
A stable bit can be stored in each green region by assigning the same spin to all sites on its border (e.g., negative spins for $`0'$ and a positive ones for $`1'$). The rest of the spins are set to $-1$ (red circles in Fig. \ref{FIG:honeycomb}(b)). Any such configuration is stable. The claim follows by taking $X_0$ uniformly distributed over such configurations and observing there are $\Omega(n)$ many green regions.
\end{IEEEproof}


\section{1-Bit Upper Bound under Gibbs Initialization}\label{SEC:positive_temp1}

This section and the next focus on the 2D grid dynamics at low but positive temperature, for which we provide two main results. The first (this section) shows that by drawing the initial configuration from the Gibbs distribution one cannot store more than a single bit for times $t\geq e^{cn^{\frac{1}{4}+\epsilon}}$. This result relies on \cite[Proposition 5.2]{Martinelli_PhaseCoex1994}, which we restate as follows.\footnote{The original claim from \cite{Martinelli_PhaseCoex1994} considers the continuous-time dynamics. Our restatement is for  discrete-time.}

Fix $\beta\in(0,\infty)$ and let $(X_t)_{t\geq 0}$ be the discrete-time grid dynamics at inverse temperature $\beta$ with $X_0\sim \pi$, where $\pi$ is the Gibbs measure (see \eqref{EQ:Hamiltonian}-\eqref{EQ:Gibbs_measure}). We use $(X_t^{\sigma})_{t\geq 0}$ to denote the dynamics initialized at $X_0=\sigma$. The case when $X_0=\boxplus$ is distinguished by denoting $Y_t:= X_t^{\boxplus}$, for all $t\geq 0$. Finally, $m(\sigma)=\frac{1}{n}\sum_{v\in\mathcal{V}_n}\sigma(v)$ is the magnetization of $\sigma\in \Omega_n$.

To restate Proposition 5.2 of \cite{Martinelli_PhaseCoex1994} we need to set up $\{(X_t^\sigma)_{t\geq 0}\}_{\sigma\in\Omega_n}$ over the same probability space. The construction is similar to that from the proof of Theorem \ref{TM:Droplet_Erosion_Time} in Appendix \ref{APPEN:Droplet_Erosion_Time_proof}. Let $\{V_t\}_{t\in\mathbb{N}}$ be an i.i.d. process with $V_1\sim\mathsf{Unif}(\mathcal{V}_n)$. Also let $\{U_t\}_{t\in\mathbb{N}}$ be an i.i.d. process with $U_1\sim\mathsf{Unif}[0,1]$. For each initial configuration $\sigma\in\Omega_n$, we construct $(X_t^\sigma)_{t\geq0}$ as follows. At each time $t\in\mathbb{N}$, $X_t^\sigma\big(\mathcal{V}_n\setminus\{V_t\}\big)=X_{t-1}^\sigma\big(\mathcal{V}_n\setminus\{V_t\}\big)$, and
\begin{equation*}
    X_t^\sigma(V_t)=\begin{cases}+1,\quad \mbox{if } U_t\leq \pi\big(+1\big|X_t(\mathcal{V}_N\setminus\{V_t\})\big)\\
    -1,\quad\mbox{otherwise}\end{cases}.
\end{equation*}
In the above, for any $\mathcal{A}\subseteq\mathcal{V}_n$, we used $X_t(\mathcal{A})$ for the restriction of $X_t$ to $\mathcal{A}$. Let $\mathbb{P}$ be the probability measure associated with this joint probability space.

\begin{proposition}[Coupling with All-Plus Phase \cite{Martinelli_PhaseCoex1994}]\label{PROP:Martinelli94}
Fix $\sqrt{n}\in\mathbb{N}$, $\epsilon\in\left(0,\frac{1}{2}\right)$ and $\gamma>0$. There exist $\beta_0,C<\infty$ such that for any $\beta\geq\beta_0$ and $t\geq n\cdot e^{C\beta n^{\frac{1}{4}+\epsilon}}$, we have
\begin{equation}
    \sum_{\substack{\sigma\in\Omega_n:\\m(\sigma)>0}} \pi(\sigma)\mathbb{P}\big(X_t^\sigma\neq Y_t\big)\leq e^{-\gamma\sqrt{n}}.\label{EQ:Martinelli94}
\end{equation}
\end{proposition}
The proposition states that if the initial configuration is distributed according to the restriction of the Gibbs measure to $\{\sigma\in\Omega_n:\,m(\sigma)>0\}$, by time $t\geq n\cdot \exp\big(C\beta n^{\frac{1}{4}+\epsilon}\big)$, the dynamics become indistinguishable from those initiated at the all-plus state, with high probability. Thus, in this time scale, the only thing the chain remembers is the magnetization of the initial configuration, not the exact starting point. Based on this we show that when $X_0\sim \pi$, $I(X_0;X_t)$ is at most 1 bit for $t\geq n\cdot \exp\big(C\beta n^{\frac{1}{4}+\epsilon}\big)$. 

\begin{theorem}[1-bit Upper Bound]\label{TM:Capacity_M94}
Let $\epsilon$, $\gamma$, $\beta_0$ and $C$ be as in Proposition \ref{PROP:Martinelli94}. For any $\beta>\beta_0$ there is $c(\beta)>0$ such~that
\begin{equation*}
    I(X_0;X_t)\leq \log2+\epsilon_n(\beta),
\end{equation*}
for all $t\geq n\cdot e^{C\beta n^{\frac{1}{4}+\epsilon}}$, where $X_0\sim\pi$ and 
\begin{align*}
    &\epsilon_n(\beta):= ne^{-c(\beta)\sqrt{n}}+n\frac{2e^{-\gamma\sqrt{n}}}{1-e^{-c(\beta)\sqrt{n}}}+h\Big(e^{-c(\beta)\sqrt{n}}\Big)\\
    &\mspace{200mu}+2h\left(\frac{2e^{-\gamma\sqrt{n}}}{1-e^{-c(\beta)\sqrt{n}}}\right),
\end{align*}
and $h:[0,1]\to[0,1]$ is the binary entropy function. In particular, $\lim_{n\to\infty}\epsilon_n(\beta)=0$ for all $\beta$ values as above.
\end{theorem}

\begin{IEEEproof}
Recall that in the phase coexistence regime (i.e., when $\beta>\beta_c$), zero magnetization is highly atypical w.r.t the Gibbs distribution \cite{Schonmann_magnetization1987}. Namely, for each $\beta>\beta_c$ there exists $c(\beta)>0$ such that 
\begin{equation}
\pi\big(\{\sigma:\,m(\sigma)=0\}\big)\leq e^{-c(\beta)\sqrt{n}}.\label{EQ:Schonmann}
\end{equation}
Fix $\beta>\beta_0$ (which, in particular, is larger than $\beta_c$) and define $E=\mathds{1}_{\big\{m(X_0)=0\big\}}$. For $t\geq n\cdot \exp^\big(C\beta n^{\frac{1}{4}+\epsilon}\big)$, consider
\begin{align*}
    I(X_0;X_t)&=I(X_0,E;X_t)\\
              &\leq H(E)+I(X_0;X_t|E)\\
              &\leq h\Big(e^{-c(\beta)\sqrt{n}}\Big)+ne^{-c(\beta)\sqrt{n}}+I(X_0;X_t|E=0)\numberthis\label{EQ:M94_UB1}
\end{align*}
where the last inequality uses \eqref{EQ:Schonmann} twice. To upper bound $I(X_0;X_t|E=0)$, let $S=\mathsf{sign}\big(m(X_0)\big)$ and observe
\begin{align*}
    I(X_0&;X_t|E=0)\\
    &\leq H(S|E=0)+I(X_0;X_t|S,E=0)\\
    &\stackrel{(a)}\leq \log2+\mathbb{P}\big(S=1\big|E=0\big)I(X_0;X_t|S=1,E=0)\\
    &\qquad+\mathbb{P}\big(S=-1\big|E=0\big)I(X_0;X_t|S=-1,E\mspace{-2mu}=0)\\
    &\stackrel{(b)}\leq \log2+2\mathbb{P}\big(S=1\big|E=0\big)I(X_0;X_t|S=1,E=0)\\
    &\stackrel{(c)}\leq \log2+I\big(X_0;X_t\big|m(X_0)>0\big)\numberthis\label{EQ:M94_UB2}
\end{align*}
where (a) is because $\mathbb{P}\big(S=0\big|E=0\big)=0$, (b) uses the symmetry w.r.t. a global spin flip of the Gibbs measure and the SIM with free boundary conditions, while (c) is because $\{S=1\}\cap\{E=0\}=\big\{m(X_0)>0\big\}$ and $\mathbb{P}\big(S=1\big|E=0\big)\leq \frac{1}{2}$.

Next, let $G=\mathds{1}_{\{X_t=Y_t\}}$ and approximate the mutual information from the RHS of \eqref{EQ:M94_UB2} as
\begin{align*}
&I\big(X_0;X_t\big|m(X_0)>0\big)\leq I\big(X_0;X_t,G\big|m(X_0)>0\big)\\
    &\leq H\big(G\big|m(X_0)>0\big)\\
    &\ \ +\mathbb{P}\big(G=0\big|m(X_0)>0\big)I\big(X_0;X_t\big|G=0,m(X_0)>0\big)\\
    &\ \ +\mathbb{P}\big(G=1\big|m(X_0)>0\big)I\big(X_0;X_t\big|G=1,m(X_0)>0\big)\\
    &\leq h\Big(\mathbb{P}\big(X_t=Y_t\big|m(X_0)\mspace{-3mu}>\mspace{-3mu}0\big)\Big)\\
    &\ \ +\mathbb{P}\big(X_t\neq Y_t\big|m(X_0)\mspace{-3mu}>\mspace{-3mu}0\big)I\big(X_0;X_t\big|X_t\neq Y_t,m(X_0)\mspace{-3mu}>\mspace{-3mu}0\big)\\
    &\ \ +\mathbb{P}\big(X_t=Y_t\big|m(X_0)\mspace{-3mu}>\mspace{-3mu}0\big)I\big(X_0;X_t\big|X_t\mspace{-2mu}=\mspace{-2mu}Y_t,m(X_0)\mspace{-3mu}>\mspace{-3mu}0\big)\mspace{-1mu}.\numberthis\label{EQ:M95_UB3}
\end{align*}

Using \eqref{EQ:Martinelli94} and \eqref{EQ:Schonmann}, we have
\begin{equation*}
    \mathbb{P}\big(X_t\neq Y_t\big|m(X_0)>0\big)\leq\frac{2e^{-\gamma\sqrt{n}}}{1-e^{-c(\beta)\sqrt{n}}}:= p_n.
\end{equation*}
Inserting this back into \eqref{EQ:M95_UB3} gives
\begin{align*}
&I\big(X_0;X_t\big|m(X_0)>0\big)\\
&\leq h(p_n)+np_n\\
&+\mathbb{P}\big(X_t=Y_t\big|m(X_0)>0\big)I\big(X_0;Y_t\big|X_t=Y_t,m(X_0)>0\big)\\
&+\mathbb{P}\big(X_t\neq Y_t\big|m(X_0)>0\big)I\big(X_0;Y_t\big|X_t\neq Y_t,m(X_0)>0\big)\\
&\leq h(p_n)\mspace{-2mu}+\mspace{-2mu}np_n\mspace{-2mu}+\mspace{-2mu}H\big(G\big|m(X_0)\mspace{-2mu}>\mspace{-2mu}0\big)\mspace{-2mu}+\mspace{-2mu}I\big(X_0;Y_t\big|m(X_0)\mspace{-2mu}>\mspace{-2mu}0\big)\\
&\leq 2h(p_n)+np_n+I\big(X_0;Y_t\big|m(X_0)>0\big).
\end{align*}


The proof is concluded by showing that $I\big(X_0;Y_t\big|m(X_0)>0\big)=0$. Indeed, $Y_t$ can be represented as a deterministic function of $Y_0=\boxplus$, $\big\{V_t\big\}_{t\in\mathbb{N}}$ and $\big\{U_t\big\}_{t\in\mathbb{N}}$. However, $\big(Y_0,\big\{V_t\big\}_{t\in\mathbb{N}},\big\{U_t\big\}_{t\in\mathbb{N}}\big)$ is independent of $X_0$, and therefore, so is $Y_t$. Thus, $I\big(X_0;X_t\big|m(X_0)>0\big)\leq 2h(p_n)+np_n$, which together with \eqref{EQ:M94_UB1} and \eqref{EQ:M94_UB2} concludes the proof.\end{IEEEproof}


\section{Storing for Exponential Time in Inverse Temperature}\label{SEC:positive_temp2}

When temperature is positive but small temperature one may consider asymptotics in $\beta$. Accordingly, we next show that a single stripe of plus-labeled sites at the bottom of the grid retains at least half of its $\sqrt{n}$ pluses for at least $e^{c\beta}$ time. This result is then generalized to stripes of width 2 inside the grid (i.e., not on the border). Together, these claims enable encoding $\Omega(\sqrt{n})$ bits into sufficiently separated monochromatic stripes and extracting the written data after $e^{c\beta}$ time via majority decoding (Theorem \ref{TM:positive_temp_achievability}).

\subsection{Single Stripe at the Bottom}

Slightly abusing notation, let $(X_t)_{t\geq 0}$ be the continuous-time grid dynamics at inverse temperature $\beta>0$ with Poisson clocks of rate 1 associated with each vertex. Whenever the clock at $v\in\mathcal{V}_n$ rings, the corresponding spin is refreshed according to $\pi_\beta\big(s\big|X_t(\mathcal{N}_v)\big)$, for $s\in\{-1,+1\}$. For each $\sigma\in\Omega_n$ and $v\in\mathcal{V}_n$, we have
\begin{equation*}
    p(\sigma,v):=\pi_\beta\big(-1\big|\sigma(\mathcal{N}_v)\big)=\phi\big(S(\sigma,v)\big),
\end{equation*}
where $\phi(a):= \frac{e^{a\beta}}{e^{a\beta}+e^{-a\beta}}$ and $S(\sigma,v):= \sum_{w:w\sim v} \sigma(w)=\big|\big\{w\sim v:\,\sigma(w)=-1\big\}\big|-\big|\big\{w\sim v:\,\sigma(w)=+1\big\}\big|$.

\begin{figure}[t!]
	\begin{center}	\includegraphics[scale = 0.45]{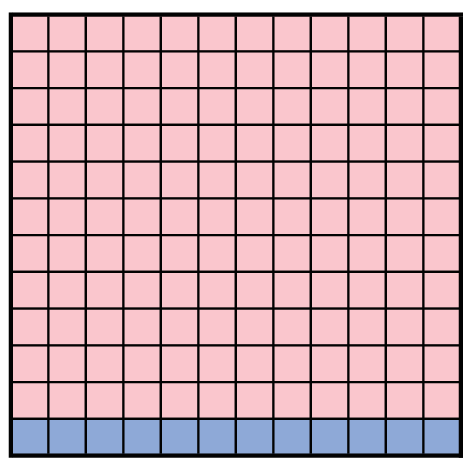}
    \caption{Initial configuration, with blue and red squares marking plus- and minus-labeled sites, respectively.} \label{FIG:One_stripe}
	\end{center}
\end{figure}

Initialize the system at $X_0=\sigma$, with $\sigma\big((i,1)\big)=+1$, for all $i\in[\sqrt{n}]$, and $\sigma(v)=-1$ otherwise (see Fig. \ref{FIG:One_stripe}). Let $\mathcal{B}:=\big\{(i,1)\big\}_{i\in[\sqrt{n}]}$, define  $N_1^{(+)}(\sigma):=\big|\big\{v\in\mathcal{B}:\,\sigma(v)=+1\big\}\big|$ and set $N_1^{(+)}(t)=N_1^{(+)}(X_t)$, for all $t\geq 0$. In words, $N_1^{(+)}(t)$ is the number of plus-labeled sites at the bottom strip after $t$~time. We have the following lower bound on the expectation of $N_1^{(+)}(e^{c\beta})$.


\begin{theorem}[Exponential Survival Time at the Bottom]\label{TM:One_Strip} Fix any $c,c'\in(0,1)$. There exist $\beta_0>0$ such that for any $\beta\geq\beta_0$ there exists $n_0\in\mathbb{N}$ such that for all $n>n_0$ and $t\leq e^{c\beta}$, we have
\begin{equation}
    \mathbb{E}\big[N_1^{(+)}(t)\big]\geq c'\sqrt{n}.\label{EQ:One_Strip}
\end{equation}
\end{theorem}

The proof is given in Appendix \ref{APPEN:One_Strip_proof}, but we outline the main idea below. The derivation has two main parts: \emph{phase separation} and \emph{analysis}. To explain their roles, we first highlight the challenges in analyzing $\mathbb{E}\big[N_1^{(+)}(t)\big]$. First, since $\beta<\infty$, pluses may spread out to the portion of $G_n$ above the bottom stripe. However, because such flips are exponentially unlikely (in $\beta$), we circumvent this complication by prohibiting minus-labeled vertices from flipping. This can only speed up the shrinkage of $N_1^{(+)}(t)$, and thus, it is sufficient to prove \eqref{EQ:One_Strip} under this restricted dynamics.

With this simplification, the main difficulty in analyzing $N_1^{(+)}(t)$ is the disorder of plus-to-minus flips inside the bottom strip $\mathcal{B}$. We distinguish between two types of flips:
\begin{itemize}
    \item \textbf{Sprinkle:} A flip of a plus-labeled vertex whose horizontal neighbours are also pluses. The probability of a sprinkle update for any $v\in\mathcal{B}\setminus\big\{(1,1),(1,\sqrt{n})\big\}$ is $\phi(-1)$.
    
    \item \textbf{Erosion:} A flip of a plus-labeled vertex with at least one horizontal neighbor that is a minus. The probability of an erosion update of $v\in\mathcal{B}\setminus\big\{(1,1),(1,\sqrt{n})\big\}$ is either $\phi(1)$ or $\phi(3)$, depending on whether $v$ has one or zero plus-labeled neighbors, respectively.
\end{itemize}

These types of updates are interleaved as the dynamics evolve. However, sprinkles have exponentially small probability (in $\beta$), while erosion updates have probability exponentially close to 1. Therefore, we expect that during a certain initial time frame the system stays close to $X_0$ with occasional occurrences of sprinkles. Each sprinkle in the bulk results in two contiguous runs of pluses (abbreviated as a `contigs') to its left and right. After a sufficient number of sprinkles, the drift of $N_1^{(+)}(t)$ is dominated by the erosion of those contigs. The first main ingredient of our proof is to show that the interleaved dynamics can indeed be separated into two pure phases of \emph{sprinkling} and \emph{erosion}. Once we are in the phase-separated dynamics, the analysis of $\mathbb{E}N_1^{(+)}(t)$ first identifies the typical length and number of contigs, and then studies how fast they are eaten up.

\subsection{Width-2 Stripe in the Bulk}

Our coding scheme also needs a result similar to Theorem~\ref{TM:One_Strip}, but for monochromatic stripes of width 2 in~the bulk of the grid. Specifically, let $X_0=\sigma$, where $\sigma\big((i,j_0)\big)=\sigma\big((i,j_0+1)\big)=-1$, for all $i\in[\sqrt{n}]$ and some $j_0\in[2:\sqrt{n}-2]$. Letting $N_{j_0}^{(+)}(t)$ be the number of pluses in this stripe after $t$ time, the following corollary holds.

\begin{corollary}[Exponential Survival Time in the Bulk]\label{CORR:One_Strip_middle} Fix any $c,'c\in(0,1)$. There exist $\beta_1>0$ such that for any $\beta\geq\beta_1$ there exists $n_1\in\mathbb{N}$ such that for all $n>n_1$ and $t\leq e^{c\beta}$, we have
\begin{equation*}
    \mathbb{E}\big[N_{j_0}^{(+)}(t)\big]\geq c'\sqrt{n}.
\end{equation*} 
\end{corollary}

This result follows by a slight modification of the argument that proves Theorem \ref{TM:One_Strip}. We describe the modification next and omit the rest of the proof. The idea is to speed up the dynamics so that they correspond to those initiated with a single bottom stripe (as in Fig. \ref{FIG:One_stripe}). We first prohibit minus-labeled sites to flip. To avoid dealing with the width dimension, every time a site is flipped we immediately also flip its vertical plus-labeled neighbor. Thus, flips occur in vertical pairs, with sprinkling and (modified) erosion probabilities as described below:
\begin{itemize}
    \item \textbf{Sprinkling:} The sprinkling rate is $\phi(-2)$ (each site has 3 plus neighbors and one minus). Note that flipping the vertical neighbor of a sprinkle-flipped site speeds up the elimination of pluses without affecting the time scale. This is since the manually flipped sites have flip probability $\phi(0)=\frac{1}{2}$ (balanced neighborhood) in the original dynamics, which is much larger than $\phi(-2)$. Sprinkling in the modified dynamics thus ties to sprinkling in the bottom stripe dynamics, as they both happen with probability exponentially small in $\beta$.
    
    \item \textbf{Erosion:} Every sprinkle produces two contigs. However, the sites at the borders of these contigs have flip rate $\phi(0)$. This is too slow compared to the bottom stripe case, where the erosion flip probability was $\phi(1)$. We therefore replace these $\phi(0)$ rates with $\phi(1)$, which matches the erosion rates with those in the bottom stripe case.
\end{itemize}
With the above tweaks one may repeat the arguments from the proof of Theorem \ref{TM:One_Strip} to arrive at Corollary \ref{CORR:One_Strip_middle}.

\subsection{Stripe-Based Achievability Scheme}


One last corollary is needed before stating the achievability result. It uses Chebyshev's inequality to translate the results of Theorem \ref{TM:One_Strip} and Corollary \ref{CORR:One_Strip_middle} to a bound on the success probability of majority decoding.

\begin{corollary}[Exponential Survival Time Probability]\label{CORR:One_Stripe_Probability} For $c\in(0,1)$, sufficiently large $\beta$ and $n$ (taken from Theorem \ref{TM:One_Strip} and Corollary \ref{CORR:One_Strip_middle}), and $t\leq e^{c\beta}$, we have
\begin{equation*}
    \mathbb{P}\left(N_{j_0}^{(+)}(t)> \frac{\sqrt{n}}{2}\right)\geq \frac{2}{3},\quad\forall j_0\in[1:\sqrt{n}-2].
\end{equation*}
By symmetry to the $j_0=1$ case, this also holds for a horizontal stripe of width 1 at the top of the grid. 
\end{corollary}
\begin{IEEEproof}
We only prove the statement for $j_0=1$. The expected value of $N_1^{(+)}(t)$ is controlled by Theorem \ref{TM:One_Strip}. Since $N_1^{(+)}(t)\leq \sqrt{n}$  almost surely (a.s.), we also have $\mathsf{var}\big(N_1^{(+)}(t)\big)\leq \big(1-c'^2\big)\sqrt{n}$. Applying Chebyshev produces
\begin{equation*}
    \mathbb{P}\left(N_1^{(+)}(t)\leq \frac{\sqrt{n}}{2}\right)\leq \frac{4(1-c'^2)}{2c'-1},
\end{equation*}
which can be made smaller than $\frac{1}{3}$ by taking $c'$ close enough to 1.\end{IEEEproof}

We follow with the information-theoretic result of this section. It shows that by coding over stripes and using majority decoding, one may store $\Omega(\sqrt{n})$ bits in the SIM at positive but low temperature for time $e^{c\beta}$. Recall that $I_n^{(\beta)}(t)$ stands for the information capacity of this system.


\begin{theorem}[Storing $\sqrt{n}$ Bits]\label{TM:positive_temp_achievability}
There exists $\beta^\star>0$ such that for any $\beta\geq\beta^\star$ there exists $n^\star\in\mathbb{N}$ such that for all $n>n^\star$, we have $I^{(\beta)}_n(t)=\Omega(\sqrt{n})$, for all $t\leq e^{c\beta}$, where $c\in(0,1)$.
\end{theorem}

\begin{IEEEproof} 
Partition the grid into monochromatic horizontal stripes (whose spins are specified later) such that:
\begin{enumerate}
    \item the top and bottom stripes are of width 1;
    \item intermediate stripes are of width 2;
    \item the stripes are separated by all-minus walls of width 2 (larger width is allowed only in the separation between the top stripe and the one right below it, in order to preserve a minimal distance of at least 2; for simplicity of notation, we assume throughout that all these widths are exactly 2).
\end{enumerate}
The partitioning is illustrated in Fig. \ref{FIG:Stripe_Scheme}. Let $K$ be the total number of such stripes, and associate an index $j\in[K]$ with each from bottom to top. Clearly, $K=\Theta(\sqrt{n})$. For $j\in[K]$, let $\mathcal{S}_j$ be the set of vertices in the $j$-th stripe (the white squares in Fig.~\ref{FIG:Stripe_Scheme}), and set $\mathcal{D}_j:= \mathcal{S}_j\cup \big\{u\in\mathcal{V}_n:\,d(u,\mathcal{S}_j)=1\big\}$.


\begin{figure}[t!]
	\begin{center}
		\begin{psfrags}
		\psfragscanon
		\psfrag{V}[][][0.8]{$\vdots$}
		\includegraphics[scale = .35]{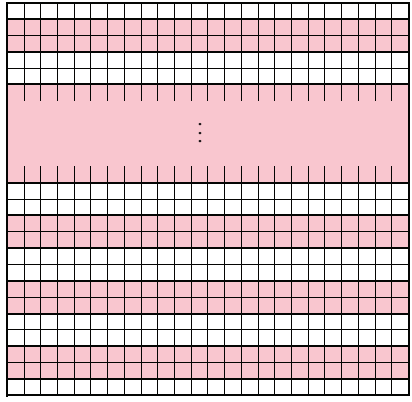}
		\caption{A partitioning of the $\sqrt{n}\times\sqrt{n}$ grid into monochromatic striped regions of width 1 on the top and bottom and width 2 in between (shown in white). The regions are separated by all-minus walls of width 2. Red squares represent negative spins, while white squares stand for unspecified spins.} \label{FIG:Stripe_Scheme}
		\psfragscanoff
		\end{psfrags}
	\end{center}
\end{figure}
		

Let $\mathcal{C}_n$ be the collection of all configurations whose topology corresponds to Fig. \ref{FIG:Stripe_Scheme}, with monochromatic spin assignments to each of the $K$ stripes. Let $X_0\sim p_{X_0}$ with $\supp(p_{X_0})=\mathcal{C}_n$ be such that $p_{X_0}$ is i.i.d. as $\mathsf{Ber}\left(0.5\right)$ across stripes. For each $\sigma\in\Omega_n$, denote the restriction of $\sigma$ to $\mathcal{S}_j$ by $\sigma^{(j)}$ . For $\mathcal{J}\subseteq[K]$, we write $\sigma^{(\mathcal{J})}$ for $\left(\sigma^{(j)}\right)_{j\in\mathcal{J}}$. Similarly, we write $\bar{\sigma}^{(j)}$ for the restriction of $\sigma$ to $\mathcal{D}_j$, and define $\bar{\sigma}^{(\mathcal{J})}$, for $\mathcal{J}\subseteq[K]$, analogously. With some abuse of notation, let $N_j^{(+)}(\sigma)$ be the number of plus labeled sites inside $\mathcal{S}_j$. Furthermore, for each $j\in[K]$, let $\psi_j:\Omega_n\to\mathcal{S}_j$ be the majority decoder inside $\mathcal{S}_j$, i.e., $\psi_j(\sigma)=\mathds{1}_{\left\{N_j^{(+)}(\sigma)\geq \frac{\sqrt{n}}{2}\right\}}-\mathds{1}_{\left\{N_j^{(+)}(\sigma)< \frac{\sqrt{n}}{2}\right\}}$.

The relation between $X_0$ and $\psi_j(X_t)$ is described by a binary channel which inputs a monochromatic stripe $X_0^{(j)}$ (all-minus or all-plus with probability 0.5 each), and outputs $+1$ if $N_j^{(+)}(t):= N_j^{(+)}(X_t)\geq \sqrt{n}/2$, and $-1$ otherwise. If $X_0=\sigma$ with $\sigma^{(j)}=\boxplus\in\{-1,+1\}^{\mathcal
{S}_j}$, then the crossover probability is $p_+^{(j)}(\sigma,t):=\mathbb{P}_{\sigma}\big(N_j^{(+)}(t)< \sqrt{n}/2\big)$, while if $X_0=\sigma'$ with $\sigma'^{(j)}=\boxminus$, then it is $p_-^{(j)}(\sigma',t):=\mathbb{P}_{\sigma'}\big(N_j^{(+)}(t)\geq \sqrt{n}/2\big)$. Note that the transition probabilities are specified by the initial configuration through the entire region outside of $\mathcal{S}_j$. Thus, for each $j\in[K]$, any $\sigma^{\mathsf{out}}_j:=\sigma^{[K]\setminus\{j\}}\in\{-1,+1\}^{\mathcal{V}\setminus\mathcal{S}_j}$ defines a binary (in general, asymmetric) channel from $\{\boxminus,\boxplus\}\subset\{-1,+1\}^{\mathcal{S}_j}$ to $\{-1,+1\}$ with the crossover probabilities given above.

For a each $j\in[K]$, let $\mathsf{T}_j:\{-1,+1\}^{\mathcal{S}_j}\to\Omega_n$ be a transformation defined by
\begin{equation*}
    \big(\mathsf{T}_j\sigma^{(j)}\big)(v)=\begin{cases}\sigma^{(j)}(v),\quad\ \ \ v\in\mathcal{S}_j,\\
    -\sigma^{(j)}(u),\quad v\notin\mathcal{S}_j
\end{cases}.
\end{equation*}
where $u$ is the bottom left vertex in $\mathcal{S}_j$.\footnote{This choice is arbitrary: $\mathsf{T}_j$ is only applied to portions of configurations from $\mathcal{C}_n$; for $\sigma\in\mathcal{C}_n$, the $\mathcal{S}_j$ regions are monochromatic.} By monotonicity of the SIC (see beginning of Appendix \ref{APPEN:Droplet_Erosion_Time_proof}), we have that for any $j\in[K]$, $t\geq 0$ and  $\sigma\in\mathcal{C}_n$, if $\sigma^{(j)}=\boxplus$, then $p_+^{(j)}(\sigma,t)\leq p_+^{(j)}(\mathsf{T}_j\sigma^{(j)},t)$, while if $\sigma^{(j)}=\boxminus$ then $p_-^{(j)}(\sigma,t)\leq p_-^{(j)}(\mathsf{T}_j\sigma^{(j)},t)$. This means that among all configurations $\sigma\in\mathcal{C}_n$ that agree on $\sigma^{(j)}$, $\mathsf{T}_j\sigma^{(j)}$ induces maximal crossover probability in the corresponding binary channel when $\sigma^{(j)}$ is transmitted. Furthermore, Corollary \ref{CORR:One_Stripe_Probability} states that for $t=t_0:= e^{c\beta}$, where $c\in(0,1)$, both these probabilities are upper bounded by $1/3$, and therefore the worst binary channel for each $\mathcal{S}_j$ is the one with crossover probability $1/3$. The latter has a positive capacity of $\mathsf{C}_{\mathsf{BSC}}(1/3)=1-h(1/3)$, with $\mathsf{Ber}(1/2)$ as the capacity achieving distribution. 

Collecting the pieces we conclude that
\begin{align*}
    I^{(\beta)}_n(t)\geq I(X_0;X_{t_0})&= \sum_{j\in[K]}I\Big(\bar{X}^{(j)}_0;X_{t_0}\Big|\bar{X}^{[j-1]}_0\Big)\\
    &\stackrel{(a)}\geq \sum_{j\in[K]}I\Big(\bar{X}^{(j)}_0;\psi_j(X_{t_0})\Big|\bar{X}^{[j-1]}_0\Big)\\
    &\stackrel{(b)}\geq K\cdot\mathsf{C}_{\mathsf{BSC}}\left(\frac{1}{3}\right)
\end{align*}
where (a) uses $I(A;B)\geq I\big(A;f(B)\big)$ for any deterministic function $f$, while (b) is because the capacity of a binary (asymmetric) channel is a monotone decreasing function of both its crossover probabilities.\end{IEEEproof} 

\section{Discussion and Future Directions}\label{SEC:summary}

This work proposed a new model for information storage in physical matter, which accounts for interparticle interactions. The idea is to relate written and read configurations through the SIM governed by Glauber dynamics. The fundamental quantity of interest was the information capacity, $I_n^{(\beta)}(t)$ (see \eqref{EQ:Information_Capacity}), which approximates the maximal number of stored bits in the operational setup. 

We first explored the 2D $\sqrt{n}\times\sqrt{n}$ grid at zero-temperature, and studied the asymptotic behaviour of $I_n(t)=\lim_{\beta\to\infty}I_n^{(\beta)}(t)$, when $n$ and $t$ grow together. It was established that $\lim_{t\to\infty}I_n(t)=\Theta(\sqrt{n})$, i.e., that order of $\sqrt{n}$ bits can be stored in the system indefinitely. Any information beyond that dissolves as $t\to\infty$. The key observation was that striped configurations (of which there are $2^{\Theta(\sqrt{n})}$ many) are stable, and every non-striped configuration is absorbed into stripes for large enough $t$. It would be very interesting to understand when the chain actually gets absorbed into stripes. More precisely, how should $t$ scale with $n$ so as to ensure that 
$S_\sigma(t):=\mathbb{P}\big(X_t\mbox{ is not a stripe}\big|X_0=\sigma\big)$ is close to 0, for any initial configuration $\sigma$? When $\sigma$ is drawn from an i.i.d. symmetric Bernoulli measure, $S_\sigma(t)$ is known as the \emph{survival probability} in the statistical physics literature. While some heuristics and numerical results are available for the survival probability (see, e.g., \cite{Spirin_Kinetic_Ising2001}), to the best of our knowledge, there are currently no rigorous bounds on it. Finding any time scale such that $\max_\sigma S_\sigma(t)$ is $o\left(\frac{1}{n}\right)$ would immediately translate (via arguments similar to those in the converse proof Theorem \ref{TM:Grid_capacity}) into an $O(\sqrt{n})$ upper bound on $I_n(t)$.



For finite time, linear codes were used to show that $I_n(t)=\Theta(n)$ up to $t\sim n/4$. A droplet-based coding scheme for superlinear time that reliably stores $\omega(\sqrt{n})$ bits ($\sqrt{n}$ is trivially achievable by coding over stripes) was then proposed. The scheme partitions the grid into appropriately spaced sub-squares of area $a(n)=o(n)$ and writes a bit onto each. The Lifshitz law \cite{Lacoin_Anisotropic2014} then implies that each sub-square retains its bit up to time $t\sim a(n)\cdot n$, implying that $I_n(t)=\Omega\left(\frac{n}{a(n)}\right)$, for $t=O\big(a(n)\cdot n\big)$. This schemes improves upon the stripe-based scheme only up to $t\sim n^{3/2}$. A question remains whether it is possible to store more than $\sqrt{n}$ bits for times longer than $n^{3/2}$. 
A potential improvement of the droplet-based scheme proposed herein nests each droplet with a growing number (with $n$) of smaller and smaller droplets. Letting smaller droplets grow at slower rates, such a scheme might enable storing $\omega(\sqrt{n})$ bits. We leave the formalization and the analysis of this scheme for future work. 





\begin{figure}[t!]
	\begin{center}	\subfloat[]{\includegraphics[scale = 0.35]{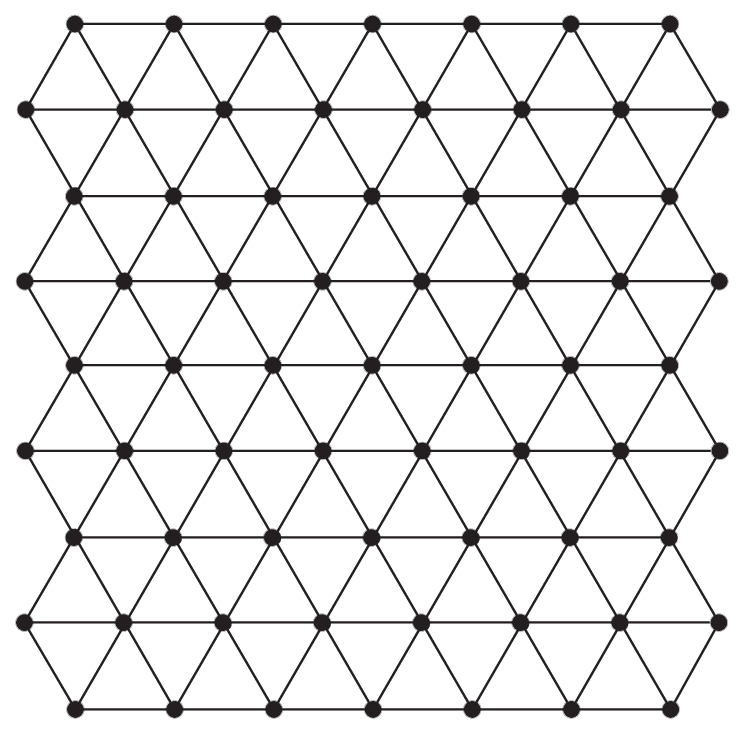}}\quad\quad\quad\quad\quad\quad
    \subfloat[]{\includegraphics[scale = 0.35]{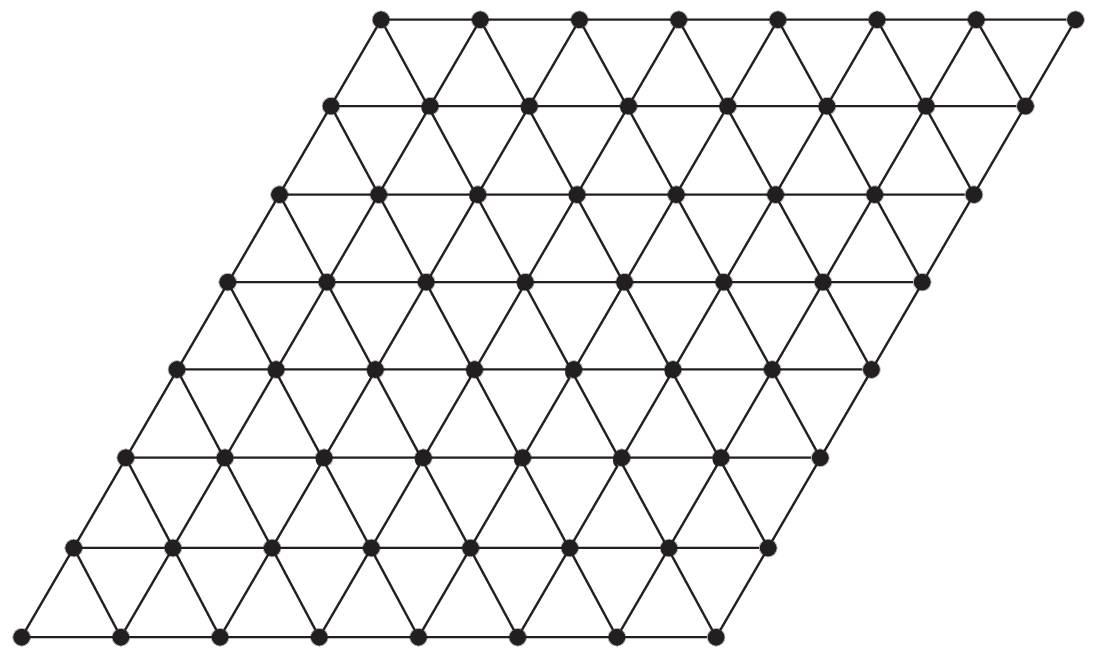}}\quad\quad
    \caption{Two arrangements of the triangular lattice in the plane.} \label{FIG:triangular}
	\end{center}
\end{figure}

It was also shown that applying an arbitrarily small external field on the zero-temperature grid dynamics abruptly boosts its storage capability. The infinite-time capacity grows from $\Theta(\sqrt{n})$ to $\Theta(n)$, which is a consequence of a tie-braking rule induced by the external field. The same holds without an external field, when the grid is replaced with the honeycomb lattice, whose structure prohibits ties to begin with. This suggests that at zero temperature both these architectures are superior to the grid without an external field for storage purposes. Switching to the triangular lattice on $n$ vertices (Fig. \ref{FIG:triangular}), the size of the stable set becomes even smaller than that of the grid. In fact, this size depends on how the triangles are arranged. Consider first the topology shown in Fig. \ref{FIG:triangular}(a), and note that only horizontally striped configurations with stripes of width at least two are stable. The stable set is, thus, effectively half the size than it is in the grid case. For the topology from Fig. \ref{FIG:triangular}(b), it can be shown that only the two ground states are stable (this follows from the even degree of the boundary vertices). We conclude that lattices with odd and small degrees are preferable for storage. The \emph{coordination number} is the chemical term that corresponds to the degree of a vertex in a graph. It refers to the number of the nearest neighbors of a central atom in a molecule. The observations above suggest that the coordination number of the molecules comprising the storage medium significantly impacts capacity at zero temperature.



We believe that the proposed model captures essential features of data storage inside matter. While most of this work focus on the zero-temperature regime, the positive (but small) temperature case is more practically appealing. Initial results for this case were provided herein. First, for $t>\exp(Cn^{1/4+\epsilon})$, we established a 1-bit upper bound on the information capacity subject to an initial configuration drawn from the Gibbs measure. This suggests that resilient configuration are not ones that are typical w.r.t. Gibbs. Finding such configuration is of great interest to us. 
While no configuration is stable at any positive temperature, stripes are quite robust when scaling $t$ with $\beta$. Specifically, we proved that the stripe-based coding scheme retains its $\sqrt{n}$ bits for $t\sim \exp(c\beta)$. This relies on a new result showing that a monochromatic stripe in a sea of opposite spins retains at least half of its original spins for the aforementioned time-scale.

One last research direction we address concerns the SIM on a 3-dimensional grid. The rich state space of this model potentially accommodates many metastable configurations, i.e., ones that retain the written bits for long time. Exploring the number and structure of 3-dimensional metastable configurations and characterizing the corresponding escape times is an interesting research avenue. This could give rise to storage schemes capable of retaining \emph{more than a single bit} for exponential times, a task that seems impossible in two dimensions.


\section*{Acknowledgements}

The authors would like to thank Yuzhou Gu for suggesting a simplified proof for Theorem \ref{TM:stable_iff_striped}, as appears in the paper. We also thank Elchanan Mossel for bringing to our attention the work of F. Martinelli \cite{Martinelli_PhaseCoex1994}.


\appendices


\section{Proof of Proposition \ref{PROP:operational_informational}}\label{APPEN:operational_informational_proof}

Let $c_n^{(t)}$ be an $\big(M^\star(n,t,\epsilon),n,t,\epsilon\big)$-code for the $\mathsf{SIC}_n(t)$. The upper bound follows by observing that
\begin{align*}
    \log M^\star(n,t,\epsilon)=H(M)
    &\stackrel{(a)}\leq I(X_0;X_t)+H(M|X_t)\\
    &\stackrel{(b)}\leq I_n(t)+h(\epsilon)+\epsilon\log M^\star(n,t,\epsilon),\label{EQ:operation_information_UB}
\end{align*}
where (a) follows by the DPI as $M\leftrightarrow X_0\leftrightarrow X_t$ forms a MC, and (b) uses Fano's inequality.

For the lower bound, we show that the $\mathsf{SIC}_n(t)$ can support roughly $\frac{n}{n_1}$ uses of a DMC with capacity $I_{n_1}(t)$. The result then follows by the finite-blocklength achievability bound from~\cite{Poluanskiy_FBL2010}. We outline the main idea here and refer the reader to the proof of Theorem \ref{TM:SupLin_Time}, where a similar construction is fully formalized. 

\begin{figure}[t!]
	\begin{center}
		\begin{psfrags}
		\psfragscanon
		\psfrag{A}[][][1]{\ \ $\mspace{-40mu}\sqrt{n_1}$}
		\psfrag{B}[][][0.8]{$\sqrt{L(n)}$}
		\psfrag{C}[][][0.5]{$\ldots$}
		\psfrag{D}[][][0.5]{$\vdots$}
		\psfrag{E}[][][0.7]{$\Ddots$}
		\includegraphics[scale = .35]{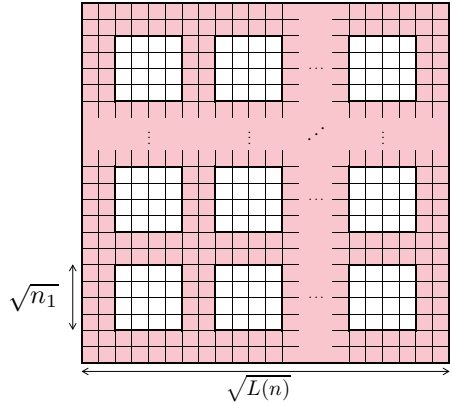}
		\caption{A partition of the $\sqrt{L(n)}\times\sqrt{L(n)}$ grid into $\sqrt{n}$ smaller $\sqrt{n}\times\sqrt{n}$ grids separated by stripes of negative spins of width at least 2 (depicted in red).} \label{FIG:Block_Coding}
		\psfragscanoff
		\end{psfrags}
	\end{center}
\end{figure}

Consider a grid of side length $\sqrt{L(n)}\mspace{-1mu}:=\mspace{-2mu} \sqrt{n/n_1}(\sqrt{n}_1\mspace{-1mu}+\mspace{-1mu}2)\mspace{-1mu}+\mspace{-1mu}2$, so that $L(n)=n+o\left(n/\sqrt{n_1}\right)$. This grid can accommodate $n/n_1$ sub-grids of size $n_1$ separated by (horizontal and vertical) walls of width at least 2 of all-minus spins (see Fig.~\ref{FIG:Block_Coding}). These walls decorrelate\footnote{For this statement to be precise one should consider the continuous-time dynamics, as defined in Section \ref{SUBSEC:Continuous_Time_Dynamics_Def}. However, as the discrete-time and the continuous-time chains are equivalent for our purposes (see Proposition \ref{PROP:discrete_continuous_relation}), this subtlety is ignored at the moment.} the $\sqrt{n_1}\times\sqrt{n_1}$ sub-grids, each has capacity $\mathsf{SIC}_{n_1}(t)$. Invoking \cite{Poluanskiy_FBL2010}, we have
\begin{equation*}
    \log M^\star\big(L(n),t,\epsilon\big)\geq \frac{n}{n_1}I_{n_1}(t)-\sqrt{\frac{nV_{n_1}(t)}{n_1(1-\epsilon)}},
\end{equation*}
where $V_{n_1}(t)=\mathsf{var}\left(\log\frac{p_{X_0,X_t}(X_0,X_t)}{p_{X_0}(X_0)p_{X_t}(X_t)}\right)$ is the dispersion of the $\mathsf{SIC}_{n_1}(t)$. Noting that $V_{n_1}(t)\leq 2n_1^2$ (see \cite[Theorem 50]{Poluanskiy_FBL2010}) concludes the proof of the lower bound.


\begin{remark}[Spatial vs. Temporal Blocks]
The proof of the lower bound can be viewed as a spatial analog of the classic argument that splits transmission over a DMC into temporal blocks. In contrast to the DMC scenario, transmitting in temporal blocks over the $\mathsf{SIC}_n(t)$ is impossible. This is since the encoder $f_n^{(t)}$ controls only the initial configuration $X_0$ and cannot access or influence the channel in subsequent time. The proposed spatial decomposition, however, exploits the Markov random field structure of the Ising model to achieve a similar block coding effect.
\end{remark}


\section{Proof of Proposition \ref{Prop:number_stable}}\label{APPEN:number_stable_proof}

Denote $k=\sqrt{n}$ and let $a_k$ be the number of valid horizontally striped configuration. Due to symmetry, we have $|\mathcal{A}_{k^2}|=2a_k$. Note that there is a bijection between horizontally striped configurations and binary strings of length~$k$, where all runs on $0$'s or $1$'s are of length at least two. Henceforth, $a_k$ is thought of as the number of such strings.

To count $a_k$, observe that a valid string of length $k+2$ either ends in a run of length exactly 2 or a run of length greater than 2. In the former case, the $(k+2)$-lengthed string can be any valid string of length $k$, followed by two characters that are opposite to the last character of the $k$-lengthed string. In the latter case, the string of length $k+2$ can be any valid string of length $k+1$ with its last character duplicated. Hence, we have the Fibonacci relation
\begin{equation*}
a_{k+2}=a_{k+1}+a_k,\quad\forall k\geq 2.    
\end{equation*}
With the initial conditions $a_2=2$ and $a_3=2$, we see that $a_k$ is $(k-1)$-th Fibonacci number (on the indexing where $f_0=0$ and $f_1=1$, respectively). Recalling that
$f_k=(\phi^k-\psi^k)/\sqrt{5}$, where $\phi=-1/\psi=(1+\sqrt{5})/2$, we obtain $|\mathcal{A}_{k^2}|=4f_{k-1}$.


\section{Proof of Theorem \ref{TM:stable_iff_striped}}\label{APPEN:stable_iff_striped_proof}

The inclusion $\mathcal{A}_n\subseteq\mathcal{S}_n$ is straightforward, since for any $\sigma\in\mathcal{A}_n$ we have $P(\sigma,\sigma)=1$. 
For the opposite inclusion, we fix $\sigma\in\mathcal{S}_n$ and construct a subgraph $\mathcal{G}_n^{(\sigma)}$ of the dual lattice, which is used for the proof. First, let $\tilde{\mathcal{G}}_n^{(\sigma)}=\left(\mathcal{V}_n,\tilde{\mathcal{E}}_n^{(\sigma)}\right)$ be an auxiliary subgraph of $\mathcal{G}_n$ with the same vertex set, but with edge set $\tilde{\mathcal{E}}_n^{(\sigma)}=\big\{\{v,u\}\in\mathcal{E}_n:\,\sigma(u)\neq\sigma(v)\big\}$. Namely, $\tilde{\mathcal{E}}_n^{(\sigma)}$ contains only edges connecting sites of opposite spins in $\sigma$.

\begin{figure}[t!]
	\begin{center}
		\begin{psfrags}
			\psfragscanon
			\psfrag{A}[][][1]{$\sqrt{n}$}
			\psfrag{B}[][][1]{$\ell$}
			\psfrag{C}[][][1]{\ }
			\psfrag{D}[][][1]{\ }
			\includegraphics[scale = .7]{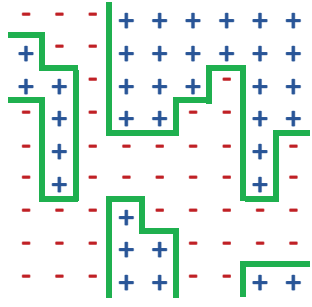}
			\caption{For a given $\sigma\in\Omega_n$, the edges $\mathcal{G}_n^{(\sigma)}$ (green lines) form contours that separate regions of opposite spins in $\sigma$.} \label{FIG:auxiliay_graph}
			\psfragscanoff
		\end{psfrags}
	\end{center}
\end{figure}

From $\tilde{\mathcal{G}}_n^{(\sigma)}$ we construct $\mathcal{G}_n^{(\sigma)}=\left(\hat{\mathcal{V}}_n,\mathcal{E}_n^{(\sigma)}\right)$, where $\hat{\mathcal{V}}_n$ is the $(\sqrt{n}+1)\times (\sqrt{n}+1)$ grid in the dual lattice that covers $\mathcal{V}_n$, i.e., $\hat{\mathcal{V}}_n=\bigcup_{i,j\in\{-1,+1\}}\big\{\mathcal{V}_n+(i/2,j/2)\big\}$. The edge set of $\mathcal{G}_n^{(\sigma)}$ is as follows: for any $x,y\in\hat{\mathcal{V}}_n$ with $||x-y||_1=1$, $\{x,y\}\in\mathcal{E}_n^{(\sigma)}$ if and only if there exists  $e\in\tilde{\mathcal{E}}_n^{(\sigma)}$ that crosses $\{x,y\}$. 
Thus, the edges of $\mathcal{G}_n^{(\sigma)}$ separate regions of opposite spins in $\sigma$ (see Fig. \ref{FIG:auxiliay_graph}), such that vertices on the boundary are never connected.


Denote the degree of a vertex $x$ of $\mathcal{G}_n^{(\sigma)}$ by $\mathsf{deg}_\sigma(x)$. Observe that for any $\eta\in\Omega_n$ and $x\in\hat{\mathcal{V}}_n$ not on the boundary, $\mathsf{deg}_\eta(x)$ is even. Furthermore, if $\sigma\in\mathcal{S}_n$, we have that  $\mathsf{deg}_\sigma(x)\in\{0,2\}$. Indeed, $\mathsf{deg}_\sigma(x)=4$ is possible only if $\sigma\big(x+(1/2,1/2)\big)=\sigma\big(x-(1/2,1/2)\big)\neq \sigma\big(x+(1/2,-1/2)\big)=\sigma\big(x+(-1/2,1/2)\big)$, which contradicts the stability of $\sigma$.

Next, note that for $x\in\hat{\mathcal{V}}_n$ with $\mathsf{deg}_\sigma(x)=2$, the two edges containing $x$ must have the same orientation (i.e., both must be vertical or horizontal). Indeed, if, for instance, the edges containing $x$ are $\{x,y\},\{x,z\}\in\mathcal{E}_n^{(\sigma)}$ with $y=x+(1,0)$ and $z=x+(0,1)$, then we have $\sigma\big(x+(1/2,1/2)\big)\neq\sigma\big(x+(1/2,-1/2)\big)=\sigma\big(x+(-1/2,1/2)\big)$, which contradicts stability. The three other cases of edges with disagreeing orientation are treated similarly. 

Finally, observe that $\mathcal{E}_n^{(\sigma)}$ does not contains parallel edges that are next to one another. Assume to the contrary that $e_1,e_2\in\mathcal{E}_n^{(\sigma)}$ are such edges. Without loss of generality let these edges be horizontal with $e_1=\{x,y\}$ and $e_2=\{x+(0,1),y+(0,1)\}$, where $x\in\hat{\mathcal{V}}_n$ and $y=x+(1,0)$. This implies $\sigma\big(x+(1/2,1/2)\big)\neq\sigma\big(x-(1/2,1/2)\big)=\sigma\big(x+(0,1)+(1/2,1/2)\big)$, which contradicts stability.

The above observations imply that each connected component of $\mathcal{G}_n^{(\sigma)}$ is a straight (horizontal or vertical) crossing of vertices in $\hat{\mathcal{V}}_n$. Clearly, horizontal and vertical crossings cannot coexists, as otherwise the vertex $x$ that is contained in them both has $\mathsf{deg}_\sigma(v)=4$, a case that was already ruled out. This implies that $\sigma$ must have the structure of monochromatic horizontal or vertical stripes of width at least two, except, the first and last stripes that may still have width 1. However, such thin stripes are impossible since their existence renders the corner vertex (e.g., $(1,1)$ and $(1,\sqrt{n})$ for a bottom stripe of width 1) unstable.

\section{Proof of Lemma \ref{LEMMA:absorbing_MC}}\label{APPEN:absorbing_MC_proof}
We construct a finite path from any $\sigma\in\mathcal{T}_n$ into $\mathcal{S}_n$. First, for each $\sigma\in\Omega_n$ and $v\in\mathcal{V}_n$, let
\begin{align*}
    &\mathcal{C}_v(\sigma):=\\
    &\left\{u\in\mathcal{V}_n:\mspace{-20mu}\begin{array}{ll}
         &  \exists m\in\mathbb{N},\, \exists\{u_k\}_{k=0}^m\subseteq\mathcal{V}_n,\,u_0=v,\,u_m=u, \\
         &u_{k-1}\sim u_k\, \sigma(u_k)=\sigma(v),\, \forall k\in[m]\end{array}\right\}
\end{align*}
be the monochromatic connected component of $v$ in $\sigma$. Define the \emph{external boundary} of $\mathcal{C}_v(\sigma)$ as
\begin{equation*}
    \partial_\mathsf{Ext}\mathcal{C}_v(\sigma):=\Big\{u\in\mathcal{V}_n\setminus\mathcal{C}_v(\sigma):\,d\big(u,\mathcal{C}_v(\sigma)\big)=1\Big\}.
\end{equation*}
We next defined certain operations on configurations that constitute the building blocks of our proof. A operation is specified by a finite sequence of single-site flips of positive probability. For any $\sigma\in\Omega_n$, let
\begin{subequations}
\begin{align}
\mathcal{N}_a(\sigma)&:=\left\{v\in\mathcal{V}_n:\,m_v(\sigma)>\frac{|\mathcal{N}_v|}{2}\right\},\\
\mathcal{N}_d(\sigma)&:=\left\{v\in\mathcal{V}_n:\,m_v(\sigma)<\frac{|\mathcal{N}_v|}{2}\right\}.
\end{align}\label{EQ:sets_definitions}%
\end{subequations}
be the sets that contain the vertices of $\mathcal{V}_n$ whose spins disagree with the majority of their neighbors under  $\sigma$. Define also $\mathcal{N}_u(\sigma):= \mathcal{V}_n\setminus\Big\{\mathcal{N}_a(\sigma)\cup\mathcal{N}_d(\sigma)\Big\}$ and let $N_a(\sigma):=\big|\mathcal{N}_a(\sigma)\big|$, $N_d(\sigma):=\big|\mathcal{N}_d(\sigma)\big|$ and $N_u(\sigma):=\big|\mathcal{N}_u(\sigma)\big|$.


\subsection{Expansion Connected Components}

Fix $\sigma\in\Omega_n$ and let $v\in\mathcal{V}_n$ be a vertex on the left border of $\mathcal{V}_n$, i.e., $v\in\mathcal{L}:=\big\{(1,j)\big\}_{j\in[\sqrt{n}]}$. Consider the following pseudo-algorithm that inputs $(\sigma,v)$ and outputs a new configuration $\eta$ reachable from $\sigma$ by a finite path.  

\begin{algorithm}
\caption{Single-flip expansion of $\mathcal{C}_v(\sigma)$}\label{ALG:evolution}
\begin{algorithmic}[1]
\State $\zeta\leftarrow\sigma$
\While{$\partial_\mathsf{Ext}\mathcal{C}_v(\zeta)\cap\big\{\mathcal{N}_u(\zeta)\cup\mathcal{N}_d(\zeta)\big\}\neq\emptyset$}
    \State Draw $w\sim\mathsf{Unif}\left(\partial_\mathsf{Ext}\mathcal{C}_v(\zeta)\cap\big\{\mathcal{N}_u(\zeta)\cup\mathcal{N}_d(\zeta)\big\}\right)$
    \State $\zeta\leftarrow\zeta^{w}$
\EndWhile 
\State $\eta\leftarrow\zeta$
\end{algorithmic}
\end{algorithm}

Since the grid is finite, Algorithm \ref{ALG:evolution} terminates in a finite number of steps. Once it does, $\partial_\mathsf{Ext}\mathcal{C}_v(\eta)$ contains no unstable nor disagreeing vertices. Let $\mathsf{E}:\Omega_n\times\mathcal{L}\to\Omega_n$ be the mapping specified by the algorithm (it can be verified that it is well defined despite the random choice of $w$ in Step 4). For any $\sigma\in\Omega_n$ and $v\in\mathcal{L}$, the connected component $\mathcal{C}_v(\eta)$, where $\eta=\mathsf{E}(\sigma,v)$, has a rectangular shape. Namely, if $v=(1,j)\in\mathcal{L}$, then there exist $1\leq j_f^{(1)}\leq j\leq j_f^{(2)}\leq \sqrt{n}$ and an $i_f\in[\sqrt{n}]$ such that $\mathcal{C}_v(\eta)=\big\{(i,j)\in\mathcal{V}_n:\,i\in[i_f],\ j\in\big[j_f^{(1)}:j_f^{(2)}\big]\big\}$. This holds because any non-rectangular connected component has flippable spins on its external boundary, which contradicts  $\partial_\mathsf{Ext}\mathcal{C}_v(\eta)\cap\big\{\mathcal{N}_u(\eta)\cup\mathcal{N}_d(\eta)\big\}=\emptyset$.

\begin{figure}[t!]
    \begin{center}
        \begin{psfrags}
            \psfragscanon
            \subfloat[]{\includegraphics[scale = 0.33]{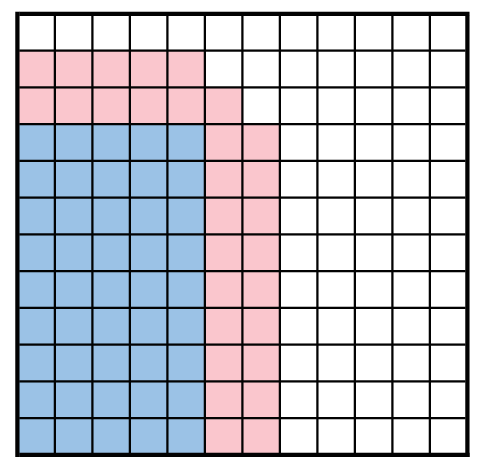}}\quad
            \subfloat[]{\includegraphics[scale = 0.33]{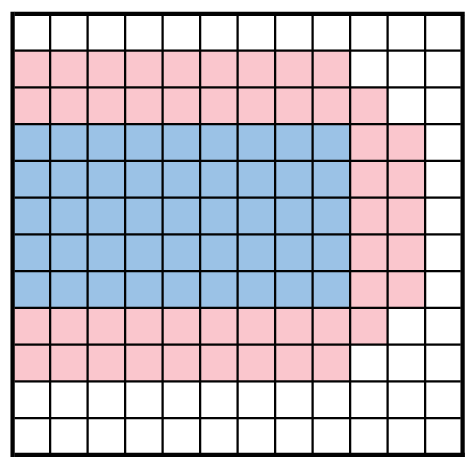}}\quad \subfloat[]{\includegraphics[scale = 0.33]{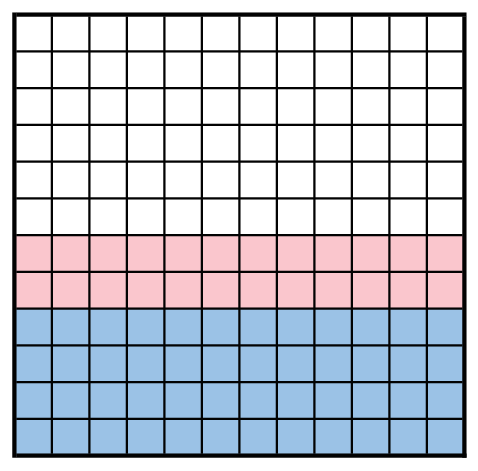}}
            \caption{Rectangular components, where blue and red squares represent opposite spins;  white squares are~unspecified.}\label{FIG:rectangles}
            \psfragscanoff
        \end{psfrags}
     \end{center}
 \end{figure}

Furthermore, observe that $\eta=\mathsf{E}(\sigma,v)$ is such that any $u\in\mathcal{V}_n$ with $d\big(u,\mathcal{C}_v(\eta)\big)\leq 2$ satisfies $\eta(u)=-\eta(v)$. Assuming $\mathcal{C}_v(\eta)\neq\mathcal{V}_n$, this statement is trivial for sites $u$ at graph distance $1$ from $\mathcal{C}_v(\eta)$. For sites with $d\big(u,\mathcal{C}_v(\eta)\big)= 2$, let $u$ be such a site but with $\eta(u)=\eta(v)$. If there exists $w\in\mathcal{C}_v(\eta)$ with $w=u-(0,2)$ (i.e., $u$ is right above $\mathcal{C}_v(\eta)$), then $w+(0,1)\in\partial_\mathsf{Ext}\mathcal{C}_v(\eta)\cap\big\{\mathcal{N}_u(\eta)\cup\mathcal{N}_d(\eta)\big\}$, in contradiction to the termination rule of Algorithm \ref{ALG:evolution}. Similarly, if $u=w+(2,0)$, for some $w\in\mathcal{C}_v(\eta)$, then the site $w+(1,0)$ disagrees with at least half of its neighbors. Finally, if $u=w+(1,1)$ for some $w\in\mathcal{C}_v(\eta)$ (the case when $u=w-(1,1)$ is treated similarly), then both $w+(0,1)$ and $w+(1,0)$ are unstable, which again leads to a contradiction. Fig. \ref{FIG:rectangles} shows three examples of possible shapes of $\mathcal{C}_v(\eta)$ with their corresponding surroundings.






\subsection{Flipping Rectangles}

Let $v=(1,j_v)\in\mathcal{L}$, $h\in[\sqrt{n}-j_v-1]$ and $\ell\in[\sqrt{n}-2]$. Define  $\mathcal{R}_{h,\ell}^{(v)}$ as the set of all $\sigma\in\Omega_n$ satisfying: (i)~$\mathcal{C}_v(\sigma)=\big\{v+(i,j):\,i\in[0:\ell],\ j\in[0:h]\big\}$; and (ii) $\mathcal{C}_v(\sigma)\cap\mathcal{N}_u(\sigma)\neq\emptyset$. Namely, $\mathcal{R}_{h,\ell}^{(v)}$ is the set of all configuration in which the connected monochromatic component of $v$ is an $h\times\ell$ rectangle with $v$ as its bottom-left corner, and such that at least two of the rectangle's sides are bounded away from the border of the grid. Further let $\bar{\mathcal{R}}_{h,\ell}^{(v)}$ and $\ubar{\mathcal{R}}_{h,\ell}^{(v)}$ be the subsets of $\mathcal{R}_{h,\ell}^{(v)}$ in which the spin at $v$ is $+1$ or $-1$, respectively.

\begin{figure}[t!]
    \begin{center}
        \begin{psfrags}
            \psfragscanon
            \psfrag{A}[][][0.9]{$\mspace{-15mu}h+4$}
            
            \psfrag{M}[][][0.9]{$\mspace{-23mu}h+2$}
            \psfrag{B}[][][0.9]{$\mspace{-5mu}\ell+2$}
            \psfrag{X}[][][0.9]{$\mspace{20mu}h+4$}
            \psfrag{Y}[][][0.9]{$\mspace{27mu}\ell+4$}
           \psfrag{F}[][][0.9]{$\mspace{-15mu}h+4$}
            \psfrag{E}[][][0.9]{$w$}
            \subfloat[]{\includegraphics[scale = 0.38]{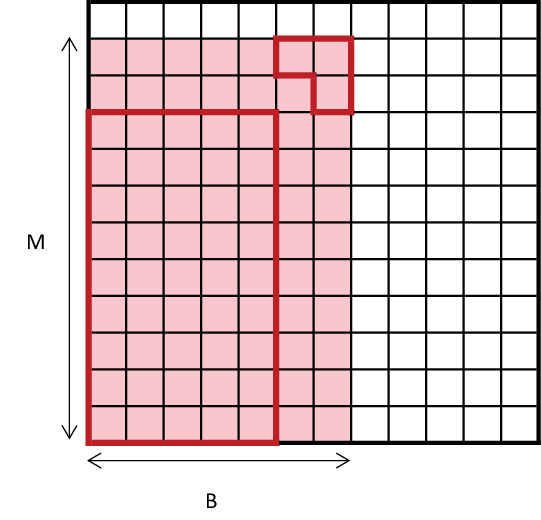}}\quad\quad
            \subfloat[]{\includegraphics[scale = 0.38]{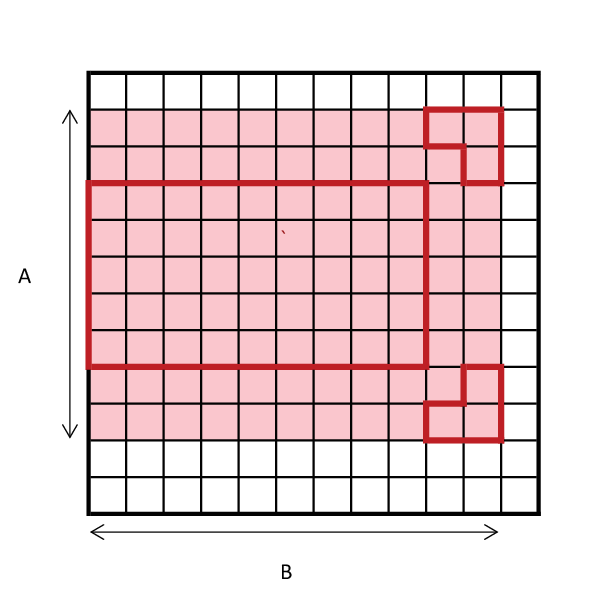}}
            \caption{Figs. (a) and (b) show, respectively, the operation of $\mathsf{F}_{h,\ell}^{(v)}$ on the configurations from Figs. \ref{FIG:rectangles}(a) and \ref{FIG:rectangles}(b). The regions outlined in red are the only ones that $\mathsf{F}_{h,\ell}^{(v)}$ affected.}\label{FIG:mapping}
            \psfragscanoff
        \end{psfrags}
     \end{center}
 \end{figure}

Let $\mathsf{F}_{h,\ell}^{(v)}:\mathcal{R}_{h,\ell}^{(v)}\to\Omega_n$ be a mapping that flips all the spins inside $\mathcal{C}_v(\sigma)$ and sets the spins at the missing corners of $\mathcal{C}_v(\sigma)\cup\partial_\mathsf{Ext}\mathcal{C}_v(\sigma)$ to $-\sigma(v)$. Formally, set  
$\mathcal{M}_v(\sigma):=\big\{\mathcal{C}_v(\sigma)\cap\mathcal{N}_u(\sigma)\big\}+\big\{(1,2),(2,1),(2,2),(1,-2),(2,-1),(2,-2)\big\}$ and define the mapping $\mathsf{F}_{h,\ell}^{(v)}$ by
\begin{equation*}
    \left(\mathsf{F}_{h,\ell}^{(v)}\sigma\right)(w)=\begin{cases}-\sigma(v),\quad w\in\mathcal{C}_v(\sigma)\cup \mathcal{M}_v(\sigma)\\
    \phantom{-}\sigma(u),\quad\mbox{otherwise}\end{cases}.
\end{equation*}
Fig. \ref{FIG:mapping} shows $\mathsf{F}_{h,\ell}^{(v)}$ applied to the configurations from Figs. \ref{FIG:rectangles}(a)-(b). Note that if $\sigma\in\mathcal{R}_{h,\ell}^{(v)}$, then $\mathcal{C}_v\big(\mathsf{F}_{h,\ell}^{(v)}\sigma\big)$ contains a rectangle of size at least $(h+2)\times(\ell+2)$.



Finally, observe that for any $\sigma\in\mathcal{R}_{h,\ell}^{(v)}$, we have $\sigma\leadsto\mathsf{F}_{h,\ell}^{(v)}\sigma$. A path $\omega:\sigma\leadsto\mathsf{F}_{h,\ell}^{(v)}\sigma$ is constructed by first flipping any unstable corner of the $h\times\ell$ rectangle, then flipping a neighbor of that flipped corner, followed by flipping their neighbors, etc. This is done until an $(h-1)\times(\ell-1)$ rectangle is reached and the process repeats itself until the entire original rectangle is flipped. Finally, if needed, the spins at sites $u\in\mathcal{M}_v(\sigma)$ are flipped as well.

We describe $\omega:\sigma\leadsto\mathsf{F}_{h,\ell}^{(v)}\sigma$ for a $\sigma\in\mathcal{R}^{(v)}_{h,\ell}$, with $v=(1,1)$ and $h,\ell\in[\sqrt{n}-2]$ (see Fig. \ref{FIG:rectangles}(a)). Set $\omega_0=\sigma$ and
\begin{equation*}
    \omega_{j\ell+i+1}=\omega_{j\ell+i}^{(\ell-i,h-j)},\quad i\in[0:\ell-1],\ j\in[0:h-1],
\end{equation*}
with (possibly) additional three steps $\omega_{h\ell+1}=\omega_{h\ell}^{(\ell+1,h+2)}$, $\omega_{h\ell+2}=\omega_{h\ell+1}^{(\ell+2,h+1)}$ and $\omega_{h\ell+3}=\omega_{h\ell+2}^{(\ell+2,h+2)}$, for any of these sites whose initial spin was $-\sigma(v)$. By similar constructions it follows that there is a finite path (of at most $h\ell+3$ positive probability, single-site flips) $\omega:\sigma\leadsto\mathsf{F}_{h,\ell}^{(v)}\sigma$, for all $\sigma\in\mathcal{R}_{h,\ell}^{(v)}$, with $v\in\mathcal{L}$ and valid lengths $h,\ell$.

\subsection{Proof of Lemma \ref{LEMMA:absorbing_MC}}

Let $\sigma\in\mathcal{T}_n$ and consider the following pseudo-algorithm that transforms $\sigma$ to a new configuration $\eta\in\mathcal{S}_n$. The transformation is described by means of the mappings $\mathsf{E}$, and $\big\{\mathsf{F}_{h,\ell}^{(v)}\big\}$ defined above. Since each mapping corresponds to a path of single-spin flips from the input to the output, Algorithm \ref{ALG:path} establishes Lemma \ref{LEMMA:absorbing_MC}. 

Given $\sigma\in\Omega_n$ and $v\in\mathcal{V}_n$ such that the monochromatic connected component $\mathcal{C}_v(\sigma)$ is a rectangle (stripes are considered rectangles), let $\mathsf{h}(v,\sigma)$ $\mathsf{l}(v,\sigma)$ be the functions that return the height and length of $\mathcal{C}_v(\sigma)$; if $\mathcal{C}_v(\sigma)$ is not a rectangle, set $\mathsf{h}(v,\sigma)=\mathsf{l}(v,\sigma)=0$. 

\begin{algorithm}
\caption{Path construction from $\mathcal{T}$ to $\mathcal{S}$}\label{ALG:path}
\begin{algorithmic}[1]

\State $\zeta\leftarrow\sigma$
\State $v\leftarrow(1,1)$
\While{$\zeta\notin\mathcal{S}_n$}
    \State $\zeta\leftarrow\mathsf{E(\zeta,v)}$ \Comment{Expand $\mathcal{C}_v(\zeta)$ to the maximal possible rectangle}
    \State $h=\mathsf{h}(v,\zeta)$ \Comment{Height of the rectangle}
    \State $\ell=\mathsf{l}(v,\zeta)$ \Comment{Length of the rectangle}
    \If{$\max(h,\ell)=\sqrt{n}\ \ \mbox{and}\ \min(h,\ell)<\sqrt{n}$} \Comment{If $\mathcal{C}_v(\zeta)$ is a stripe}
        \If{$\ell<\sqrt{n}$} \Comment{Horizontal stripe}
            \State $v\leftarrow(1,h+1)$ \Comment{Move $v$ above stripe}
        \EndIf

        \If{$\ell<\sqrt{n}$} \Comment{Vertical stripe}
            \State $v\leftarrow(\ell+1,1)$ \Comment{Move $v$ to the right of the stripe}
        \EndIf
    \EndIf

    \If{$\max(h,\ell)<\sqrt{n}$} \Comment{If $\mathcal{C}_v(\zeta)$ is a rectangle bounded away from the top and right borders}
        \State $\zeta\leftarrow\mathsf{F}_{h,\ell}^{(v)}\zeta$ \Comment{Flip the rectangle to create a larger rectangle with an opposite spin}
    \EndIf
\EndWhile 

\If{$\mathsf{h}\big(\zeta,(1,1)\big)=1$ or $\mathsf{l}\big(\zeta,(1,1)\big)=1$} \Comment{If the first stripe is of width 1}

    \If{$\mathsf{h}\big(\zeta,(1,1)\big)=1$} \Comment{Horizontal stripe of height 1}
        \State $\zeta(i,1)\leftarrow-\zeta(i,1),\quad \forall i\in[\sqrt{n}]$ \Comment{Flip the spins at the bottom horizontal stripe of width 1}
    \EndIf

    \If{$\mathsf{l}\big(\zeta,(1,1)\big)=1$} \Comment{Vertical stripe of length 1}
        \State $\zeta(1,j)\leftarrow-\zeta(1,j),\quad \forall j\in[\sqrt{n}]$ \Comment{Flip the spins at the leftmost vertical stripe of width 1}
    \EndIf
    
\EndIf

\State $\eta\leftarrow\zeta$ \Comment{Output}
\end{algorithmic}
\end{algorithm}

Algorithm \ref{ALG:path} transforms $\sigma\in\mathcal{T}_n$ to $\eta\in\mathcal{S}_n$ by first expanding the connected component of $v=(1,1)$ to an $h\times\ell$ rectangle. If the rectangle is a stripe, i.e., if $h=\sqrt{n}$ or $\ell=\sqrt{n}$ (if both equalities hold then we are done), then $v$ is recast as the bottom-left corner of the rest of the grid and the process repeats. Otherwise, we use $\mathsf{F}_{h,\ell}^{(1,1)}$ to flip the rectangle's spins. This produces a configuration where the connected component of $(1,1)$ contains a rectangle of size at least $(h+2)\times(\ell+2)$. The finiteness of $\mathcal{V}_n$ ensures that Algorithm \ref{ALG:path} terminates after finitely many steps.

\begin{figure*}[t!]
    \begin{center}
        \begin{psfrags}
            \psfragscanon
           
           \psfrag{S}[][][0.9]{\ \ $v=(1,1)$} 
           \psfrag{T}[][][0.9]{\ \ $v=(1,1)$}
           \psfrag{U}[][][0.9]{\ \ $v=(1,1)$}
           \psfrag{V}[][][0.9]{\ \ $v=(1,1)$}
           \psfrag{W}[][][0.9]{\ \ $v=(1,8)$}
           \psfrag{X}[][][0.9]{\ \ $v=(1,8)$}
           \psfrag{Y}[][][0.9]{\ \ $v=(1,10)$}
           \psfrag{Z}[][][0.9]{\ \ $v=(1,10)$}
           \psfrag{A}[][][0.9]{\ \ $\mathsf{E}(v,\zeta)$}
           \psfrag{B}[][][0.9]{\ \ $\mathsf{F}^{(v)}_{3,8}\zeta$}
           \psfrag{C}[][][0.9]{\ \ \ \ \ \ $\mathsf{E}(v,\zeta)$}
           \psfrag{D}[][][0.8]{\ \ \ \ \ $v\mspace{-3mu}\leftarrow\mspace{-3mu}(1,\mspace{-2mu}8)$}
           \psfrag{E}[][][0.9]{\ \ \ $\mathsf{E}(v,\zeta)$}
           \psfrag{F}[][][0.8]{\ \ \ \ \ $v\mspace{-3mu}\leftarrow\mspace{-3mu}(1,\mspace{-2mu}10)$}
           \psfrag{G}[][][0.9]{\ \ $\mathsf{E}(v,\zeta)$}\includegraphics[scale = 0.4]{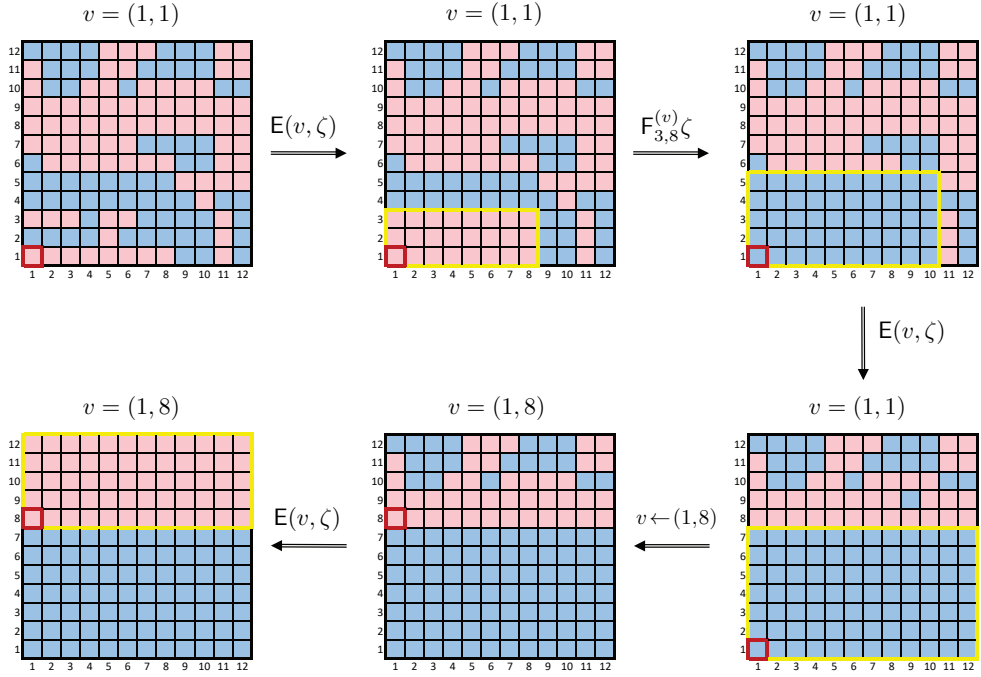}
            \caption{Illustration of Algorithm \ref{ALG:path}. In each figure, the square outlined in red represents the current $v$ (which is also indicated on the top of each grid). The yellow line encloses the portion of the grid that was modified from the previous step.}\label{FIG:algorithm}
            \psfragscanoff
        \end{psfrags}
     \end{center}
 \end{figure*}

The configuration $\zeta$ in Step 13 is striped. However, its connected component of $(1,1)$ may be a stripe of width 1
, which is unstable. We address this in Steps 14-18, and produce the final configuration $\eta\in\mathcal{S}_n$. 
An example of the algorithm's operation is shown in Fig. \ref{FIG:algorithm}. Denoting by $\mathsf{A}:\Omega_n\to\Omega_n$ the mapping described by Algorithm \ref{ALG:path}, we have that for each $\sigma\in\mathcal{T}_n$, $\mathsf{A}\sigma\in\mathcal{S}_n$ and $\sigma\leadsto\mathsf{A}\sigma$, which proves Lemma \ref{LEMMA:absorbing_MC}.


\section{Proof of Proposition \ref{PROP:discrete_continuous_relation}}\label{APPEN:discrete_continuous_relation_proof}

Set $E_{\delta,t}:=\mathds{1}_{\big\{|N_t-t|\leq \delta t\big\}}$. For the lower bound, let $X_0^{(c)}\sim p^\star_{X_0}$ attain the maximum in $I_n^{(c)}(t)$. We have
\begin{align*}
    I_n^{(c)}(t)
    &\stackrel{(a)}\leq I\left(X_0;X_{N_t}\Big|E_{\delta,t}=1\right)+n\cdot\mathbb{P}\Big(|N_t-t|> \delta t\Big)\\
    &\stackrel{(b)}\leq I\left(X_0;X_{\lceil(1-\delta)t\rceil}\right)+n\cdot\mathbb{P}\Big(|N_t-t|> \delta t\Big)\numberthis\label{EQ:Discrete_Continuous_LB}
\end{align*}
where (a) uses the independence of $X_0^{(c)}$ and $(N_t)_{t\geq 0}$ together with $I\left(X_0^{(c)};X_t^{(c)}\Big|E_{\delta,t}=0\right)\leq \log|\Omega_n|$, while (b) is due to the DPI and the independence of $(X_t)_{t\in\mathbb{N}_0}$ and $(N_t)_{t\geq 0}$. 

Fixing $\epsilon\in\left(0,1\right)$ and setting $\delta=t^{-\frac{1-\epsilon}{2}}$, the Chernoff bound for Poisson tails gives
\begin{equation}
    \mathbb{P}\big(|N_t-t|> \delta t\big)\leq 2e^{-\frac{\sqrt{t}}{2\left(\sqrt{t}+1\right)}t^{\epsilon}}=e^{-\Theta(t^{\epsilon})}.\label{EQ:Poisson_Tail}
\end{equation}
Combining \eqref{EQ:Discrete_Continuous_LB} and \eqref{EQ:Poisson_Tail} and maximizing the RHS of \eqref{EQ:Discrete_Continuous_LB} over all $p_{X_0}$, we obtain
\begin{equation*}
    I_n(t)\geq I_n^{(c)}\big((1+o(1))t\big)-n\cdot e^{-\Theta(t^{\epsilon})},\quad \forall t\in\mathbb{N}_0.
\end{equation*}

For the upper bound, let $X_0^{(c)}\sim\tilde{p}_{X_0}$ attain the maximum in $I_n\big((1+\delta)t\big)$, where $\delta$ is as above. Consider
\ \\
\begin{align*}
    &I_n^{(c)}(t)\\
    &\geq I\big(X_0^{(c)};X_t^{(c)}\big|E_{\delta,t}\big)-I\big(X_0^{(c)};E_{\delta,t}\big| X_t^{(c)}\big)\\
    &\stackrel{(a)}\geq\mathbb{P}\big(|N_t-t|\mspace{-2mu}\leq\mspace{-2mu} \delta t\big)I\big(X_0;X_{N_t}\big|E_{\delta,t}\mspace{-2mu}=\mspace{-2mu}1\big)\mspace{-2mu}-\mspace{-2mu}h\big(\mathbb{P}\big(E_{\delta,t}=0\big)\big)\\
    &\stackrel{(b)}\geq\mathbb{P}\big(|N_t-t|\leq \delta t\big) I\left(X_0;X_{\lfloor(1+\delta)t\rfloor}\right)-h\big(\mathbb{P}\big(E_{\delta,t}=0\big)\big)\\
    &=\mathbb{P}\big(|N_t-t|\leq \delta t\big) I_n\big(\lfloor(1+\delta)t\rfloor\big)-h\big(\mathbb{P}\big(E_{\delta,t}=0\big)\big)
\end{align*}
where (a) is because $I\big(X_0^{(c)};E_{\delta,t}\big| X_t^{(c)}\big)\leq H(E_{\delta,t})=h\big(\mathbb{P}(E_{\delta,t})\big)$ with $h$ denoting the binary entropy, and (b) uses the DPI and the independence of the discrete-time dynamics and the Poisson process. Using \eqref{EQ:Poisson_Tail}, this becomes
\begin{equation*}
     I_n\big(t)\leq \left(1-e^{-\Theta(t^\epsilon)}\right)^{-1}\bigg[I_n^{(c)}\big((1-o(1))t\big)+h\left(e^{-\Theta(t^{\epsilon})}\right)\bigg],
\end{equation*}
for any $t\in\mathbb{N}_0$, which concludes the proof.


\section{Equivalent Representation of Continuous-Time Dynamics}\label{APPEN:continuout_time_equivalence}

Any continuous-time MC $(Y_t)_{t\geq 0}$ on a finite state space $\Xi$ with kernel $Q$, is characterized by its transition rates $\{c_{x,y}\}_{\substack{x,y\in\mathcal{S}\\\mspace{-13mu}x\neq y}}$, which satisfy $Q\big(Y_{t+h}=y|Y_t=x\big)=c_{x,y}h+o(h)$. The generator $\mathfrak{L}_Y$ associated with $(Y_t)_{t\geq 0}$ operates on functions $f:\Xi\to\mathbb{R}$ as $\mathfrak{L}_Yf(x)=\sum_{y\in\Xi}c_{x,y}\big(f(y)-f(x)\big)$.

This representation simplifies for Glauber dynamics. By definition (Section \ref{SUBSEC:Continuous_Time_Dynamics_Def}), if $\sigma,\eta\in\Omega_n$ with $\sigma\neq\eta$, then
\begin{align*}
    \mathbb{P}\Big(X_{t+h}^{(c)}&=\eta\Big|X_t^{(c)}=\sigma\Big)\\
    &=\mathbb{P}_\sigma\Big(X_h^{(c)}=\eta\Big)\\
    &=\sum_{k=1}^\infty\frac{e^{-h}h^k}{k!}P^k(\sigma,\eta)\\
    &=P(\sigma,\eta)\cdot h+o(h)
\end{align*}
Recall that $P(\sigma,\eta)>0$ if an only if $\eta=\sigma^v$, for some $v\in\mathcal{V}_n$, and that
\begin{equation*}
    P(\sigma,\sigma^v)=\begin{cases}\frac{1}{n},\quad\mspace{8mu}m_v(\sigma)<\ell_v(\sigma)\\\frac{1}{2n},\quad m_v(\sigma)=\ell_v(\sigma)\\0,\quad\mspace{12mu}m_v(\sigma)>\ell_v(\sigma)\end{cases}.
\end{equation*}
Denoting $c_{v,\sigma}:= P(\sigma,\sigma^v)$, for each $\sigma\in\Omega_n$ and $v\in\mathcal{V}_n$, the generator $\mathfrak{L}$ of $\big(X_t^{(c)}\big)_{t\geq 0}$ reduces to
\begin{equation*}
    \mathfrak{L}f(\sigma)=\sum_{v\in\mathcal{V}_n}c_{v,\sigma}\big[f(\sigma^v)-f(\sigma)\big],
\end{equation*}
as given in \eqref{EQ:Droplet_generator}. Accordingly, any $v\in\mathcal{V}_n$ whose spin disagrees with the majority of its neighbors (i.e., $m_v(\sigma)<\ell_v(\sigma)$) has flip rate $\frac{1}{n}$; sites with a balanced neighborhood ($m_v(\sigma)=\ell_v(\sigma)$) flip with rate $\frac{1}{2n}$; sites whose spin agrees with the majority of their neighbors ($m_v(\sigma)>\ell_v(\sigma)$) cannot flip.


\section{Proof of Theorem \ref{TM:Droplet_Erosion_Time}}\label{APPEN:Droplet_Erosion_Time_proof}

Before starting we formalize the framework. Consider the continuous-time Markov process on $\Omega_{\mathbb{Z}^2}:=\{-1,+1\}^{\mathbb{Z}^2}$, with generator $\mathfrak{L}$ acting on functions $f:\Omega_{\mathbb{Z}^2}\to\mathbb{R}$ as
\begin{equation}
    \mathfrak{L}f(\sigma)=\sum_{v\in\mathbb{Z}^2}r_{v,\sigma}\big[f(\sigma^v)-f(\sigma)\big],\label{EQ:generator}
\end{equation}
where $r_{v,\sigma}$ is the flip rate at vertex $v$ when the system is at state $\sigma$, given by (see Remark \ref{REM:Speedup_Slowdown})
\begin{equation}
    r_{v,\sigma}=\begin{cases}1,\quad\mspace{8mu}m_v(\sigma)<\ell_v(\sigma)\\\frac{1}{2},\quad\ m_v(\sigma)=\ell_v(\sigma)\\0,\quad\mspace{12mu}m_v(\sigma)>\ell_v(\sigma).\end{cases}\label{EQ:rates}
\end{equation}
With some abuse of notation, in the above, $m_v(\sigma)$ and $\ell_v(\sigma)$ are defined as in Section \ref{SUBSEC:Definitions} but w.r.t. $\mathbb{Z}^2$ (rather than the $\sqrt{n}\times\sqrt{n}$ grid). To simplify notation, we reuse $(X_t)_{t\geq0}$ for the considered Markov process and $\tau$ for the stopping time of interest (rather than $\big(\bar{X}_t^{(c)}\big)_{t\geq0}$ and $\bar{\tau}$, respectively).

A key property of Glauber dynamics used throughout the proof is its monotonicity. It enables comparing the evolution of the MC under different initializations. For each $\sigma\in\Omega_{\mathbb{Z}^2}$, let $(X_t^\sigma)_{t\geq0}$ be the continuous-time MC described by \eqref{EQ:generator}-\eqref{EQ:rates} initiated from $X_0=\sigma$. We first set up the different $(X_t^\sigma)_{t\geq0}$, for $\sigma\in\Omega_{\mathbb{Z}^2}$, over the same probability space.\footnote{Whenever it is clear from the context, the superscript $\sigma$ is omitted.} Let $\big\{T^{(v)}\big\}_{v\in\mathbb{Z}^2}$ be a collection of independent Poisson clock processes, i.e., for each $v\in\mathbb{Z}^2$, $T^{(v)}=\big\{T^{(v)}_n\big\}_{n\in\mathbb{N}_0}$, where $T^{(v)}_0=0$ and $T^{(v)}_{n+1}-T^{(v)}_n$ are i.i.d. $\mathsf{Exp}(1)$ random variables. Also let $\big\{U^{(v)}\big\}_{v\in\mathbb{Z}^2}$ be a collection of i.i.d. Rademacher processes, i.e., $U^{(v)}=\big\{U^{(v)}_n\big\}_{n\in\mathbb{N}_0}$ are i.i.d. with $U^{(v)}_n\sim2\cdot \mathsf{Ber}\left(0.5\right)-1$, for all $n\in\mathbb{N}_0$. Now, for $\sigma\in\Omega_{\mathbb{Z}^2}$, we construct $(X_t^\sigma)_{t\geq0}$ as follows:
\begin{itemize}
    \item For each $v\in\mathbb{Z}^2$, $\big(X_t^\sigma(v)\big)_{t\geq 0}$ is constant on the intervals $\big\{\big[T^{(v)}_n,T^{(v)}_{n+1}\big)\big\}_{n\in\mathbb{N}_0}$
    \item For each $v\in\mathbb{Z}^2$ and $n\in\mathbb{N}$, $X^\sigma_{T^{(v)}_n}(v)$ equals the spin of the majority of its neighbours, if a strict majority exists; otherwise, $X^\sigma_{T^{(v)}_n}(v)=U^{(v)}_n$. Note that, a.s., different sites will not update at the same time.
\end{itemize}
It is readily verified that the above setup conforms with \eqref{EQ:generator}-\eqref{EQ:rates}. We reuse $\mathbb{P}$ and $\mathbb{E}$ to denote, respectively, the probability measure and the expectation over this probability space.

Monotonicity is now stated as follows. Consider the partial ordering of $\Omega_{\mathbb{Z}^2}$ where $\sigma \geq \sigma'$ if and only if $\sigma(v)\geq \sigma'(v)$, for all $v\in\mathbb{Z}^2$. If $\sigma \geq \sigma'$, we have that $\mathbb{P}$-a.s., $X_t^{\sigma}\geq X_t^{\sigma'}$, for all $t\geq 0$. Thus, when initiated at two comparable configurations $\sigma\geq \sigma'$, the coupled chains preserve this ordering as $t$ grows. 



\subsection{Upper Bound}

We show that $\mathbb{P}_\rho\big(\tau\leq c\ell^2\big)\geq 1-\exp\left(-\left(\frac{c}{12}-\frac{1}{3}\right)\ell\right)$, for any $c>4$. Since the hitting time of $\boxminus$ is invariant to the placement of the initial droplet, we set  $X_0=\rho\in\Omega_{\mathbb{Z}^2}$ with $\rho(v)=+1$ for all $v\in[\ell]^2$, and $\rho(v)=-1$ otherwise.



To simplify the analysis, we stochastically dominate $(X_t)_{t\geq0}$ by a process that erodes the initial droplet only from a single corner. For $\sigma\in\Omega_{\mathbb{Z}^2}$ and $i,j\in\mathbb{Z}$, define
\begin{align*}
    \sigma^{i\uparrow}\big((i',j')\big)&=\begin{cases}
                -1, &i'=i\ \mbox{and}\ j'\geq \ell+1\\                \sigma\big((i',j')\big), &\mbox{otherwise}
    \end{cases},\\ \sigma^{j\rightarrow}\big((i',j')\big)&=\begin{cases}
                -1, &j'=j\ \mbox{and}\ i'\geq \ell+1\\                \sigma\big((i',j')\big), &\mbox{otherwise}
    \end{cases}.
\end{align*}
Thus, $\sigma^{i\uparrow}$ and $\sigma^{j\rightarrow}$ coincide with $\sigma$ everywhere, except possibly on $\{(i,j)\}_{j\geq \ell+1}$ or $\{(i,j)\}_{i\geq \ell+1}$, respectively, where they are set to $-1$. Let $\mathsf{Q}:\Omega_{\mathbb{Z}^2}\to\Omega_{\mathbb{Z}^2}$ be a transformation described by
\begin{equation*}
    \mathsf{Q}(\sigma)=\begin{cases}
                \sigma^{i\uparrow}\mspace{9mu},\quad \mbox{if } \sigma((i,\ell))=-1\\
                \sigma^{j\rightarrow},\quad \mbox{if } \sigma((\ell,j))=-1\\
                
                \sigma,\quad\quad\ \mbox{otherwise}
    \end{cases}
\end{equation*}
In words, $\mathsf{Q}$ finds the minus-labeled sites in $\big\{(i,\ell)\big\}_{i\in[\ell]}$ and $\big\{(\ell,j)\big\}_{j\in[\ell]}$, and vertically or horizontally continues them to an all-minus semi-infinite stripe upwards or rightwards, respectively.

Consider a modified dynamics $(\bar{Y}_t)_{t\geq0}$ on $\Omega_{\mathbb{Z}^2}$ with $\bar{Y}_0=\bar{\rho}$, where $\bar{\rho}\big((i,j)\big)=+1$ if and only if $i,j\geq 1$. Let $\bar{\tau}$ be the first time the spin at $(\ell,\ell)$ is flipped to $-1$, and set $\bar{X}_t:=\bar{Y}_{t\wedge\bar{\tau}}$. When the Poisson clock at $v\in\mathbb{Z}^2$ rings in the $\bar{X}_t$ dynamics, the spin at $v$ is refreshed and then $\mathsf{Q}$ is applied to the obtained configuration. Consequently, $(\bar{X}_t)_{t\geq0}$ coincides with $(X_t)_{t\geq0}$, when $X_0=\bar{\rho}$, except for times when the refreshed site $v=(i,j)$ has $i=\ell$ or $j=\ell$. In this case, the $\bar{X}_t$ dynamics transforms all the spins above or to the right of $v$ to $-1$. Observe that if $\bar{X}_t\big((i,j))=-1$, for some $t\geq 0$ and $i,j\in[\ell]^2$, then $\bar{X}_t\big((i',j')\big)=-1$, for all $i'\leq i$ and $j'\leq j$. Thus, $X_{\bar{\tau}}(v)=-1$, for all $v\in[\ell]^2$. Denote the probability measure, expectation, and generator associated with $(\bar{X}_t)_{t\geq0}$ by $\bar{\mathbb{P}}$, $\bar{\mathbb{E}}$, and $\bar{\mathfrak{L}}$, respectively.

One can couple $(X_t)_{t\geq0}$ and $(\bar{X}_t)_{t\geq0}$ so that $\bar{X}_t\geq X_t$, for all $t\geq 0$. Clearly, $\bar{X}_0=\bar{\rho}\geq\rho=X_0$, while for $t>0$, the coupling uses the same Poisson clocks for all sites whose neighborhoods in both dynamics coincide, and independent clocks otherwise. If up to time $t>0$ only sites inside $[\ell-1]^2$ were updated, then $\bar{X}_t\geq X_t$ trivially follows from monotonicity. Furthermore, $\bar{X}_t\geq X_t$ still holds after sites $(i,j)\in[\ell]^2$, with $i=\ell$ or $j=\ell$, are updated in the $(\bar{X}_t)_{t\geq 0}$ dynamics (after which $\mathsf{Q}$ is applied). Indeed, assume that at $t>0$ the spin $\bar{X}_t(i,\ell)$, for some $i\in[\ell]$, is flipped. The aforementioned coupling gives $\bar{X}_t\big((i',j')\big)=X_t\big((i',j')\big)=-1$, for all $i'\leq i$ and $j'\leq \ell$. Since $X_t\big((i,j')\big)=-1$, for all $j\geq \ell+1$ and $t\geq 0$, even after $\mathsf{Q}$ is applied, we have $\bar{X}_t\geq X_t$. The argument that accounts for sites $(\ell,j)$, with $j\in[\ell]$, is symmetric. Consequently, $\tau\leq\bar{\tau}$ a.s., and so
\begin{equation}
    \mathbb{P}\big(\tau> c\ell^2\big)\leq \bar{\mathbb{P}}\big(\bar{\tau}> c\ell^2\big).\label{EQ:Droplet_UB_monotone_prob}
\end{equation}


For $\sigma\in\Omega_{\mathbb{Z}^2}$, let $\mathcal{C}(\sigma):=\big\{v\in[\ell]^2:\,\sigma(v)=+1\big\}$, and define $N(\sigma):=|\mathcal{C}(\sigma)|$. To evaluate $\bar{\mathbb{E}}[\bar{\tau}]$, we study the quantity $\phi(t):=\bar{\mathbb{E}}\big[N(t)]$. Recalling that $\bar{\mathfrak{L}}$ is the generator of $(\bar{X}_t)_{t\geq0}$, we have
\begin{equation*}
    \frac{d\phi(t)}{dt}=\bar{\mathbb{E}}\bar{\mathfrak{L}}\big[N(\bar{X}_t)\big].
\end{equation*}
Furthermore, since $N(\sigma)$ depends only on the values of $\sigma$ inside $[\ell]^2$, while $\mathsf{Q}$ may alter configurations only outside $[\ell]^2$, for any $\sigma\in\Omega_{\mathbb{Z}^2}$, we have
\begin{equation}
    \bar{\mathfrak{L}}N(\sigma)=\sum_{v\in[\ell]^2}r_{v,\sigma}\big[N(\sigma^v)-N(\sigma)\big].\label{EQ:bar_generator_def}
\end{equation}

\begin{figure}[t!]
	\begin{center}
		\begin{psfrags}
			\psfragscanon
			\psfrag{A}[][][1]{$\mspace{30mu}\gamma(\sigma)$}
			\psfrag{B}[][][1]{$\mspace{-3mu}\ell$}
			\includegraphics[scale = .8]{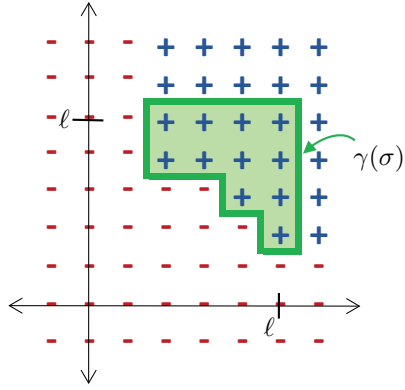}
			\caption{A typical shape of $\gamma(\sigma)$ is depicted by the green line. The region $\mathcal{D}(\sigma)$ is colored in light green.} \label{FIG:Gamma}
			\psfragscanoff
		\end{psfrags}
	\end{center}
\end{figure}

Now, let $\mathcal{D}(\sigma):=\bigcup_{v\in\mathcal{C}(\sigma)}\mathcal{B}_\infty\left(v,1/2\right)$, where $\mathcal{B}_\infty(v,r)$ is the $L^\infty$-ball of radius $r$ centered at $v$. Denote the boundary of $\mathcal{D}(\sigma)$ by $\gamma(\sigma)$ and note that the for all configurations $\sigma$ reachable from $\bar{\rho}$ in the $(\bar{X}_t)_{t\geq 0}$ dynamics (except the final configuration $\bar{X}_{\bar{\tau}}$, for which we account separately) $\gamma(\sigma)$ is a simple path. Furthermore, if $v_1(\sigma)$ and $v_2(\sigma)$ are, respectively, the bottom-right and top-left vertices in $\mathcal{C}(\sigma)$, then the shape of $\gamma(\sigma)$ is as follows: it goes directly up from $v_1(\sigma)+\left(1/2,-1/2\right)$ to $\left(\ell+1/2,\ell+1/2\right)$; from there, it proceeds directly left to $v_2(\sigma)+\left(-1/2,1/2\right)$; and finally it descends from $v_2(\sigma)+\left(-1/2,1/2\right)$ to $v_1(\sigma)+\left(1/2,-1/2\right)$ by downwards and rightwards moves over segments of positive integer lengths. A typical shape of $\gamma(\sigma)$ is shown in Fig. \ref{FIG:Gamma}.


We use $\gamma(\sigma)$ to count the number of flippable plus and minus spins, which in turn allows to upper bound $\bar{\mathfrak{L}}N(\sigma)$. Similarly to \eqref{EQ:sets_definitions}, for any $\sigma\in\Omega_{\mathbb{Z}^2}$ define $\mathcal{N}_a(\sigma):=\big\{v\in\mathbb{Z}^2:\,m_v(\sigma)>|\mathcal{N}_v|/2\big\}$ and $\mathcal{N}_d(\sigma):=\big\{v\in\mathbb{Z}^2:\,m_v(\sigma)<|\mathcal{N}_v|/2\big\}$ as the sets of vertices that agree or disagree, respectively, with the majority of their neighbors. Also let $\mathcal{N}_u(\sigma):= \mathbb{Z}^2\setminus\big\{\mathcal{N}_a(\sigma)\cup\mathcal{N}_d(\sigma)\big\}$ be the set of undecided vertices. Denote the sizes of these sets by $N_a(\sigma)$, $N_d(\sigma)$, and $N_u(\sigma)$. We also use $\mathcal{N}_a^+(\sigma)=\big\{v\in\mathcal{N}_a(\sigma):\,\sigma(v)=+1\big\}$, $\mathcal{N}_a^-(\sigma):=\mathcal{N}_a(\sigma)\setminus\mathcal{N}_a^+(\sigma)$, and analogously define $\mathcal{N}_d^+(\sigma),\ \mathcal{N}_d^-(\sigma),\ \mathcal{N}_u^+(\sigma)$ and $\mathcal{N}_u^-(\sigma)$; the cardinalities of these sets are denoted by $N_a^+(\sigma)$, $N_a^-(\sigma)$, etc. The next lemma, which is a consequence of Hopf's Umlaufsatz, is used to relate these cardinalities.

\begin{lemma}\label{LEMMA:Hopf_general}
Let $\sigma\in\Omega_{\mathbb{Z}^2}$ be a configuration with $L+1$ connected monochromatic components $\big\{\mathcal{C}_i(\sigma)\big\}_{i\in[0:L]}$ of size at least $2$, such that $\cup_{i\in[0:L]}\mathcal{C}_i(\sigma)=\mathbb{Z}^2$ and $\mathcal{C}_0(\sigma)$ is the unique infinite component which is labeled $-1$. Assume that for each $i,j\in[L]$, $d\big(\mathcal{C}_i(\sigma),\mathcal{C}_j(\sigma)\big)\geq 2$, where $d:\mathbb{Z}^2\times\mathbb{Z}^2\to\mathbb{N}_0$ is the graph distance. Let $\mathcal{E}(\sigma)$ be the collection of edges that separate cells of opposite polarity. The configuration $\sigma$ is said to be shard-free if there does not exist two parallel edges in $\mathcal{E}(\sigma)$ at $L^1$ distance 1. Assuming that $\sigma$ is shard-free, we have $N_u^+(\sigma) - N_u^-(\sigma) + 2 N_d^+(\sigma) - 2N_d^-(\sigma) = 4L$.
\end{lemma}
The proof of Lemma \ref{LEMMA:Hopf_general} is given in Appendix \ref{APPEN:Hopf_general_proof}. As a corollary, we have the following relation.

\begin{corollary}[Relation between $N_u^+(\bar{X}_t)$ and $N_u^-(\bar{X}_t)$]\label{CORR:Hopf}
For any $t\geq 0$, $N_d^+(\bar{X}_t)=N_d^-(\bar{X}_t)=0$ a.s.. Also, given $t<\bar{\tau}$, we have $N_u^+(\bar{X}_t)-N_u^-(\bar{X}_t)=1$, a.s.. 
\end{corollary}

We proceed with upper bounding the RHS of \eqref{EQ:Droplet_UB_monotone_prob}. For any $\lambda>0$, we have \begin{equation}
    \bar{\mathbb{P}}\big(\bar{\tau}> c\ell^2\big)=\bar{\mathbb{P}}\Big(N\big(\bar{X}_{c\ell^2}\big)\geq 1\Big)\leq e^{-\lambda}\bar{\mathbb{E}}e^{\lambda N\big(\bar{X}_{c\ell^2}\big)}.\label{EQ:Droplet_EXP_Markov} 
\end{equation}
Denoting $f:= e^{\lambda N}$ and $g(t):=\bar{\mathbb{E}}\big[f(\bar{X}_t)\big]$, we next show that $g(c\ell^2)$ decays exponentially with $\ell$. To do so, first observe that
\begin{equation}
    \frac{dg(t)}{dt}=\bar{\mathbb{E}}\big[\bar{\mathfrak{L}}f(\bar{X}_t)\big]=\bar{\mathbb{E}}\left[\sum_{v\in[\ell]^2}r_{v,\bar{X}_t}\big[f(\bar{X}_t^v)-f(\bar{X}_t)\big]\right].\label{EQ:derivatice_generator_relation}
\end{equation}
For any $t\geq 0$, we have
\begin{align*}
    &\bar{\mathbb{E}}\big[\bar{\mathfrak{L}}f(\bar{X}_t)\big]\\
    &\stackrel{(a)}=\bar{\mathbb{E}}\left[\sum_{v\in[\ell]^2}r_{v,\bar{X}_t}\big[f(\bar{X}_t^v)-f(\bar{X}_t)\big]\mathds{1}_{\{\bar{\tau}>t\}}\right]\\
    &=\bar{\mathbb{E}}\Bigg[\bigg(N_u^+(\bar{X}_t)\cdot f(\bar{X}_t)\frac{e^{-\lambda}-1}{2}\\
    &\qquad\qquad\qquad\qquad+N_u^-(\bar{X}_t)\cdot f(\bar{X}_t)\frac{e^{\lambda}-1}{2}\bigg)\mathds{1}_{\{\bar{\tau}>t\}}\Bigg]\\
    &\stackrel{(b)}\leq\bar{\mathbb{E}}\Bigg[f(\bar{X}_t)\bigg(\frac{\lambda}{2}\big(N_u^-(\bar{X}_t)-N_u^+(\bar{X}_t)\big)+\frac{\lambda^2}{4}N_u^+(\bar{X}_t)\\
    &\qquad\qquad\qquad\qquad\qquad+\frac{\lambda^2}{2(1-\lambda)}N_u^-(\bar{X}_t)\bigg)\mathds{1}_{\{\bar{\tau}>t\}}\Bigg]\\
    &\stackrel{(c)}\leq\left(-\frac{\lambda}{2}+\frac{\lambda^2\ell}{4}+\frac{\lambda^2(\ell-1)}{2(1-\lambda)}\right)g(t)\numberthis\label{EQ:expected_generator_UB}
\end{align*}
where (a) is because $r_{v,\bar{X}_t}\mathds{1}_{\{\bar{\tau}\leq t\}}=0$ a.s., (b) uses $e^{-\lambda}\leq 1-\lambda+\lambda^2/2$ and $e^\lambda\leq 1+\lambda+\lambda^2/(1-\lambda)$, while (c) invokes Corollary~\ref{CORR:Hopf} and the fact that $N_u^+(\bar{X}_t)\leq \ell$ and $N_u^-(\bar{X}_t)\leq \ell-1$, a.s..

Substituting $\lambda=1/(3\ell)$ into the RHS of \eqref{EQ:expected_generator_UB}, further gives $\bar{\mathbb{E}}\big[\bar{\mathfrak{L}}f(\bar{X}_t)\big]\leq -g(t)/(12\ell)$. Combined with \eqref{EQ:derivatice_generator_relation}, this produces
\begin{equation*}
    \frac{dg(t)}{dt}\leq -\frac{g(t)}{12\ell},
\end{equation*}
and since $g(0)=e^{\ell/3}$, we obtain
\begin{equation*}
    g(t)\leq e^{-\frac{t}{12\ell}+\frac{\ell}{3}}.
\end{equation*}
Inserting the latter into \eqref{EQ:Droplet_EXP_Markov} and using  \eqref{EQ:Droplet_UB_monotone_prob} finally gives
\begin{equation*}
   \mathbb{P}\big(\tau> c\ell^2\big)\leq e^{-\frac{1}{3\ell}}g(c\ell^2)\leq e^{-\left(\frac{c}{12}-\frac{1}{3}\right)\ell}.
\end{equation*}


\subsection{Lower Bound}
We next show that there exist $c,\gamma>0$ such that $\mathbb{P}_\rho\big(\tau\geq c\ell^2\big)\geq 1-e^{-\gamma\ell}$. To approximate this probability, we may switch to an auxiliary dynamics that speeds up the droplet's erosion. Furthermore, $\tau$ may be replaced with any earlier stopping time.

\begin{figure}[t!]
	\begin{center}
		\begin{psfrags}
			\psfragscanon
		    \psfrag{D}[][][1]{$\mspace{-3mu}-k$}
		    \psfrag{B}[][][1]{$\mspace{-3mu}k$}
		    \psfrag{C}[][][1]{$\mspace{-3mu}-k$}
			\includegraphics[scale = .65]{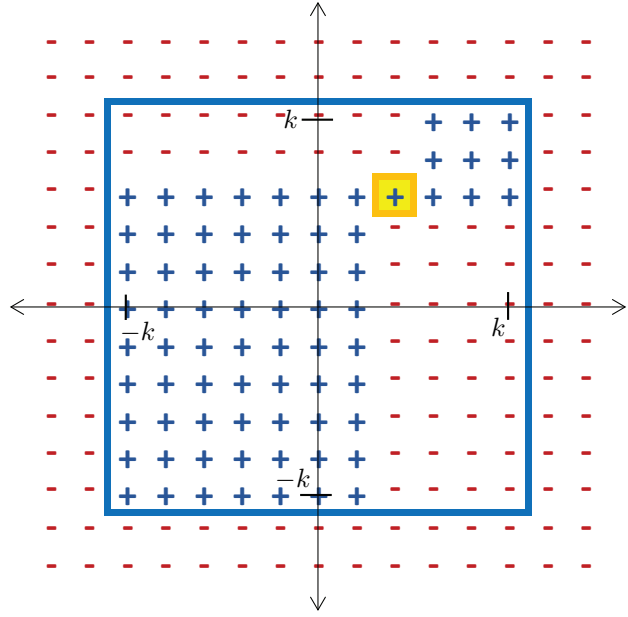}
			\caption{An illustration of $\eta_{v_0}$, for $k=5$ and $v_0=(2,3)$. The unstable site $v_0$ is highlighted in yellow. The blue line is the contour of the original square droplet of pluses. If the clock at $v_0$ rings, $\mathcal{C}(\eta_{v_0})$ seizes to be connected.} \label{FIG:bad_droplet}
			\psfragscanoff
		\end{psfrags}
	\end{center}
\end{figure}

For convenience, assume that $\ell=2k+1$, for some $k\in\mathbb{N}$. 
Furthermore, due to translation invariance of the droplet dynamics over $\mathbb{Z}^2$, assume that the droplet is centered at $(0,0)\in\mathbb{Z}^2$ and set $\mathcal{C}(\sigma):=\big\{v\in[-k:k]^2:\,\sigma(v)=+1\big\}$. The main challenge in analyzing $(X_t)_{t\geq0}$ is that while $\mathcal{C}(\rho)$ is a connected components, it can split into two disconnected components as the dynamics evolve. For instance, consider flipping the spin at  $v_0=(i_0,j_0)\in[-(k-2):k-2]^2$ in the $\eta_{v_0}$ configuration shown Fig. \ref{FIG:bad_droplet} (and reachable from $\rho$):
\begin{equation*}
    \eta_{v_0}(v)=\begin{cases} +1,\quad v\in\left\{\begin{array}{lll}
         &\mspace{-25mu}[i_0+1:k]\times[j_0:k]\\
         &\mspace{-25mu}[-k:i_0-1]\times[-k:j_0]\\
         &\mspace{-25mu}\big\{(i_0,j_0)\big\}
    \end{array}\right.\\
    -1, \quad\mbox{otherwise}    \end{cases}\mspace{-15mu},
\end{equation*}


We circumvent this issue by introducing several modifications to the original dynamics. First, define $\tilde{\rho}\in\Omega_{\mathbb{Z}^2}$ as
\begin{equation*}
    \tilde{\rho}(v)=\begin{cases} +1,& v\in[-k:k]^2\cap \mathcal{B}_1(0,k+1)\\
    -1,&\mbox{otherwise}.
    \end{cases}
\end{equation*}
Next, let $m=\lfloor 3k/4\rfloor$ and set
$$\mathcal{T}_m=\big\{(i,0)\big\}_{\substack{i\in[-m:m]\\j\in[-1:1]}}\cup\big\{(0,j)\big\}_{\substack{i\in[-1:1]\\j\in[-m:m]}}.$$
Finally, define the transformation  $\mathsf{E}:\Omega_{\mathbb{Z}^2}\to\Omega_{\mathbb{Z}^2}$ as
\begin{equation*}
    \mathsf{E}(\sigma)=\begin{cases} \sigma^v,\quad \mbox{if } \exists!\ v\in\mathcal{N}_d(\sigma)\\
    \sigma\mspace{8mu},\quad\mbox{otherwise}\end{cases},
\end{equation*}
whose role is to eliminate unstable sites.


Consider an auxiliary dynamics $(\tilde{Y}_t)_{t\geq0}$ on $\Omega_{\mathbb{Z}^2}$ initiated at  $\bar{Y}_0=\tilde{\rho}\in\Omega_{\mathbb{Z}^2}$. For $t>0$, whenever the Poisson clock at $v\in\mathbb{Z}^2$ rings, update the current spin at $v$, and then apply $\mathsf{E}$ to the obtained configuration. Let $\tilde{\tau}$ be the first time when the spin of any of the sites in $\mathcal{T}_m$ is flipped, and  set $\tilde{X}_t=\tilde{Y}_{t\wedge\tilde{\tau}}$, for all $t\geq 0$. The probability measure, expectation and generator associated with $(\tilde{Y}_t)_{t\geq0}$ are denoted by $\tilde{\mathbb{P}}$, $\tilde{\mathbb{E}}$ and $\tilde{\mathfrak{L}}$, respectively. Like before, since $\tilde{\rho}\leq \rho$ and $\tilde{\tau}<\tau$ $\mathbb{P}$-a.s., we have
\begin{equation*}
    \mathbb{P}_\rho\big(\tau<c\ell^2\big)\leq\mathbb{P}_{\tilde{\rho}}\big(\tilde{\tau}<c\ell^2\big).
\end{equation*}
Furthermore, getting the desired upper bound on the RHS above with $k$ instead of $\ell$ is sufficient as $2k\leq \ell \leq 3k$. 

Using a coupling argument similar to that presented in the proof of the upper bound, we can define both chains over the same probability space, such that $\tilde{Y}_t\leq X_t$, for any $t\geq 0$. Thus, to prove $\mathbb{P}_\rho\big(\tau\geq c\ell^2\big)\geq 1-e^{-\gamma\ell}$, it suffices to establish
\begin{equation*}
    \tilde{\mathbb{P}}\big(\tilde{\tau}<ck^2\big)\leq e^{-\gamma k}.
\end{equation*}

To that end, we control the difference between the number of unstable plus- and minus-labeled sites, which again relies on Hopf's Umlaufsatz. Define 
\begin{equation*}
    \mathcal{D}(\sigma):=\bigcup_{v\in\mathcal{C}(\sigma)}\mathcal{B}_\infty\left(v,\frac{1}{2}\right).
\end{equation*}
and denote its boundary by $\gamma(\sigma)$. 
To apply Hopf's Umlaufsatz, we split the set of unstable plus-labeled sites, $\mathcal{N}_u^+(\sigma)$, into two non-intersecting subsets:
\begin{subequations}
\begin{align}
    \mathcal{N}_1^+\mspace{-2mu}(\sigma)&\mspace{-2mu}:=\mspace{-2mu}\left\{\mspace{-3mu}v\mspace{-2mu}\in\mspace{-2mu}\mathcal{N}_u^+\mspace{-2mu}(\sigma)\mspace{-2.5mu}:\mspace{-25mu}\begin{array}{lcl}
         & \sigma\big(v\mspace{-2mu}+\mspace{-2mu}(1,0)\big)\mspace{-2.5mu}=\mspace{-2mu}\sigma\big(v\mspace{-2mu}-\mspace{-2mu}(1,0)\big)\mspace{-2.5mu}=\mspace{-2mu}+\mspace{-2mu}1 \\
         &\mspace{10mu}\mbox{or}\\
         & \sigma\big(v\mspace{-2mu}+\mspace{-2mu}(0,1)\big)\mspace{-2.5mu}=\mspace{-3mu}\sigma\big(v\mspace{-2mu}-\mspace{-2mu}(0,1)\big)\mspace{-2.5mu}=\mspace{-3mu}+\mspace{-2mu}1
    \end{array}
    \mspace{-9mu}\right\}\label{EQ:droplet_LB_N1set}\\
    \mathcal{N}_2^+(\sigma)&:=\mathcal{N}_u^+(\sigma)\setminus\mathcal{N}_1^+(\sigma).
\end{align}
\end{subequations}
The set $\mathcal{N}_1^+(\sigma)$ contains all the sites of positive spin whose plus-labeled neighbors are either above and below them or to their left and right; all the rest are in $\mathcal{N}_2^+(\sigma)$. Furthermore, for any $\sigma\in\Omega_{\mathbb{Z}^2}$, let
\begin{equation*}
\mathcal{M}_u^+(\sigma):= \Big\{v\in\mathcal{N}_u^+(\sigma):\,\mathcal{N}_d(\sigma^v)\neq \emptyset\Big\}
\end{equation*}
be the set of sites in $\sigma$ whose flipping creates at least one unstable site in the obtained configuration.

\begin{lemma}[Properties of $\mathcal{C}(\tilde{X}_t)$]\label{LEMMA:Regularity}
For any $t\geq 0$, the following holds a.s.:
\begin{enumerate}
    \item $\mathcal{N}_d^+(\tilde{X}_t)=\mathcal{N}_d^-(\sigma)=\emptyset$;
    \item $\gamma(\tilde{X}_t)$ is a simple and closed curve in $\mathbb{R}^2$;
    \item For every    $v\in\mathcal{M}_u^+(\tilde{X}_t)$, we have $\big|\mathcal{C}\big(\mathsf{E}(\tilde{X}_t^v)\big)\big|=|\mathcal{C}(\tilde{X}_t)|-2$. Furthermore, $M_u^+(\tilde{X}_t):=\big|\mathcal{M}_u^+(\tilde{X}_t)\big|\leq 8$;
    \item $\mathcal{N}_1^+(\tilde{X}_t)=\emptyset$.
\end{enumerate}
\end{lemma}
See Appendix \ref{APPEN:Regularity_proof} for a proof. Together with Lemma \ref{LEMMA:Hopf_general}, Lemma \ref{LEMMA:Regularity} implies that for all $t\geq 0$, we a.s. have
\begin{equation}
    N_u^+(\tilde{X}_t)-N_u^-(\tilde{X}_t)=4.\label{EQ:Droplet_LB_Hopf}
\end{equation}

Recall that the dynamics $(\tilde{X}_t)_{t\geq0}$ stops when a spin inside $\mathcal{T}_m$ is flipped. Thus, the first site $(i,j)\in\mathcal{T}_m$ to flip must have $i\in\{-1,1\}$ and $j\in\{-m,m\}$. Assume without loss of generality that $(1,m)$ is the first to become unstable. For this to happen, all the plus-labeled sites inside $\mathcal{G}_l=\big\{(i,j):\,i\leq 0, j\geq m \big\}$ or $\mathcal{G}_r=\big\{(i,j):\,i\geq 2, j\geq m \big\}$ must have already flipped. Consequently, if $\tilde{\tau}<ck^2$, then 
\begin{equation*}
    N(\tilde{X}_0)-N(\tilde{X}_{ck^2})\geq |\mathcal{G}_r|\geq \frac{\left(\frac{1}{5}k\right)^2}{2}=\frac{k^2}{50},
\end{equation*}
and we have
\begin{align*}
    \tilde{\mathbb{P}}\big(\tilde{\tau}<ck^2\big)&\leq \tilde{\mathbb{P}}\left( N(\tilde{X}_0)-N(\tilde{X}_{ck^2})\geq \frac{k^2}{50}\right)\\
    &\leq e^{-\frac{\lambda k^2}{50}}\tilde{\mathbb{E}}\left[e^{\lambda \big(N(\tilde{\rho})-N(\tilde{X}_{ck^2})\big)}\right],\numberthis\label{EQ:Droplet_LB_EXP_Markov} 
\end{align*}
for any $\lambda>0$. Setting $f_\lambda(\sigma):= \exp\big(\lambda\big( N(\tilde{\rho})-N(\sigma)\big)\big)$ and $g_\lambda(t):=\tilde{\mathbb{E}}\big[f_\lambda(\tilde{X}_t)\big]$, we use the fact that $$\frac{dg_\lambda(t)}{dt}=\tilde{\mathbb{E}}\big[\tilde{\mathfrak{L}}f_\lambda(\tilde{X}_t)\big]$$
to approximate the expected value. Expanding the RHS
\begin{align*}
    &\tilde{\mathbb{E}}\big[\tilde{\mathfrak{L}}f_\lambda(\tilde{X}_t)\big]\\
    &\mspace{-3mu}\stackrel{(a)}=\tilde{\mathbb{E}}\left[\sum_{v\in\mathbb{Z}^2}r_{v,\tilde{X}_t}\big[f_\lambda(\tilde{X}_t^v)-f_\lambda(\tilde{X}_t)\big]\mathds{1}_{\{\bar{\tau}>t\}}\right]\\
    &\mspace{-3mu}=\tilde{\mathbb{E}}\bigg[f_\lambda(\tilde{X}_t)\bigg(M_u^+(\tilde{X}_t)\frac{e^{2\lambda}-1}{2}\\
    &\mspace{-3mu}\qquad\qquad\qquad\qquad+\big(N_u^+(\tilde{X}_t)-M_u^+(\tilde{X}_t)\big)\frac{e^{\lambda}-1}{2}\\
    &\mspace{-3mu}\qquad\qquad\qquad\qquad\qquad\qquad+N_u^-(\tilde{X}_t)\frac{e^{-\lambda}-1}{2}\bigg)\mathds{1}_{\{\bar{\tau}>t\}}\bigg]\\
    &\mspace{-3mu}\stackrel{(b)}=\tilde{\mathbb{E}}\bigg[f_\lambda(\tilde{X}_t)\bigg(\frac{\lambda\big(M_u^+(\tilde{X}_t)+N_u^+(\tilde{X}_t)-N_u^-(\tilde{X}_t)\big)}{2}\\
    &\mspace{-3mu}\qquad\quad\quad\mspace{-5mu}+\frac{\lambda^2\big(M_u^+(\tilde{X}_t)+N_u^+(\tilde{X}_t)+N_u^-(\tilde{X}_t)\big)}{2(1-\lambda)}\bigg)\mathds{1}_{\{\bar{\tau}>t\}}\bigg]\\ &\mspace{-3mu}\stackrel{(c)}\leq\mspace{-3mu}\tilde{\mathbb{E}}\mspace{-3mu}\left[\mspace{-1mu}f_\lambda\mspace{-1mu}(\mspace{-1mu}\tilde{X}_t\mspace{-1mu})\mspace{-4mu}\left(\mspace{-4mu}6\lambda\mspace{-3mu}+\mspace{-3mu}\frac{\lambda^2\mspace{-2mu}\big(\mspace{-1mu}M_u^+\mspace{-3mu}(\mspace{-1mu}\tilde{X}_t\mspace{-1mu})\mspace{-3mu}+\mspace{-3mu}N_u^+\mspace{-3mu}(\mspace{-1mu}\tilde{X}_t\mspace{-1mu})\mspace{-3mu}+\mspace{-3mu}N_u^-\mspace{-3mu}(\mspace{-1mu}\tilde{X}_t\mspace{-1mu})\big)}{2(1-\lambda)}\mspace{-3mu}\right)\mspace{-4.5mu}\mathds{1}_{\mspace{-2mu}\{\bar{\tau}>t\}}\mspace{-3.5mu}\right]\\
    &\mspace{-3mu}\stackrel{(d)}\leq\bigg(6\lambda+\frac{4\lambda^2(k+2)}{1-\lambda}\bigg)g_\lambda(t).\numberthis\label{EQ:droplet_LB_expected_generator_UB}
\end{align*}
where:\\
(a) is because $c_{v,\tilde{X}_t}\mathds{1}_{\{\bar{\tau}\leq t\}}=0$;\\
(b) follows by $e^{2\lambda}\leq 1+2\lambda+2\lambda^2/(1-\lambda)$ and $e^{-\lambda}\leq e^\lambda\leq 1+\lambda+\lambda^2/(1-\lambda)$;\\
(c) follows by Claim 3) from Lemma \ref{LEMMA:Regularity} and \eqref{EQ:Droplet_LB_Hopf};\\
(d) is since 
$N_u(\tilde{X}_t)\leq \big|\partial_\mathsf{Int}\mathcal{C}(\tilde{\rho})\big|+\big|\partial_\mathsf{Ext}\mathcal{C}(\tilde{\rho})\big|\leq 8(k+1)$.

\vspace{2mm}

Inserting $\lambda=1/k$ into the RHS of \eqref{EQ:droplet_LB_expected_generator_UB} and denoting $g(t):= g_{1/k}(t)$, we see the there exists $c'>0$ such that $$\frac{dg(t)}{dt}\leq \frac{c'}{k}g(t).$$
Solving with the boundary condition $g(0)=1$ gives $g(t)\leq \exp(\frac{c'}{k}t)$. With this proxy, we upper bound the RHS of \eqref{EQ:Droplet_LB_EXP_Markov} as
\begin{equation*}
    \tilde{\mathbb{P}}\big(\tilde{\tau}<ck^2\big)\leq e^{-\frac{k}{50}}g(ck^2)\leq e^{-\left(\frac{1}{50}-cc'\right)k}. 
\end{equation*}
Taking $c\in\left(0,\frac{1}{50c'}\right)$ and denoting $\gamma:= \frac{1}{50}-cc'>0$ concludes the proof.



\section{Proof of Theorem \ref{TM:One_Strip}}\label{APPEN:One_Strip_proof}

An alternative yet equivalent description of the continuous-time dynamics uses Poisson thinning. 
The idea is to absorb the the update probability into the rate of the Poisson clock associated with each vertex. Thus, each $v\in\mathcal{V}_n$ is associated with a time-varying Poisson process $\big(N^{(v)}_t\big)_{t\geq 0}$ whose rate $\lambda_v(t)$ depends on the instantaneous neighborhood of $v$ through
\begin{equation}
    \lambda_v(t)=\begin{cases}\phi\big(S(X_t,v)\big),\quad\quad\ \ X_t(v)=+1\\
    1-\phi\big(S(X_t,v)\big),\quad X_t(v)=-1\end{cases}.\label{EQ:flip_rate_original}
\end{equation}
With the thinned rates, each time $\big(N^{(v)}_t\big)_{t\geq 0}$ jumps, the spin at $v$ flips. To see the equivalence between the two descriptions, just consider the generator of each dynamics.

\begin{figure}[t!]
	\begin{center}
		\begin{psfrags}
		\psfragscanon
		\psfrag{A}[][][1]{\ \ \ $t$}
		\psfrag{B}[][][1]{${\color{red!85!black}\lambda_v(t)}$}
		\psfrag{C}[][][1]{$0$}
		\psfrag{D}[][][1]{$T_w\wedge T_u$}
		\psfrag{E}[][][1]{$T_w\vee T_u$}
		\psfrag{F}[][][1]{\ \ $T_v$}
		\psfrag{G}[][][1]{${\color{red!85!black}\mspace{-25mu}\phi(-1)}$}
		\psfrag{H}[][][1]{${\color{red!85!black}\mspace{-10mu}\phi(1)}$}
		\psfrag{I}[][][1]{${\color{red!85!black}\phi(3)}$}
		\includegraphics[scale = 1.1]{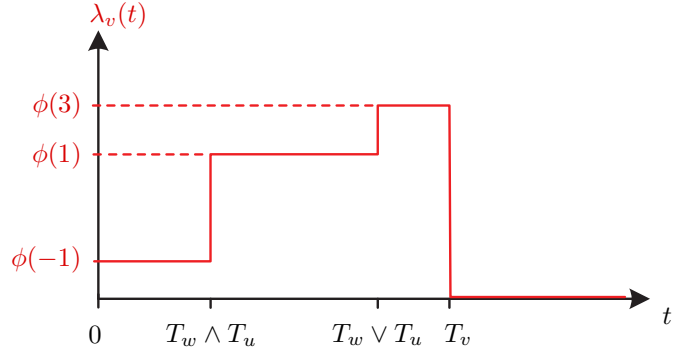}
		\caption{A sample path of the flip rate process $\lambda_v(t)$ of a vertex 
		$v$ with two horizontal neighbors $u$ and $w$. The flip times (from $+1$ to $-1$) of $u$, $v$ and $w$ are denoted by $T_u$, $T_v$ and $T_w$. The illustration assumes $T_v>T_u\vee T_w$.} \label{FIG:rate_evolution}
		\psfragscanoff
		\end{psfrags}
	\end{center}
\end{figure}

We first restrict minus-labeled sites from flipping. With some abuse of notation, we replace \eqref{EQ:flip_rate_original} with \begin{equation}
\lambda_v(t)=\phi\big(S(X_t,v)\big)\mathds{1}_{\big\{X_t(v)=+1\big\}},\label{EQ:clock_rates_fixed}    
\end{equation}
thus assigning flip rate 0 to minus-labeled sites. With this change, until the spin at $v\in\mathcal{B}$ flips to $-1$, $\lambda_v(t)$ is a piece-wise constant monotonically non-decreasing process that jumps when $v$'s horizontal neighbors flip their spins. For example, any any $v\in\mathcal{B}_\mathsf{Int}:=\mathcal{B}\setminus\big\{(1,1),(\sqrt{n},1)\big\}$ has rate
\begin{equation}
    \lambda_v(t)=\begin{cases}\phi(-1),& N_v^{(-)}(t)=1\ \mbox{and}\ X_t(v)=+1\\
    \phi(1),&  N_v^{(-)}(t)=2\ \mbox{and}\ X_t(v)=+1\\
    \phi(3),&  N_v^{(-)}(t)=3\ \mbox{and}\ X_t(v)=+1\\
    0,& X_t(v)=-1\end{cases}\label{EQ:flip_rate_simple}
\end{equation}
where $N_v^{(-)}(t)=\big|\big\{u\in\mathcal{N}_v:\,X_t(u)=-1\big\}\big|$. A sample path of $\lambda_v(t)$ for the case when $v$ flips after both its horizontal neighbors is shown in Fig. \ref{FIG:rate_evolution}. Further abusing notation, we still use $(X_t)_{t\geq 0}$ for the restricted dynamics under which flips of minus-labeled sites are prohibited. By monotonicity of the SIM, the expected number of pluses in $\mathcal{B}$ w.r.t. the restricted dynamics cannot be higher than the original one.


Our next step is to replace  $(X_t)_{t\geq 0}$ with a new dynamics with two pure (sprinkling and erosion) phases, at least up to a certain point. To establish terminology, any plus-labeled site with $N_v^{(-)}(t)=i$, where $i=1,2,3$, is referred to as a site of Type $i$. When the spin at $v$ is $-1$, its Type is 0. Accordingly, a \emph{sprinkle} is a flip of a Type 1 site; flips of Type 2 or 3 sites are \emph{erosion} flips; Type 0 sites never flip. Our goal is, for any fixed $s>0$, to construct from $\big(X_t\big)_{t\geq 0}$ a new process $\big(X_t^{(s)}\big)_{t\geq 0}$ over the same probability space, such that 
\begin{enumerate}
    \item For $t\in[0,s)$, only Type 1 sites are flipped in $X_t^{(s)}$ (sprinkling phase).
    
    \item For $t\in[s,2s)$, only Type 2 or 3 sites are flipped in $X_t^{(s)}$ (erosion phase).
    
    \item $X_s\geq X_{2s}^{(s)}$.
\end{enumerate}
Here $s>0$ is a design parameter that we tune for the purpose of the proof. Provided that such a process exists, setting $s=t_0:= \exp(c\beta)$ and letting $\hat{N}_\mathcal{B}^{(+)}(t):=\big|\big\{v\in\mathcal{B}:\,X_t^{(t_0)}(v)=+1\big\}\big|$ be the number of pluses in $\mathcal{B}$ w.r.t. $\big(X_t^{(t_0)}\big)_{t\geq 0}$, we have
\begin{equation}
\mathbb{E}\big[N_1^{(+)}(t_0)\big]\geq \mathbb{E}\big[\hat{N}_\mathcal{B}^{(+)}(2t_0)\big].\label{EQ:sufficient_one_strip}
\end{equation}
Note that due to monotonicity $\mathbb{E}\big[N_1^{(+)}(t)\big]\geq \mathbb{E}\big[N_1^{(+)}(t_0)\big]$, for any $t\leq t_0$, proving  \eqref{EQ:sufficient_one_strip} is sufficient to establish Theorem \ref{TM:One_Strip}. Since the RHS of \eqref{EQ:sufficient_one_strip} corresponds to the phase-separated dynamics, it allows a much simpler per-phase analysis.

\textbf{Construction of Phase-Separated Dynamics:} 
Fix $s>0$, and consider first an auxiliary dynamics $\big(\tilde{X}_t^{(s)}\big)_{t\in[0,2s)}$ constructed as follows. 
For $t\in[0,2s)$ and any $v\in\mathcal{V}_n$, define 
\begin{equation*}
    \tilde{N}_t^{(v)}=\begin{cases}
    N_t^{(v)},& t\in[0,s)\\
    N_{t-s}^{(v)},& t\in[s,2s)
    \end{cases},
\end{equation*}
i.e., $\big(\tilde{N}_t^{(v)}\big)_{t\in[0,2s)}$ consecutively repeats the sequence of $\big(N_t^{(v)}\big)_{t\in[0,s)}$ clock rings twice. The process $\big(\tilde{N}_t^{(v)}\big)_{t\in[0,2s)}$ specifies the clock rings for $\big(\tilde{X}_t^{(s)}\big)_{t\in[0,2s)}$, but as explained next, not all rings result in a flip.

Specifically, assume $k_s$ spin flips occurred in $(X_t)_{t\geq 0}$ during $[0,s)$. Let $t_1<t_2<\ldots<t_{k_s}$ be the flip times and $v_1,v_2,\ldots,v_{k_s}$ be the corresponding sites. Note that the $v_i$'s, for $i\in[k_s]$, are all distinct since only plus-labeled sites may flip (minus-labeled sites have flip rate 0). We define the clock ring times $\{\tau_j\}_{j\in[2k_s]}$ and their corresponding sites $\{u_j\}_{j\in[2k_s]}$ for the $\big(\tilde{X}_t^{(s)}\big)_{t\in[0,2s)}$ dynamics as follows: 
\begin{align*}
    \tau_j&=\begin{cases}t_j,& j\in[k_s]\\
    t_{j-k_s}+s,& j\in[k_s+1:2k_s]\end{cases}\\ u_j&=\begin{cases}v_j,& j\in[k_s]\\
    v_{j-k_s},& j\in[k_s+1:2k_s]\end{cases}.
\end{align*}
Some of the flips corresponding to these clock rings are blocked as described next:
\begin{itemize}
    \item For $j\in[k_s]$, if in the original $\big(X_t\big)_{t\in[0,s)}$ we have $N_{u_j}^{(-)}(\tau_j)\in\{2,3\}$, i.e., the $j$-th update was an erosion flip, then it is blocked with probability 1. Sprinkling updates, for which $N_{u_j}^{(-)}(\tau_j)=1$, are allowed.
    
    \item For $j\in[k_s+1:k_s]$, if in the original dynamics $N_{u_j}^{(-)}(\tau_j-s)=1$ (sprinkle), then it is blocked with probability 1. Erosion flips, for which $N_{u_i}^{(-)}(\tau_j-s)\in\{2,3\}$, go through.
\end{itemize}
By construction, $\big(\tilde{X}_t^{(s)}\big)_{t\in[0,2s)}$ is a phase-separated dynamics. It reorders the flips in $\big(X_t\big)_{t\in[0,s)}$ (in a stretched~time interval) to have only Type 1 updates during $[0,s)$, and only updates of Types 2 or 3 in $[s,2s)$. Clearly, $X_s=\tilde{X}^{(s)}_{2s}$. Note that while the two phases of $\big(\tilde{X}_t^{(s)}\big)_{t\in[0,2s)}$ have the same clock rings, the rates of the rings during the 2nd phase are incorrect. Specifically, since $\tilde{X}_s^{(s)}\leq X_0$, the rates of $\big(N_t^{(v)}\big)_{t\in[0,s)}$, for each $v\in\mathcal{B}$, are (possibly) too low to reflect the neighborhood of $v$ at times $t\in[s,2s)$. 


Our final step in the construction is to fix the flip rates at the 2nd phase. Consider a third dynamics $\big(X_t^{(s)}\big)_{t\in[0,2s)}$, with a ring process $\big(M_t^{(v)}\big)_{t\in[0,2s)}$ defined as follows. Let $\big\{\big(L_t^{(v)}\big)_{t\geq 0}\big\}_{v\in\mathcal{B}}$
be a collection of independent Poisson processes (amongst themselves and independent of $\big\{\big(N_t^{(v)}\big)_{t\geq 0}\big\}_{v\in\mathcal{B}}$), each with rate
\begin{equation*}
    \eta_v(t):= \Big[\phi\big(S(\tilde{X}_{t+s}^{(s)},v)\big)-\phi\big(S(X_t,v)\big)\Big]\mathds{1}_{\big\{\tilde{X}_{t+s}(v)=+1\big\}}.
\end{equation*}
Since $\tilde{X}_s^{(s)}\leq X_0$, and because so long as $v$ has a positive spin its flip rate is monotonically non-decreasing with $t$, we have that $\eta_v(t)\geq 0$. Furthermore, $\eta_v(t)+\lambda_v(t)=\phi\big(S(\tilde{X}_{t+s}^{(s)},v)\big)\mathds{1}_{\big\{\tilde{X}_{t+s}(v)=+1\big\}}$ (see \eqref{EQ:clock_rates_fixed}), which is the correct rate of clock rings for $v$ in the 2nd phase of the $\big(\tilde{X}_t^{(s)}\big)_{t\in[0,2s)}$ process. Defining
\begin{equation*}
    M_t^{(v)}=\begin{cases}
    N_t^{(v)},&t\in[0,s)\\
    N_{t-s}^{(v)}+L_t^{(v)},& t\in[s,2s)
    \end{cases}
\end{equation*}
as the the ring process for $\big(X_t^{(s)}\big)_{t\in[0,2s)}$, we see that it is Poisson with rate 
\begin{equation*}
    \mu_v(t)=\phi\big(S(\tilde{X}_t^{(s)},v)\big)\mathds{1}_{\big\{\tilde{X}_t(v)=+1\big\}},\quad\forall t\in[0,2s)
\end{equation*}
as desired. To preserve the phase-separated property, we apply the same flip-blocking rule on $\big(X_t^{(s)}\big)_{t\in[0,2s)}$. Namely, for $t\in[0,s)$, flips associated of Type 2 or 3 sites are suppressed, while for $t\in[s,2s)$, we block flips of Type 1. 


To conclude, $\big(X_t^{(s)}\big)_{t\in[0,2s)}$ is a phase-separated dynamics that satisfies $X_{2s}^{(s)}\leq \tilde{X}_{2s}^{(s)}=X_s$ (since $M_t^{(v)}\geq \tilde{N}_t^{(v)}$). On account of \eqref{EQ:sufficient_one_strip}, we now focus on bounding $\mathbb{E}\big[\hat{N}_\mathcal{B}^{(+)}(2t_0)\big]$ for $\hat{N}_\mathcal{B}^{(+)}(t):=\big|\big\{v\in\mathcal{B}:\,X_t^{(t_0)}(v)=+1\big\}\big|$.



\vspace{2mm}
\textbf{Analysis of Phase-Separated Dynamics:} 

\indent{1) \underline{Sprinkling Phase}:} Set $X_0^{(t_0)}\big((1,1)\big)=X_0^{(t_0)}\big((\sqrt{n},1)\big)=-1$, which only further decreases $\mathbb{E}\big[\hat{N}_\mathcal{B}^{(+)}(2t_0)\big]$. Doing so prohibits $(2,1)$ and $(\sqrt{n}-1,1)$ from flipping during the 1st phase, while ensuring that every $v\in\hat{\mathcal{B}}:=\mathcal{B}_\mathsf{Int}\setminus\big\{(2,1),(\sqrt{n}-1,1)\big\}$ has the same neighborhood at $t=0$.


At the end of Phase 1, $X_{t_0}^{(t_0)}(\mathcal{B})$ comprises runs of plus-labeled vertices separated by minus-labeled sprinkles that occurred during the first phase. We refer to these contiguous runs of pluses as contigs. For $v=(i,1)\in\mathcal{B}$ with $X_{t_0}^{(t_0)}(v)=-1$, we denote by $L_i$ the length of the contig that begins at $v$ (including $v$), namely:
\begin{align*}
    &L_i:= 1+\\
    &\mspace{-4mu}\max\mspace{-3mu}\left\{\mspace{-2mu}j\mspace{-2mu}\in\mspace{-2mu}[i\mspace{-2mu}+\mspace{-2mu}1\mspace{-3mu}:\mspace{-3mu}\sqrt{n}]\mspace{-2mu}:\forall \ell\mspace{-2mu}\in\mspace{-2mu}[i\mspace{-2mu}+\mspace{-2mu}1\mspace{-3mu}:\mspace{-3mu}j]\mspace{-17mu}\begin{array}{lll}
    &X_{t_0}^{(t_0)}\big((\ell,1)\big)\mspace{-2mu}=\mspace{-2mu}+\mspace{-2mu}1\\
    &\qquad\quad\mbox{and}\\
    &X_{t_0}^{(t_0)}\mspace{-2mu}\big((j\mspace{-2mu}+\mspace{-2mu}1,1)\big)\mspace{-2mu}=\mspace{-2mu}-\mspace{-2mu}1
    \end{array} \mspace{-8mu} \right\}\mspace{-3mu}.
\end{align*}
The flip time of any $v\in\hat{\mathcal{B}}$ with plus-labeled horizontal neighbors is  $\mathsf{Exp}\big(\phi(-1)\big)$. However, if either of $v$'s horizontal neighbors flip to $-1$ before $v$ itself, the spin at $v$ stays $+1$ until the end of Phase 1. We may speed up the disappearance of pluses by allowing the sites in $\hat{\mathcal{B}}$ to flip independently after $\mathsf{Exp}\big(\phi(-1)\big)$ time during~$[0,t_0)$. With this relaxation, the probability that $v\in\hat{\mathcal{B}}$ flips its spin until time $t_0$ is
\begin{equation*}
    \mathbb{P}\Big(\mathsf{Exp}\big(\phi(-1)\big)\leq t_0\Big)=1-e^{-\phi(-1)t_0}:= p(\beta),
\end{equation*}
where the dependence of $p(\beta)$ on $\beta$ is through the exponent $$\phi(-1)t_0=\frac{e^{(c-1)\beta}}{e^{-\beta}+e^{\beta}}.$$
As $c<1$ (in fact, $c<2$ suffices here), $p(\beta)$ can be made arbitrarily small by increasing $\beta$.

Under the independent flip times, the contigs' lengths are mutually independent truncated geometric random variable. Namely, $L_i=G_i\wedge (\sqrt{n}-1-i)$, where $\{G_i\}$ are i.i.d. $\mathsf{Geo}\big(p(\beta)\big)$. Consider the total lengths of the first $M$ contigs $L_M:=\sum_{i=1}^M L_i$. For an appropriate choice of $M$, we show that the probability of $L_M$ exceeding the right corner $v_r=(\sqrt{n},1)$ of $\mathcal{B}$ can be made arbitrarily small with $n$, for any $\beta>0$. Setting $\mu:= \mathbb{E}\big[\sum_{i=1}^MG_i\big]= M/p(\beta)$, we have
\begin{align*}
    \mathbb{P}\Big(L_M\geq \sqrt{n}\Big)&\leq \mathbb{P}\left(\sum_{i=1}^M G_i\geq \sqrt{n}\right)\\
    &=\mathbb{P}\left(\sum_{i=1}^M G_i\geq \frac{p(\beta)\sqrt{n}}{M}\mu\right).
\end{align*}
Taking $M=\frac{p(\beta)}{2}\sqrt{n}$ and bound the tail probability of a sum of geometric random variables (see, e.g., \cite[Theorem 2.1]{Janson_Geo_Tail2018}) to obtain
\begin{equation*}
    \mathbb{P}\Big(L_M\geq \sqrt{n}\Big)\leq e^{-\frac{1}{2}p(\beta)(1-\ln2)\sqrt{n}}:= \theta_n(\beta).
\end{equation*}
For any $\beta>0$, $\theta_n(\beta)$ can be made arbitrarily small by taking $n$ large enough. This alleviates the need to deal with boundary effect when considering the first $M$ contigs formed after the sprinkling phase.

We next analyze the length of each such contig. On the event that $\big\{L_M\leq \sqrt{n}-1\big\}$, $L_i$ has a geometric distribution with parameter $p(\beta)$. Let $G\sim\mathsf{Geo}\big(p(\beta)\big)$ and denote $\ell(\beta):=\mathbb{E}[G]=1/p(\beta)$. For any $\alpha\in(0,1)$, the Paley-Zygmund inequality gives $\mathbb{P}\big(G\geq \alpha\ell(\beta)\big)\geq \frac{(1-\alpha)^2}{2-p(\beta)}$. This further implies
\begin{align*}
    \mathbb{P}\big(L_i\geq \alpha\ell(\beta)\big)
    &\geq \mathbb{P}\left(\big\{G>\alpha\ell(\beta)\big\}\cap\big\{L_M\leq \sqrt{n}-1\big\}\right)\\
    &\geq \big(1-\theta_n(\beta)\big)\frac{(1-\alpha)^2}{2-p(\beta)}.\numberthis\label{EQ:Chebyshev_length_LB}
\end{align*}


The last step in the analysis of Phase 1 is to show that \eqref{EQ:Chebyshev_length_LB} implies there are many contigs of length at least $\ell(\beta)$. We extend the notation of $L_i$ to all $i\in[\sqrt{n}]$, by setting $L_i=0$ for all $i$ with $X_{t_0}^{(t_0)}\big((i,1)\big)=+1$.
The number of contigs whose length is at least $\alpha\ell(\beta)$ is $N:=\sum_{i=1}^{\sqrt{n}}\mathds{1}_{\big\{L_i>\alpha\ell(\beta)\big\}}$, and by \eqref{EQ:Chebyshev_length_LB}, we have
\begin{equation}
    \mathbb{E}[N]=\sum_{i=1}^{\sqrt{n}}\mathbb{P}\Big(L_i>\alpha\ell(\beta)\Big)\geq \frac{\big(1-\theta_n(\beta)\big)(1-\alpha)^2}{2-p(\beta)}\sqrt{n}.\label{EQ:Expected_Number_good_contigs}
\end{equation}
Furthermore, since $N \leq \frac{\sqrt{n}}{\alpha\ell(\beta)}$, $\mathbb{E}[N^2]$ is upper bounded by $(p^2(\beta)n)/\alpha^2$. Using Paley-Zygmund inequality, we obtain 
\begin{align*}
    &\mathbb{P}\left(N>\gamma \frac{\big(1-\theta_n(\beta)\big)(1-\alpha)^2}{2-p(\beta)}\sqrt{n}\right)\\
    &\qquad\geq(1-\gamma)^2\frac{\alpha^2(1-\alpha)^4}{p^2(\beta)\big(2-p(\beta)\big)^2},\quad\forall\gamma\in(0,1).\numberthis\label{EQ:Number_good_contigs_LB}
\end{align*}


Based on \eqref{EQ:Number_good_contigs_LB} 
we approximate from below the expected number of pluses at time $2t_0$ by
\begin{align*}
    \mathbb{E}\left[\hat{N}_\mathcal{B}^{(+)}(2t_0)\right]&\geq \mathbb{P}\big(N\geq \gamma \mathbb{E}[N]\big)\mathbb{E}\Big[\hat{N}_\mathcal{B}^{(+)}(2t_0)\Big|N>\gamma\mathbb{E}N\Big]\\
    &\geq \mathbb{P}\Big(N\geq \gamma \mathbb{E}N\Big)\cdot\gamma\mathbb{E}[N] \cdot\mathbb{E}\Big[\hat{N}^{(+)}_{\ell_0}(t_0)\Big],
\end{align*}
where $\hat{N}^{(+)}_{\ell_0}(t_0)$ is the number of pluses that survived for $t_0$ time in a single contig of length $\ell_0:=\alpha\ell(\beta)$ during the 2nd phase of the dynamics. Plugging \eqref{EQ:Expected_Number_good_contigs}-\eqref{EQ:Number_good_contigs_LB} into the above, we obtain
\begin{align*}
    &\mathbb{E}\left[\hat{N}_\mathcal{B}^{(+)}(2t_0)\right]\\
    &\geq (1-\gamma)^2\frac{\alpha^2(1-\alpha)^4}{p^2(\beta)\big(2-p(\beta)\big)^2}\cdot \gamma\frac{\big(1-\theta_n(\beta)\big)(1-\alpha)^2}{2-p(\beta)}\\
    &\mspace{255mu}\times\sqrt{n}\cdot \mathbb{E}\Big[\hat{N}^{(+)}_{\ell_0}(t_0)\Big]\\
    &=\gamma(1-\gamma)^2\frac{\alpha^2(1-\alpha)^6\big(1-\theta_n(\beta)\big)}{p^2(\beta)\big(2-p(\beta)\big)^3}\sqrt{n}\cdot \mathbb{E}\Big[\hat{N}^{(+)}_{\ell_0}(t_0)\Big].\numberthis\label{EQ:1Phase_done_Exp_LB}
\end{align*}

\indent{2) \underline{Erosion Phase}:} It remains to analyze the expected value of $\hat{N}^{(+)}_{\ell_0}(t_0)$. During the second phase, each contig is eaten from both sides with speed $\phi(1)$. For any $\lambda\in(0,1)$, consider the probability:
\begin{align*}
    &\mathbb{P}\left(\hat{N}^{(+)}_{\ell_0}(t_0)>\lambda\ell_0\right)\\
    &\geq 1-\mathbb{P}\left(\left\lceil\frac{1-\lambda}{2}\ell_0\right\rceil\ \mbox{pluses eaten from each side}\right).\numberthis\label{EQ:2Phase_Prob}
\end{align*}
The event that exactly $k_0:=\left\lceil(1-\lambda)\ell_0/2\right\rceil$ pluses from a given side have flipped in $t_0$ time is $\big\{\sum_{i=1}^{k_0}T_i\leq t_0\big\}$, where the $T_i$'s are i.i.d. exponential random variables with $\mathbb{E}T_i=1/\phi(1)$. Observe that
\begin{align*}
    \frac{1}{2}k_0\mathbb{E}[T_1]&\geq\frac{1-\lambda}{4}\ell_0\big(1+e^{-2\beta}\big)\\
    &=\frac{(1-\lambda)\alpha}{4p(\beta)}\big(1+e^{-2\beta}\big)\\
    &\stackrel{(a)}\geq \frac{(1-\lambda)\alpha}{4}e^{(2-c)\beta}\big(1+e^{-2\beta}\big),\numberthis\label{EQ:lambda_constraint}
\end{align*}
where (a) is because $t_0=e^{c\beta}$ and $p(\beta)=1-e^{-\phi(-1)t_0}\leq e^{-(2-c)\beta}$. Recalling that $c<1$, we see that for any $\beta$ sufficiently large $t_0\leq \frac{1}{2}k_0\mathbb{E}[T_1]$. Consequently, we obtain
\begin{align*}
    \mathbb{P}&\left(\left\lceil\frac{(1-\lambda)}{2}\ell_0\right\rceil\ \mbox{pluses eaten from each side}\right)\\
    &\qquad\qquad\leq  \mathbb{P}\left(\sum_{i=1}^{k_0}T_i\leq t_0\right)\\
    &\qquad\qquad\leq\mathbb{P}\left(\sum_{i=1}^{k_0}T_i\leq \frac{1}{2}k_0\mathbb{E}[T_1]\right)\\
    &\qquad\qquad\leq e^{-\kappa \frac{\alpha(1-\lambda)}{2p(\beta)}}.\numberthis\label{EQ:Exponential_tail_bound}
\end{align*}
where the last step bounds the tail probability of a sum of exponential random variables (see, e.g., \cite[Theorem 5.1 Part (iii)]{Janson_Geo_Tail2018}) for $\kappa=\left(\ln2 -1/2\right)>0$. Denoting the RHS of \eqref{EQ:Exponential_tail_bound} by $s(\beta)$ and inserting back to \eqref{EQ:2Phase_Prob}, we have
\begin{equation}
    \mathbb{P}\left(\hat{N}^{(+)}_{\ell_0}\big(t_0\big)>\lambda\ell_0\right)\geq  1-s(\beta).\label{EQ:2Phase_Prob_final}
\end{equation}

Collecting the pieces and proceeding from \eqref{EQ:1Phase_done_Exp_LB}, we have
\begin{align*}
    &\mathbb{E}\left[\hat{N}_\mathcal{B}^{(+)}(2t_0)\right]\\
    &\geq \gamma(1-\gamma)^2\frac{\alpha^2(1-\alpha)^6\big(1-\theta_n(\beta)\big)}{p^2(\beta)\big(2-p(\beta)\big)^3}\sqrt{n}\cdot \mathbb{E}\Big[\hat{N}^{(+)}_{\ell_0}\big(t_0)\big)\Big]\\
    &\stackrel{(a)}\geq\gamma(1-\gamma)^2\frac{\alpha^2(1-\alpha)^6\big(1-\theta_n(\beta)\big)}{p^2(\beta)\big(2-p(\beta)\big)^3}\sqrt{n}\cdot\big(1-s(\beta)\big)\lambda\ell_0\\
    &\stackrel{(b)}\geq \gamma(1-\gamma)^2\frac{\alpha^2(1-\alpha)^6\big(1-\theta_n(\beta)\big)}{8p^3(\beta)\big(2-p(\beta)\big)^3}\big(1-s(\beta)\big)\lambda\sqrt{n}\numberthis\label{EQ:Both_Phases_final}
\end{align*}
where (a) uses \eqref{EQ:2Phase_Prob_final}, while (b) substitutes $\ell_0=\alpha/p(\beta)$ and uses $(2-p(\beta))<2$. To conclude the proof we tune the parameters $\alpha$, $\gamma$ and $\lambda$ as follows. Let $\gamma=p^2(\beta)\in(0,1)$ (see \eqref{EQ:Number_good_contigs_LB}) and set $\lambda=8p(\beta)/\alpha^2$. Now increase $\beta$ sufficiently so that $\lambda\in(0,1)$. Since $\lambda$ can be made arbitrarily small with $\beta$, under this choice $t_0\leq 0.5k_0\mathbb{E}[T_1]$ still holds (see \eqref{EQ:lambda_constraint}). Together with \eqref{EQ:sufficient_one_strip}, the bound from \eqref{EQ:Both_Phases_final} gives
\begin{equation}
    \mathbb{E}\mspace{-2mu}\left[\mspace{-2mu}N_1^{(+)}(t_0)\mspace{-2mu}\right]\mspace{-3mu}\geq\mspace{-3mu} \left(1\mspace{-2mu}-\mspace{-2mu}p^2(\beta)\right)^2\mspace{-2mu}\big(1-\theta_n(\beta)\big)\mspace{-3mu}\big(1-s(\beta)\big)\mspace{-2mu}(1-\alpha)^6\mspace{-2mu}\sqrt{n},\label{EQ:Both_Phases_super_final}
\end{equation}
for $t_0=e^{c\beta}$ and any $c<1$.  

To complete the proof of \eqref{EQ:One_Strip}, fix any $c'\in(0,1)$ and recall that $p(\beta)$ and $s(\beta)$ decay to 0 as $\beta$ increases (independently of $n$), and that $\lim_{n\to\infty}\theta_n(\beta)=0$, for all $\beta>0$. Setting $\alpha\in(0,1)$ as a sufficiently small constant, taking $\beta$ large enough and adjusting $n$ produces the result. 





\section{Proof of Lemma \ref{LEMMA:Hopf_general}}\label{APPEN:Hopf_general_proof}

Since $\big\{\mathcal{C}_i(\sigma)\big\}_{i\in[L]}$ are at graph distance of at least 2 from one another, it suffices to show that 
\begin{equation}
    N_u^+(\sigma_i) - N_u^-(\sigma_i) + 2 N_d^+(\sigma_i) - 2N_d^-(\sigma_i) = 4\label{EQ:Hopf_general_i}
\end{equation}
for each $i\in[L]$, where $\sigma_i\in\Omega_{\mathbb{Z}^2}$ is the configuration that agrees with $\sigma$ on $\mathcal{C}_i(\sigma)$ and is $-1$ everywhere else.

To evaluate the LHS of \eqref{EQ:Hopf_general_i} let $\gamma_i(s)$, $s\in[0,s_f]$, be a smooth counter-clockwise parameterization\footnote{The particular parameterization in use is of no consequence.} of $\gamma(\sigma_i)$ which slightly rounds its corners. We have that $\gamma_i(s)$ is a continuously differentiable curve with 
$\gamma_i(0)=\gamma_i(s_\mathsf{f})$.

Consider the 2-dimensional Gauss map from $\gamma_i(s)$ to the unit circle, described by the trajectory of the unit vector $v(s)=\frac{\gamma_i'(s)}{|\gamma_i'(s)|}$. As one travels along $\gamma_i(s)$, the $v(s)$ vector travels on the circle. The trajectory of $v(s)$ is mostly constant with quick sweeps happening when we approach the slightly smoothed corners of $\gamma_i(s)$. Sweeps of $v(s)$ are reveal unstable or disagreeing plus- or minus-labeled sites in $\sigma_i$ as follows. 
\begin{itemize}
\item a single sweep in the counter-clockwise direction corresponds to a site in $N_u^+(\sigma_i)$;
\item a double sweep in the counter-clockwise direction corresponds to a site in $N_d^+(\sigma_i)$;
\item a single sweep in the clockwise direction corresponds to a site in $N_u^-(\sigma_i)$;
\item a double sweep in the clockwise direction corresponds to a site in $N_d^-(\sigma_i)$.
\end{itemize}
The only way a triple sweep can occur is if $\mathcal{C}_i(\sigma)$ is just a single site of positive spin (in which case we have four counter-clockwise consecutive sweeps). This case is excluded by the assumption that $\big|\mathcal{C}_i(\sigma)\big|\geq 2$.

Since the index of the simple closed curve is 1, overall the vector $v(t)$ has to complete exactly one full rotation on a unit circle. Hopf's Umlaufsatz thus produces \eqref{EQ:Hopf_general_i}, which, in turn, implies the result of Lemma \ref{LEMMA:Hopf_general}.

\section{Proof of Corollary \ref{CORR:Hopf}}\label{APPEN:Hopf_proof}

The result is trivial for the final configuration (i.e., when $(\bar{X}_t)_{t\geq 0}$ stops). Indeed, denoting this configuration by $\zeta\in\Omega_{\mathbb{Z}^2}$, where $\zeta\big((i,j)\big)=+1$ if and only if $i,j\geq \ell+1$, we have $\mathcal{N}_u^+(\zeta)=\big\{(\ell+1,\ell+1)\big\}$ and $\mathcal{N}_d^+(\zeta)=\mathcal{N}_d^-(\zeta)=\mathcal{N}_u^-(\zeta)=\emptyset$. Now, assuming $\bar{X}_t=\sigma\neq\zeta$, define $\bar{\sigma}\in\Omega_{\mathbb{Z}^2}$ as
\begin{equation*}
    \bar{\sigma}\big((i,j)\big)=\begin{cases}\sigma\big((i,j)\big),& i,j\leq\ell\\
    -1,& \mbox{otherwise}\end{cases},
\end{equation*}
and apply Lemma \ref{LEMMA:Hopf_general} with $L=1$, $\mathcal{C}_1(\bar{\sigma})=\mathcal{C}(\bar{\sigma})=\mathcal{C}(\bar{\sigma})\cap[\ell]^2$ and $\mathcal{C}_0(\bar{\sigma})=\mathbb{Z}^2\setminus\mathcal{C}_1(\bar{\sigma})$. This gives
\begin{equation}
    N_u^+(\bar{\sigma}) - N_u^-(\bar{\sigma}) + 2 N_d^+(\bar{\sigma}) - 2N_d^-(\bar{\sigma}) = 4.\label{EQ:Hopf_general_applied}
\end{equation}

Let $v_1(\bar{\sigma})$ and $v_2(\bar{\sigma})$ be the bottom-right and top-left vertices in $\mathcal{C}_1(\bar{\sigma})$. We decompose $\gamma(\bar{\sigma})$ into three parts $\gamma_1(\bar{\sigma})$, $\gamma_2(\bar{\sigma})$ and $\gamma_3(\bar{\sigma})$, where: (i) $\gamma_1(\bar{\sigma})$ is a straight vertical segment from $v_1(\bar{\sigma})+\left(1/2,-1/2\right)$ to $\left(\ell+1/2,\ell+1/2\right)$; (ii) $\gamma_2(\bar{\sigma})$ is a straight horizontal segment between $\left(\ell+1/2,\ell+1/2\right)$ and $v_2(\bar{\sigma})+\left(-1/2,1/2\right)$; and (iii) $\gamma_3(\bar{\sigma})$ connects $v_2(\bar{\sigma})+\left(-1/2,1/2\right)$ back to $v_1(\bar{\sigma})+\left(1/2,-1/2\right)$ by downwards and rightwards moves over segments of positive integer length. Note that every flippable site in $\bar{\sigma}$ is adjacent from the top of from the right to a portion of $\gamma_3(\bar{\sigma})$. In particular, the shape of $\gamma_3$ implies that $\mathcal{N}_d^+(\bar{\sigma})=\mathcal{N}_d^-(\bar{\sigma})=\emptyset$, which  further gives
\begin{equation}
 \mathcal{N}_d^+(\sigma)=\mathcal{N}_d^-(\sigma)=\emptyset.\label{EQ:LB_gamma_analysis_emptyset}   
\end{equation}

Finally, observe that $N_u^+(\bar{\sigma})=N_u^+(\sigma)+3$, which follows because $\bar{\sigma}$ sets all the positive spins in $\sigma$ that are in the 1st quadrant outside of $[\ell]^2$ to $-1$. This adds 3 sites (namely, $v_1(\sigma)$, $v_2(\sigma)$ and $(\ell,\ell)$) to $\mathcal{N}_u^+(\bar{\sigma})$ that did not exist in the original $\mathcal{N}_u^+(\sigma)$. Inserting that, along with \eqref{EQ:LB_gamma_analysis_emptyset}, into \eqref{EQ:Hopf_general_applied} completes the proof.

\section{Proof of Lemma \ref{LEMMA:Regularity}}\label{APPEN:Regularity_proof}

Claim (1), that $\mathcal{N}_d^+(\tilde{X}_t)=\mathcal{N}_d^-(\tilde{X}_t)=\emptyset$, is immediate. Indeed, $\mathcal{N}_d^+(\tilde{X}_t)=\emptyset$ holds due to the application of $\mathsf{E}$ after each flip, while $\mathcal{N}_d^-(\tilde{X}_t)=\emptyset$ is a consequence of  $\mathcal{N}_d^-(\tilde{\rho})=\emptyset$ and the evolution pattern.

To prove the 2nd statement, we show that for any $t\geq 0$, if $\gamma(\tilde{X}_t)$ is simple and closed, then $\gamma(\tilde{X}_t^v)$ is simple and closed, for any flippable site $v\in\mathcal{N}_u(\tilde{X}_t)$. First note that $\gamma(\tilde{X}_0)=\gamma(\tilde{\rho})$ is simple and closed. Next, fix any $t\geq 0$ and assume $\tilde{X}_t=\sigma$ is such that $\gamma(\sigma)$ is simple and closed. Let $v\in\mathcal{N}_u(\sigma)$ and consider the following. 

If $v\in\mathcal{N}_u^-(\sigma)$, we have $\mathcal{D}(\sigma)\subseteq\mathcal{D}(\sigma^v)$, which implies that $\gamma(\sigma^v)$ stays simple and closed. To account for $v\in\mathcal{N}_u^+$, we define the top, bottom, right and left frames of $\mathcal{C}(\tilde{X}_0)$ as 
\begin{align*}
&\mathcal{F}_t:=\big\{(i,k+1)\big\}_{i\in[-k:k]}\quad,\quad \mathcal{F}_b:=\big\{(i,-k-1)\big\}_{i\in[-k:k]},\\ &\mathcal{F}_r:=\big\{(k+1,j)\big\}_{j\in[-k:k]}\quad,\quad \mathcal{F}_l:=\big\{(-k-1,j)\big\}_{j\in[-k:k]},
\end{align*}
respectively. Furthermore, define $\mathcal{T}(\sigma)$, $\mathcal{B}(\sigma)$, $\mathcal{R}(\sigma)$, and $\mathcal{L}(\sigma)$ as the set $\left\{v\in\mathcal{C}(\sigma):\,v=\argmin_{u\in\mathcal{C(\sigma)}}d(u,A)\right\}$ with $A$ replaced by $\mathcal{F}_t$, $\mathcal{F}_b$, $\mathcal{F}_r$, and $\mathcal{F}_l$, respectively. 
These sets contain the sites in $\mathcal{C}(\sigma)$ closest to each of the above defined frames. The corners of these sets are
\begin{align*}
    v_t^{(r)}&:=\argmin_{u\in\mathcal{T}(\sigma)}d\big(u,\mathcal{R}(\sigma)\big)\ ;\ v_t^{(l)}:=\argmin_{u\in\mathcal{T}(\sigma)}d\big(u,\mathcal{L}(\sigma)\big)\\
    \ v_b^{(r)}&:=\argmin_{u\in\mathcal{B}(\sigma)}d\big(u,\mathcal{R}(\sigma)\big)\ ;\ v_b^{(l)}:=\argmin_{u\in\mathcal{B}(\sigma)}d\big(u,\mathcal{L}(\sigma)\big)\\
    v_r^{(t)}&:=\argmin_{u\in\mathcal{R}(\sigma)}d\big(u,\mathcal{T}(\sigma)\big)\ ;\ v_r^{(b)}:=\argmin_{u\in\mathcal{R}(\sigma)}d\big(u,\mathcal{B}(\sigma)\big)\\
    v_l^{(t)}&:=\argmin_{u\in\mathcal{L}(\sigma)}d\big(u,\mathcal{T}(\sigma)\big)\ ;\  v_l^{(b)}:=\argmin_{u\in\mathcal{L}(\sigma)}d\big(u,\mathcal{B}(\sigma)\big).
    \end{align*}
Note that since $\mathcal{N}_d^+(\sigma)=\emptyset$, these corners are unique and different from one another. 

\begin{figure}[t!]
	\begin{center}
		\begin{psfrags}
			\psfragscanon
		    \psfrag{B}[][][1]{$\mspace{-3mu}k$}
		    \psfrag{C}[][][1]{$\mspace{-3mu}-k$}
		    \psfrag{D}[][][1]{$\mspace{-3mu}-k$}
		    \psfrag{E}[][][1]{$\mspace{-4mu}v_r^{(t)}$}
		    \psfrag{F}[][][1]{$\mspace{-10mu}v_t^{\mspace{-2mu}(\mspace{-2mu}r\mspace{-2mu})}$}
		    \psfrag{G}[][][1]{$\mspace{15mu}v_t^{(l)}$}
		    \psfrag{H}[][][1]{$\mspace{4mu}v_l^{\mspace{-3mu}(t)}$}
		    \psfrag{I}[][][1]{$\ \ \ v_l^{(b)}$}
		    \psfrag{J}[][][1]{$\ \ v_b^{(l)}$}
		    \psfrag{K}[][][1]{$v_b^{(r)}$}
		    \psfrag{L}[][][1]{$\mspace{-9mu}v_r^{(b)}$}
		    \psfrag{P}[][][1]{$\ \ \ \ \ \ \ \ {\color{green!65!black}\bm{\gamma_{r\to t}(\sigma)}}$}
		    \psfrag{Q}[][][1]{$\mspace{-18mu}{\color{red!65!black}\bm{\gamma_{t\to l}(\sigma)}}$}
		    \psfrag{R}[][][1]{$\mspace{-18mu}{\color{NavyBlue!65!black}\bm{\gamma_{l\to b}(\sigma)}}$}
		    \psfrag{S}[][][1]{$\ \ \ \ \ \ \ \ \ \  {\color{Orange!95!black}\bm{\gamma_{b\to r}(\sigma)}}$}\includegraphics[scale = .6]{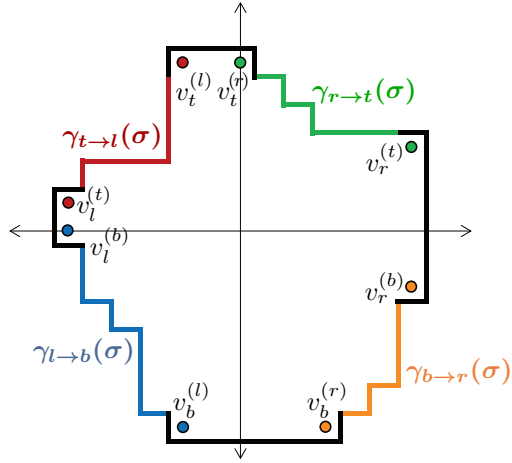}
			\caption{The portions $\gamma_{r\to t}(\sigma)$, $\gamma_{t\to l}(\sigma)$, $\gamma_{l\to b}(\sigma)$ and $\gamma_{b\to r}(\sigma)$ are shown by the green, red, blue and orange curves, respectively. The remaining pieces of $\gamma(\sigma)$ are colored in black. The corner of $\mathcal{R}(\sigma)$, $\mathcal{T}(\sigma)$, $\mathcal{L}(\sigma)$ and $\mathcal{B}(\sigma)$ that define the aforementioned pieces of $\gamma(\sigma)$ are shown by circles  colored as the corresponding piece of $\gamma(\sigma)$.} \label{FIG:gamma_parts}
			\psfragscanoff
		\end{psfrags}
	\end{center}
	\vspace{-3mm}
\end{figure}

We consider the following four pieces of the curve $\gamma(\sigma)$ (See Fig. \ref{FIG:gamma_parts} for an illustration):
\begin{itemize}
    \item $\gamma_{r\to t}(\sigma)$ is the piece from $v_r^{(t)}+\left(-1/2,1/2\right)$ to $v_t^{(r)}+\left(1/2,-1/2\right)$.
    
    \item $\gamma_{t\to l}(\sigma)$ is the piece from $v_t^{(l)}-\left(1/2,1/2\right)$ to $v_l^{(t)}+\left(1/2,1/2\right)$.
    
    \item $\gamma_{l\to b}(\sigma)$ is the piece from $v_l^{(b)}+\left(1/2,-1/2\right)$ to $v_b^{(l)}+\left(-1/2,1/2\right)$.
    
    \item $\gamma_{b\to r}(\sigma)$ is the piece from $v_b^{(r)}+\left(1/2,1/2\right)$ to $v_r^{(b)}-\left(1/2,1/2\right)$.
\end{itemize}


Lastly, we define the sets of vertices outlined by each piece of $\gamma(\sigma)$. Let:
\begin{align*}
    \mathcal{C}_{r\to t}(\sigma)&:=\big\{v\in\mathcal{C}(\sigma):\,\left|\mathcal{B}_\infty\left(v,1/2\right)\cap\gamma_{r\to l}(\sigma)\right|>1\big\}\\
    \mathcal{C}_{t\to l}(\sigma)&:=\big\{v\in\mathcal{C}(\sigma):\,\left|\mathcal{B}_\infty\left(v,1/2\right)\cap\gamma_{t\to l}(\sigma)\right|>1\big\}\\
    \mathcal{C}_{l\to b}(\sigma)&:=\big\{v\in\mathcal{C}(\sigma):\,\left|\mathcal{B}_\infty\left(v,1/2\right)\cap\gamma_{l\to b}(\sigma)\right|>1\big\}\\
    \mathcal{C}_{b\to r}(\sigma)&:=\big\{v\in\mathcal{C}(\sigma):\,\left|\mathcal{B}_\infty\left(v,1/2\right)\cap\gamma_{b\to r}(\sigma)\right|>1\big\}.
    \end{align*}
To clarify, the set $\mathcal{C}_{r\to t}(\sigma)$, for instance, contains all vertices whose surrounding $L^\infty$ ball of radius $1/2$ contributes at least one side to $\gamma_{r\to t}(\sigma)$ (in which case the cardinality of the intersection is that of the continuum). However, we want to exclude the corners $v_t^{(r)}$ and $v_r^{(t)}$ from belonging to $\mathcal{C}_{r\to t}(\sigma)$, which holds because, e.g., $\mathcal{B}_\infty\left(v_t^{(r)},1/2\right)\cap\gamma_{r\to t}(\sigma)=\big\{v_t^{(r)}+\left(1/2,-1/2\right)\big\}$. With these definitions, the internal boundary of $\mathcal{C}(\sigma)$ uniquely decomposes as
\begin{align*}
    \partial_\mathsf{Int}\mathcal{C}(\sigma)=\mathcal{C}_{r\to t}(\sigma)&\uplus\mathcal{T}(\sigma)\uplus\mathcal{C}_{t\to l}(\sigma)\uplus\mathcal{L}(\sigma)\\
    &\uplus\mathcal{C}_{l\to b}(\sigma)
    \uplus\mathcal{B}(\sigma)\uplus\mathcal{C}_{b\to r}(\sigma)\uplus\mathcal{R}(\sigma).
\end{align*}

We are now ready to prove that $\gamma(\sigma^v)$ is simple and connected, for any $v\in\mathcal{N}_u^+(\sigma)$. Fix $v\in\mathcal{N}_u^+(\sigma)$ and assume, without loss of generality, that $v\in\mathcal{C}_{r\to t}(\sigma)\uplus\mathcal{T}(\sigma)$. Two cases are considered.

First, assume  $v\in\mathcal{T}(\sigma)$ and $|\mathcal{T}(\sigma)|=2$. Recalling that  $\mathcal{T}(\sigma)$ always contains $v_t^{(r)}$ and $v_t^{(l)}$, we have $\mathcal{T}(\sigma)=\big\{v_t^{(r)},v_t^{(l)}\big\}$. Assume $v=v_t^{(r)}$ and denote $u:= v_t^{(l)}$. Flipping $v$ will cause $u$, for which $\sigma(u)=+1$, to disagree with the negative spins of 3 of its 4 neighbors in $\sigma^v$. As $\mathcal{N}_d^+(\sigma)=\emptyset$, we get $\mathcal{N}_d^+(\sigma^v)=\{u\}$. By definition of the auxiliary dynamics, since $\mathcal{N}_d^+(\sigma^v)$ is non-empty and contains a single element, the transformation $\mathsf{E}$ is then be applied. This produces $\mathsf{E}(\sigma^v)=(\sigma^v)^u$ as the next state of the dynamics. Since  $\mathcal{C}\big((\sigma^v)^u\big)=\mathcal{C}(\sigma)\setminus\mathcal{T}(\sigma)$, the resulting curve $\gamma\big((\sigma^v)^u\big)$ remains connected and simple.

The remaining case is when $v\in\mathcal{T}(\sigma)$ with $|\mathcal{T}(\sigma)|\geq3$ or $v\in\mathcal{C}_{r\to t}(\sigma)$. These two scenarios are treated together because both satisfy $\mathsf{E}(\sigma^v)=\sigma^v$. To see that $\gamma(\sigma^v)$ is simple and connected, assume that the contrary is true. This can happen only if $\sigma\big(v-(1,1)\big)=-1$ (any other situation contradicts that fact that the original $\gamma(\sigma)$ is closed and simple). Since $\sigma(v)=+1$, having a negative spin at $v-(1,1)$ implies that $\sigma(u)=-1$, for all $u=v-(i,j)$ with $i,j\geq 1$. Furthermore, since $v\in\mathcal{N}_u^+(\sigma)$ with $\sigma\big(v+(1,0)\big)=\sigma\big(v+(0,1)\big)=-1$, we have that $\sigma(w)=-1$ for all $w\in\big\{v+(i,j):\,i,j\geq 0\big\}\setminus\{v\}$.

\begin{figure}[t!]
	\begin{center}
		\begin{psfrags}
			\psfragscanon
		    \psfrag{A}[][][1]{$\mspace{6mu}k$}
		    \psfrag{B}[][][1]{$\mspace{10mu}k$}
		    \psfrag{C}[][][1]{$\mspace{-10mu}-k$}
			\psfrag{D}[][][1]{$\mspace{-6mu}-k$}
		   
			\includegraphics[scale = .5]{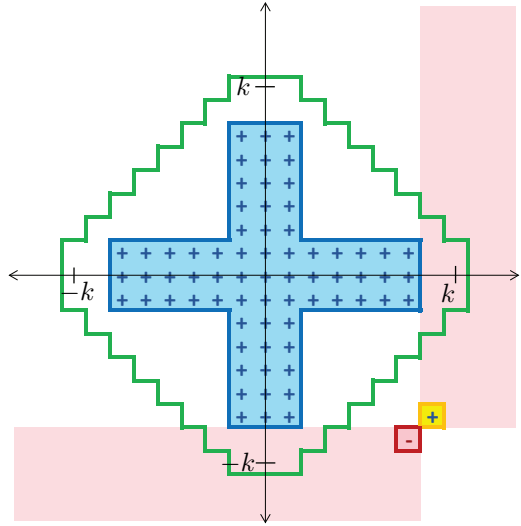}
			\caption{An illustration of the obtained contradiction for $k=8$, $m=3k/4=6$.The site $v$ (here shown for case (i) when $i_v\geq m+1$ and $j_v\leq -m$) and $u$ are outlined by the yellow and red borders, respectively. The pink regions enclose sites that must have a negative spin in order for $v\in\mathcal{N}_u^+(\sigma)$ and $\sigma(u)=-1$ to hold. The original rhombus-shaped plus-labeled component $\mathcal{C}(\tilde{\rho})$ is enclosed by the green curve. The set $\mathcal{T}_m$ is shown in blue.} \label{FIG:droplet_contradiction}
			\psfragscanoff
		\end{psfrags}
	\end{center}
	\vspace{-3mm}
\end{figure}

By definition of the $(\tilde{X}_t)_{t\geq 0}$ dynamics, we have  $\mathcal{T}_m\subseteq\mathcal{C}(\sigma)$.\footnote{Recall: $\mathcal{T}_m=\big\{(i,j)\big\}_{\substack{i\in[-m:m]\\j\in[-1:1]}}\cup\big\{(i,j)\big\}_{\substack{i\in[-1:1]\\j\in[-m:m]}}$, with $m=3k/4$.} Denoting $v=(i_v,j_v)$, the fact that $\mathcal{T}_m\subseteq\mathcal{C}(\sigma)$ implies that one of the following two assertions hold: (i) $i_v\geq m+1$ and $j_v\leq -m$; or (ii) $i_v\leq -m$ and $j_v\geq m+1$. However, as $m=\frac{3}{4}k$, in both cases we get that $v\notin\mathcal{C}(\tilde{\rho})$, which is a contradiction. This idea is illustrated in Fig.~\ref{FIG:droplet_contradiction}. This concludes the proof that if $\gamma(\sigma)$ is a simple and closed curve in $\mathbb{R}^2$, then so is $\gamma(\sigma^v)$, for all $v\in\mathcal{N}_u(\sigma)$.


For Claim (3), since $$\mathcal{M}_u^+(\sigma)\subseteq\left\{v_r^{(b)},v_r^{(t)},v_t^{(r)},v_t^{(l)},v_l^{(t)},v_l^{(b)},v_b^{(l)},v_b^{(l)}\right\},$$
we have $\big|\mathcal{M}_u^+(\sigma)\big|\leq 8$. If $\mathcal{M}_u^+(\sigma)\neq\emptyset$, the fact that flipping $v\in\mathcal{M}_u^+(\sigma)$ results in $\big|\mathcal{C}\big(\mathsf{E}(\sigma^v)\big)\big|=\big|\mathcal{C}(\sigma)\big|-2$ was already made while proving Claim~(2). 

Lastly, we show that $\mathcal{N}_1^+(\sigma)=\emptyset$ (see \eqref{EQ:droplet_LB_N1set}). Thus, we need to demonstrate that there are no unstable plus-labeled sites with disagreeing neighbors located only horizontally (to the right and left) or only vertically (above and below). This follows by combining Claims (1) and (2). Fix $t\geq 0$ and let $\tilde{X}_t=\sigma$. By Claim (2), $\gamma(\sigma)$ is simple and closed, and, in particular, its interior $\mathcal{D}(\sigma)$ is a connected set. Assume in contradiction that there exists $v\in\mathcal{N}_1^+(\sigma)$. Without loss of generality further assume that $v\in\mathcal{C}_{r\to t}(\sigma)\uplus\mathcal{T}(\sigma)$, which implies that $\sigma\big(v+(1,0)\big)=\sigma\big(v-(1,0)\big)=-1$. As argued in the proof of Claim (2), $\sigma\big(v+(1,0)\big)=-1$ implies that $\sigma\big(v+(1,0)+(i,j)\big)=-1$, for all $i,j\geq 0$. Similarly, because $\sigma\big(v-(1,0)\big)=-1$, we have $\sigma\big(v-(1,0)+(i,j)\big)=-1$, for all $i\leq0$ and $j\geq 0$. Denote $v=(i_v,j_v)$ and let $j_t=\max\big\{j\geq j_v:\,(i_v,j)\in\mathcal{C}(\sigma)\big\}$ (note that this set is non-empty because it always contains $v+(0,1)$). The site $(i_v,j_t)\in\mathcal{C}(\sigma)$, but it has three neighbors with negative spin (to its right, its left and above it). Therefore, $(i_v,j_t)\in\mathcal{N}_d^+(\sigma)$, which contradicts $\mathcal{N}_d^+(\sigma)=\emptyset$.

\bibliographystyle{IEEEtran}
\bibliography{ref}

\end{document}